\documentclass{article}

\usepackage[utf8]{inputenc}
\usepackage[T1]{fontenc}

\usepackage{mathpazo} 

\usepackage[margin=1.0in]{geometry}
\usepackage[bottom]{footmisc}

\usepackage{graphicx}
\usepackage[skip=0pt, labelfont=bf, font=footnotesize]{caption}
\usepackage[capposition=top]{floatrow}
\usepackage{tikz}

\usepackage{booktabs}
\usepackage{longtable}
\usepackage{tabularx}
\usepackage{threeparttable}
\usepackage{threeparttablex}
\usepackage{makecell}
\usepackage{etoolbox}
\BeforeBeginEnvironment{tabular}{\small}

\newcounter{magicrownumbers}
\newcommand\rownumber{\stepcounter{magicrownumbers}\arabic{magicrownumbers}. }

\usepackage{pdflscape} 

\usepackage{authblk}

\usepackage{xcolor}
\usepackage{nameref}

\usepackage{booktabs} 
\usepackage{nicefrac}  
\usepackage{microtype}
\usepackage[utf8]{inputenc}

\usepackage{wrapfig}
\usepackage{multirow}

\usepackage{amssymb}
\usepackage{amsthm}
\usepackage{bbm}
\usepackage{bbm}
\usepackage{array}
\usepackage[colorlinks=true,urlcolor=purple,
linkcolor=purple,citecolor=purple]{hyperref}
\usepackage{amsmath}
\usepackage{mathtools}
\usepackage{float}

\usepackage{lscape}
\usepackage{graphicx}

\usepackage{booktabs}
\usepackage{longtable}
\usepackage{threeparttable}
\usepackage{threeparttablex}

\usepackage{mathrsfs}
\usepackage{graphicx}
\usepackage{color}
\usepackage[numbers,sort&compress]{natbib}

\usepackage{algorithm}
\usepackage{algpseudocode}
\usepackage{thmtools,thm-restate} 

\makeatletter
\def\paragraph{\@startsection{paragraph}{4}%
  \z@\z@{-\fontdimen2\font}%
  {\normalfont\bfseries}}
\makeatother

\usepackage{tikz}
\usetikzlibrary{decorations.markings, backgrounds, calc, positioning, shapes, shadows, arrows, fit, automata}
\usepackage{tikz-cd}
\usetikzlibrary{arrows.meta}
\tikzset{>={Latex}}

\newcommand{\beginsupplement}{%
        \setcounter{table}{0}
        \renewcommand{\thetable}{S\arabic{table}}%
        \setcounter{figure}{0}
        \renewcommand{\thefigure}{S\arabic{figure}}%
     }

\newcounter{desccount}

\newcommand{\descref}[1]{\hyperref[#1]{#1}}

\renewcommand{\P}{\mathbb{P}}

\renewcommand{\emptyset}{\varnothing}

\newcommand{\im}{\operatorname{im}}

\newcommand{\rank}{\operatorname{rank}}

\makeatletter
\newcommand{\pushright}[1]{\ifmeasuring@#1\else\omit\hfill$\displaystyle#1$\fi\ignorespaces}
\newcommand{\pushleft}[1]{\ifmeasuring@#1\else\omit$\displaystyle#1$\hfill\fi\ignorespaces}
\makeatother

\newcommand{\revision}[1]{\textcolor{black}{#1}}


\newcommand{\equalcontrib}{\textsuperscript{*}}

\title{Opening Knowledge Gaps Drives Scientific Progress}

\author[1]{Kara Kedrick\equalcontrib}
\author[2,3]{Wenlong Yang\equalcontrib}
\author[4]{Thomas Gebhart}
\author[2]{Yang Wang\textsuperscript{\dag}}
\author[5]{Russell J. Funk\textsuperscript{\ddag}}

\affil[1]{Institute for Complex Social Dynamics, Carnegie Mellon University}
\affil[2]{School of Public Policy and Administration, Xi'an Jiaotong University}
\affil[3]{College of Public Administration, Huazhong University of Science and Technology}
\affil[4]{Computer Science and Engineering, University of Minnesota}
\affil[5]{Carlson School of Management, University of Minnesota}

\date{}

\usepackage{adjustbox}

\begin{document}
\maketitle

\begingroup
\renewcommand{\thefootnote}{}
\footnotetext{We thank the National Science Foundation (grant nos. 1829168, 1932596, 2318172, and 2449660 to R.J.F.; grant no. 2449659 to K.K.), the Alfred P. Sloan Foundation (grant no. G-2024-25123 to R.J.F. and T.G.), and the National Natural Science Foundation of China (grant no. 72004177 to Y.W.) for financial support of work related to this project. R.J.F.\ is also grateful for many helpful conversations related to this work with Lori Ziegelmeier, Jason Owen-Smith, Tanisha Dodla, Nadezhda Dominguez Salinas, Gavin Engelstad, Jingyi Guan, Floyd Liu, Lucia Luo, Frances McConnell, Tam Nguyen, Adam Schroder, and Ethan Scheelk.}
\endgroup

\begingroup
\renewcommand{\thefootnote}{}
\footnotetext{\textsuperscript{\dag}Corresponding author: \texttt{yang.wang@xjtu.edu.cn}}
\footnotetext{\textsuperscript{\ddag}Corresponding author: \texttt{rfunk@umn.edu}}
\footnotetext{\textsuperscript{*}These authors contributed equally.}
\endgroup

\begin{abstract}
Knowledge production is often viewed as an endogenous process in which discovery arises through the recombination of existing theories, findings, and concepts. Yet not all recombinations are equally valuable, and identifying those that may prove most generative remains challenging. We propose that a crucial form of recombination occurs when linking concepts creates knowledge gaps---empty regions in the conceptual landscape that focus scientific attention on unexplored connections. Using computational topology, we develop a method to identify knowledge gaps in science at scale. In this framework, gaps form when new connections complete cycles in evolving networks of scientific concepts. We apply the method to 34,363,623 papers in Microsoft Academic Graph spanning 1900--2020. The main analyses use samples ranging from 8,643,571 to 31,983,201 papers. Compared with papers that introduce novel concept pairs without opening gaps, gap-opening papers have greater long-term citation impact, are more disruptive of prior work, and exhibit greater delayed recognition. Gap-opening papers receive fewer citations within five years of publication but more over twenty years, consistent with scientific communities needing time to recognize and develop newly opened research territories. Evidence for the generative capacity of gap-opening papers---as indicated by high citation impact and disruption---replicates in five additional databases (OpenAlex, the PubMed Knowledge Graph, Dimensions, American Physical Society journals, and United States patents) and across eight concept vocabularies, including several derived directly from text. Together, these results reveal a distinctive generative role for gap opening in scientific discovery.

\end{abstract}

\vspace{1em}

\pagebreak

\section*{Significance Statement}
When scientists connect concepts in novel ways, some combinations prove far more generative than others. We focus on combinations that create structural gaps in the knowledge landscape---empty regions that focus attention on unexplored connections. Using computational topology, we introduce a method to identify these gaps and apply it to millions of papers spanning 120 years. Papers that open gaps have greater long-term citation impact and are more disruptive than papers that combine concepts without opening one. Together, these findings identify gap opening as a structurally distinctive and unusually generative form of recombination.
\pagebreak

\section{Introduction}
 
Scientific knowledge is widely thought to grow through an endogenous process in which existing theories, empirical findings, and conceptual frameworks lay the foundation for future discoveries \citep{kuhn1962,lakatos_musgrave_1970,laudan1978}. Central to this advancement are scientific works that recombine existing ideas in novel ways \citep{weitzman1998recombinant,uzzi2013atypical,hargadon1997technology,foster2015tradition,rzhetsky2015choosing, fleming2001}. Yet not all recombinations are equally valuable for advancing understanding. While some combinations have limited influence, others fundamentally reshape what is known, opening new research territories or resolving longstanding questions. Prior research has shown, for instance, that the most highly cited papers often emerge when atypical combinations are strategically integrated with conventional ones \citep{uzzi2013atypical}, suggesting that atypicality alone does not guarantee exceptional scientific impact. What, then, distinguishes highly generative conceptual recombinations---those that stimulate sustained future research---from those with more limited influence?

We propose that a crucial form of recombination occurs when linking concepts creates knowledge gaps---empty regions in the conceptual landscape that focus scientific attention on unexplored connections and signal promising directions for investigation. This perspective builds on the widespread recognition that knowledge gaps play a central role in scientific practice. Researchers routinely frame their contributions in terms of addressing ``gaps in the literature,'' ``gaps in understanding,'' or ``gaps in knowledge'' (e.g., \citep{costin1971,LaVail1972,hall2002,bodelier2004}), suggesting that scientists view the identification and filling of such conceptual voids as valuable work. Yet how specific papers create such gaps within the broader landscape of knowledge remains largely unstudied.

In the study of social networks, sociologists have long recognized that the absence of a connection can be as consequential as its presence. Weak ties link otherwise disconnected groups and give their members access to novel information \citep{granovetter1973strength,hansen1999search}; brokers who span the structural holes between clusters gain advantages in information access and control \citep{burt1992structural,burt2004structural,ahuja2000collaboration,fleming2007collaborative,hargadon1997technology,obstfeld2005social}; and in small-world networks, a single tie between distant clusters can dramatically shorten paths throughout the network \citep{watts1999small,uzzi2005collaboration}. These traditions largely study the connection itself---the bridging tie and the position of the actors who hold it. We study instead what a connection encloses. When a first-time pairing of concepts closes a cycle, the interior of that cycle stays unfilled---a gap---and it is this gap, not any single tie, that marks an unexplored region of the conceptual landscape and attracts subsequent research. We identify when and where such gaps form as networks evolve, directly and at scale, using tools from computational topology.

Several theoretical perspectives suggest that gap opening may represent a particularly influential form of recombination. When papers identify knowledge gaps, they may delineate clear boundaries between what is known and what remains unexplored, strengthening connections among existing ideas while signaling that particular areas remain unfilled. Psychological research supports this view, indicating that the recognition of knowledge gaps triggers cognitive processes essential to scientific discovery, including curiosity and information seeking \citep{Loewenstein1994,Kang2009,kedrick2023}. Once gaps are identified, scientists may prioritize research aimed at addressing them \citep{Sharot2020,Golman2018,Torunsky2025}. Additionally, the formation of gaps may indicate that the concepts along their boundaries form coherent, meaningful units. Defining these boundaries can minimize the cognitive effort required to understand how ideas relate to one another, offering robust frameworks upon which other researchers can build without redefining fundamental relationships \citep{deJong2010}.

However, despite these theoretical reasons to expect gap-opening papers to be influential, the empirical question remains open. For gap-opening papers to be generative---stimulating future research and new trajectories---the gaps they create must be both recognized and valued by the scientific community. This recognition may vary across fields; for instance, the physical sciences tend to maintain broadly shared lists of unsolved problems, making newly opened gaps easier to locate within existing priorities, whereas the social sciences are characterized by weaker consensus about which questions most need answering \citep{cole1994}. Moreover, the significance of newly opened knowledge gaps may not be immediately apparent because they create unexplored territories that lack established evaluation criteria, whereas other recombinatory work can be more readily assessed within existing frameworks. The value of gap-opening work may emerge only over time, as communities develop tools to explore these new territories and recognize the opportunities they present for contributions.

To empirically investigate whether gap-opening papers demonstrate distinctive generative capacity, we analyze 34,363,623 journal articles and conference proceedings from Microsoft Academic Graph (MAG) spanning 1900 to 2020. We model scientific knowledge as evolving networks of concepts \citep{kedrick2024cp,hofstra2020paradox}, where concepts are represented by Microsoft Academic Graph's level-3 fields---fine-grained research topics within broader scientific disciplines---and concepts are connected when they first co-occur in the same paper. We assess robustness to data source and taxonomy by repeating the analysis in five additional databases---OpenAlex, the PubMed Knowledge Graph, Dimensions, American Physical Society (APS) journals, and United States patents---using eight concept vocabularies ranging from curated hierarchies to concepts extracted directly from text.

To systematically identify papers that open genuine knowledge gaps, we introduce a method using persistent homology, a technique from topological data analysis that detects and measures topological features across different scales in data. Persistent homology has been successfully applied across diverse domains including neuroscience, dynamical systems analysis, and machine learning \citep{reimann2017cliques,gardner2022toroidal,myers2019,bhaskar2020,carriere2020perslay,love2023topological}. When applied to concept networks, this technique can identify holes---regions where concepts remain unconnected despite being linked to common neighboring concepts, precisely the structural signature of knowledge gaps. Persistent homology allows us to systematically detect when new concept combinations create such holes, distinguishing papers that open topologically significant gaps from those that simply introduce novel concept pairs.

Our analysis reveals that gap-opening papers have greater long-term citation impact, are more disruptive---subsequent work builds on them while relying less on their predecessors---and exhibit greater delayed recognition than both papers introducing novel concept pairs without opening gaps and papers making no novel concept pairs. The citation-impact and disruption results persist when the comparison is restricted to papers introducing novel concept pairs and replicate across five additional databases and eight concept vocabularies. Among gap openers, larger gaps are followed by more research drawing on the concepts that define their boundaries and remain open longer. Gap-opening papers also emerge disproportionately from smaller, less established teams working in close proximity. Taken together, these patterns highlight the distinctive \emph{generative capacity} of gap-opening papers. By linking the structural properties of concept networks to long-term impact, our study shows that opening knowledge gaps is a key mechanism driving scientific progress.

\section{Results}

\subsection{Prevalence and characteristics of gap-opening papers}

\begin{figure}[!htbp]
 \centering
 \includegraphics[width=\textwidth]{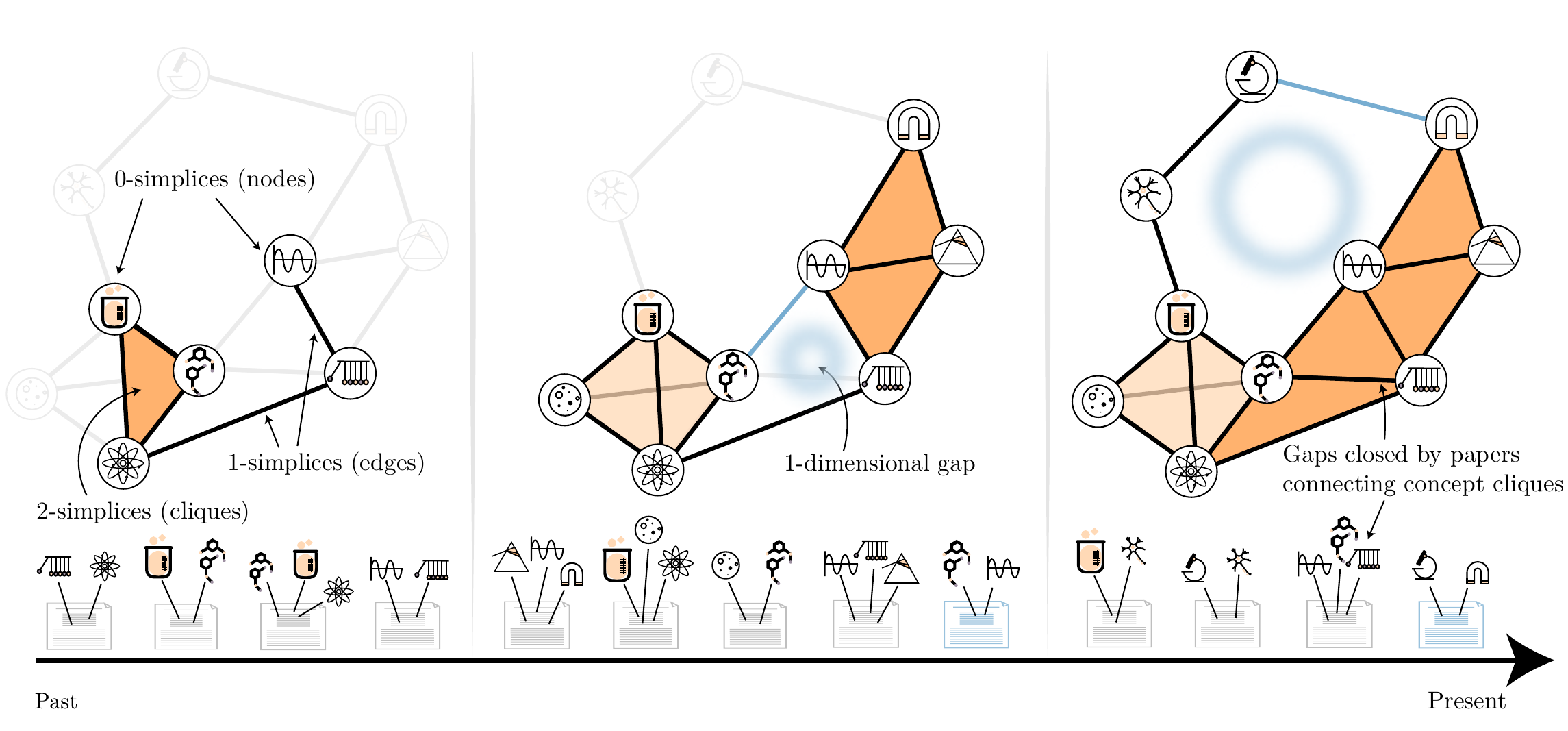}
 \caption{\textbf{How papers create topological gaps in concept networks.} A concept network is shown across three time points, with the underlying structure held fixed (gray) to highlight key structural changes. Each node represents a concept, and edges connect concepts that co-occur in the same paper. New edges and filled cliques are highlighted progressively in black and orange, while gap-opening edges are shown in blue. The left panel depicts an early stage of network development, with simple pairwise links representing novel concept pairs. As additional papers connect previously unlinked concepts, the network evolves to include more complex structures, including cliques (orange triangles) and gaps (cycles, highlighted by blue circles). A paper opens a gap when it introduces an edge (blue) that completes a cycle without filling its interior. In the middle and right panels, \emph{gap-opening papers} give rise to one-dimensional gaps (cycles). Papers that are \emph{non-gap openers (with novel pair)} are those that first introduce the black edges into the network. By contrast, \emph{non-gap openers (without novel pair)} combine concepts that are already connected.}
 \label{fig:1}
\end{figure}

To identify papers that open knowledge gaps and assess their generative capacity, we analyzed the topological evolution of concept networks over time. Fig.~\ref{fig:1} illustrates this process, showing how concepts initially connect through simple pairwise links representing locally novel concept pairs. As new papers link previously unconnected concepts, the network evolves to exhibit more complex structures, including cliques (i.e., fully connected subgroups of concepts) and cycles (i.e., closed loops where concepts are connected in a ring). Three mutually connected concepts form a filled triangle, while four form a solid tetrahedron. Such structures are a natural consequence of continued recombination among concepts. Our study focuses on papers that create cycles, or knowledge gaps. The middle panel in Fig.~\ref{fig:1} highlights a blue edge, which represents a combination introduced by a gap-opening paper. This combination opens a gap by establishing a non-local connection, giving rise to topologically significant features---specifically, a hole or gap manifested as a hollow quadrilateral. Similarly, in the right panel, another blue edge denotes a combination introduced by a gap-opening paper, and this combination results in the formation of a gap represented by a hollow pentagon.

We categorize papers into three types. \emph{Gap openers} introduce new links between two concepts that create gaps in the concept network (Fig.~\ref{fig:1}). \emph{Non-gap openers (with novel pair)} recombine two existing concepts in new ways, but these combinations do not create gaps in the concept network. \emph{Non-gap openers (without novel pair)} do not recombine existing concepts in novel ways. Note that both gap openers and non-gap openers (with novel pair) represent forms of knowledge recombination, yet gap openers introduce structurally significant connections that alter the topology of the concept network.

\revision{To illustrate, consider four papers our method identifies as gap openers. Adleman's \citeyearpar{adleman1994} demonstration that DNA molecules could solve an instance of the Hamiltonian path problem brought molecular biology to bear on a combinatorial problem, opening the field of DNA computation; a \emph{PNAS} commentary described the experiment as marking ``the birth of a new field'' \citep{chen2000}. Iijima's \citeyearpar{iijima1991} report established carbon nanotubes as a major object of modern materials research, catalyzing sustained work on their synthesis, properties, and technological applications \citep{devolder2013}. Marcus et al.'s \citeyearpar{marcus1993} Penn Treebank brought statistical language analysis together with large-scale syntactic annotation, providing shared data infrastructure that helped drive the emergence of data-intensive computational linguistics. Binnig, Quate, and Gerber's \citeyearpar{binnig1986} invention of the atomic force microscope extended nanoscale imaging beyond conductive materials, making a broad range of physical and biological systems accessible to direct study and becoming one of the foundational instruments of nanoscience and nanotechnology \citep{giessibl2003}. Substantively, each paper opened a durable line of inquiry; topologically, each introduced a link that completed a cycle in its field's concept network.}

\revision{The Adleman and Marcus cases show this topological structure more concretely. Within MAG's computer science network, earlier work formed an indirect path from computational work on DNA through DNA sequencing, image processing, and Boolean circuits to research on Hamiltonian paths. Adleman's paper directly connected the two endpoints, closing this five-concept path into a cycle. The gap persisted for just over two years, until Ogihara and Ray's ``Simulating Boolean circuits on a DNA computer'' connected DNA computing directly to Boolean circuits in 1997, adding a link across the cycle's interior and closing the gap \citep{ogihara1997}. Similarly, before the Penn Treebank appeared, MAG records an indirect path from statistical language tagging through bigram models and natural-language understanding to large annotated corpora. Marcus et al.'s paper directly connected those endpoints, closing this four-concept path into a cycle. In both cases, the resulting cycle surrounded an unfilled region of the concept network---a knowledge gap.}

\revision{The key distinction is therefore not whether a paper connects distant areas. A bridging connection may span distant areas without closing a pre-existing path into a cycle. In both examples, the endpoints of the new connection were already connected indirectly; adding the direct connection closed that path into a loop. Gap opening depends on this topological position, not on dyadic novelty alone.}

\begin{figure}[htbp]
 \centering
 \includegraphics[width=\textwidth]{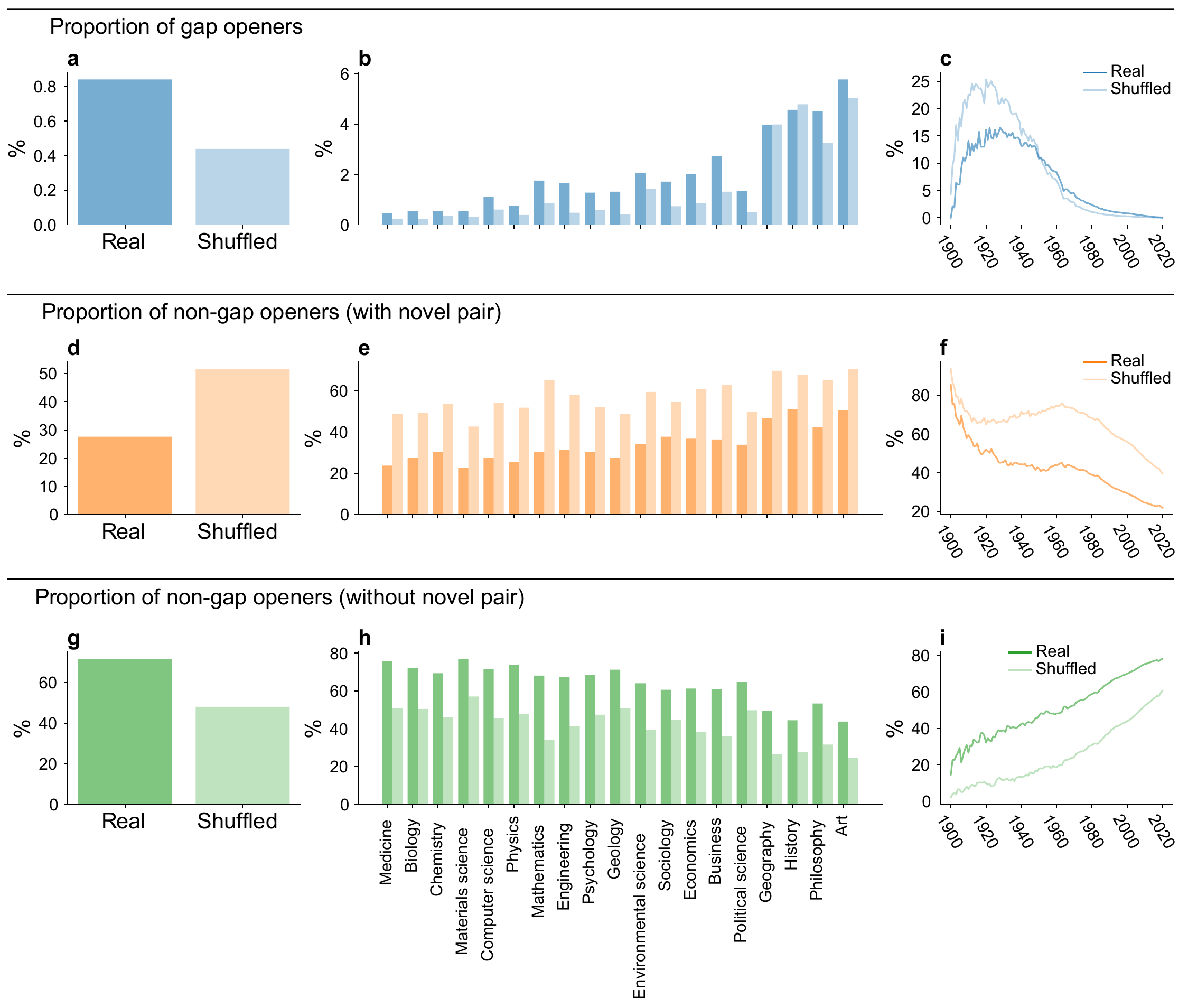}
 \caption{\textbf{Prevalence of gap-opening papers: comparisons across fields, time, and random baselines.} \textbf{(a,d,g)} Proportion of each paper type in real data (dark bars) versus randomized networks (light bars): (a) gap openers (n = 289,041), (d) non-gap openers (with novel pair) (n = 9,484,465), (g) non-gap openers (without novel pair) (n = 24,590,117). \textbf{(b,e,h)} Proportions across 19 scientific fields for (b) gap openers (see Fig.~\ref{fig:s1} for detailed analysis of cross-field variation), (e) non-gap openers (with novel pair), (h) non-gap openers (without novel pair). \textbf{(c,f,i)} Trends from 1900--2020 for (c) gap openers, (f) non-gap openers (with novel pair), (i) non-gap openers (without novel pair), showing real data (dark curves) versus randomized baselines (light curves).}
 \label{fig:2}
\end{figure}

\textbf{\emph{How prevalent are gap-opening papers?}} Gap openers account for only 0.84\% of all articles, yet this proportion is twice that of a random model in which concept labels are shuffled across papers (Fig.~\ref{fig:2}a). This suggests that gap formation is not a byproduct of network growth alone. Non-gap openers (with novel pair) make up more than 25\% of all articles, yet their proportion is nearly half that of the random case (Fig.~\ref{fig:2}d). The remaining papers---non-gap openers (without novel pair)---constitute over 70\% of the corpus, compared to nearly 50\% in the random case (Fig.~\ref{fig:2}g). These proportions are broadly consistent across fields (Fig.~\ref{fig:2}b,e,h), though notable variation exists---the share of gap openers is lowest in Medicine, Biology, and Chemistry and highest in Geography, History, Philosophy, and Art.

\revision{The cross-field variation in prevalence of gap openers (Fig.~\ref{fig:2}b) appears to reflect how different fields produce and absorb new knowledge rather than differences in innovative activity. Fast-moving fields such as Medicine and Computer Science generate large absolute numbers of gap openers, but their much larger publication volumes yield small proportions; slower-moving fields such as History and Art publish far less, so gap openers make up a larger share of their corpora (Fig.~\ref{fig:s1}a). Gaps also close more quickly in fast-moving fields (Fig.~\ref{fig:s1}b), reflecting how quickly fields fill in new connections. This pattern is consistent with well-established differences in how knowledge accumulates across fields---from the cumulative, sequential progress characteristic of the physical and biological sciences to the more pluralistic, open-ended development of the social sciences and humanities \citep{biglan1973,popper1963,kolb1981,cole1994,cole1995}. Importantly, gap openers are more highly cited than within-field controls in most fields (Fig.~\ref{fig:s1}d), consistent with gap opening capturing a meaningful contribution across fields of differing pace and scale. Full results, including analyses of publication volume, gap duration, concept counts, and within-field citation comparisons, appear in the SI Appendix.} 

The proportion of gap openers in both real and random cases grows substantially from 1900 to 1930, followed by a steady decline through 2020 (Fig.~\ref{fig:2}c). Non-gap openers (with novel pair) follow a distinct trajectory, first decreasing, then rebounding, and finally decreasing again (Fig.~\ref{fig:2}f), while non-gap openers (without novel pair) increase steadily over the full period (Fig.~\ref{fig:2}i). The rarity and decline of gap-opening papers is consistent with several, potentially complementary, explanations. As scientific fields mature, concept networks become denser and publication volumes grow, which increases the difficulty of opening new gaps while accelerating the rate at which existing gaps are closed---the same dynamics that shape cross-field variation in the prevalence of gap openers (Fig.~\ref{fig:s1}a). At the same time, the decline aligns with broader evidence of slowing scientific progress \citep{Park2023PapersAP, cowen2019rate, bloom2020ideas, jones2009burden, arbesman2011quantifying, cowen2011great, gordon2017rise,wu2026innovation}, particularly given that these papers demonstrate distinctive generative capacity (as we show below).

\begin{figure}[htbp]
 \centering
 \includegraphics[width=\textwidth]{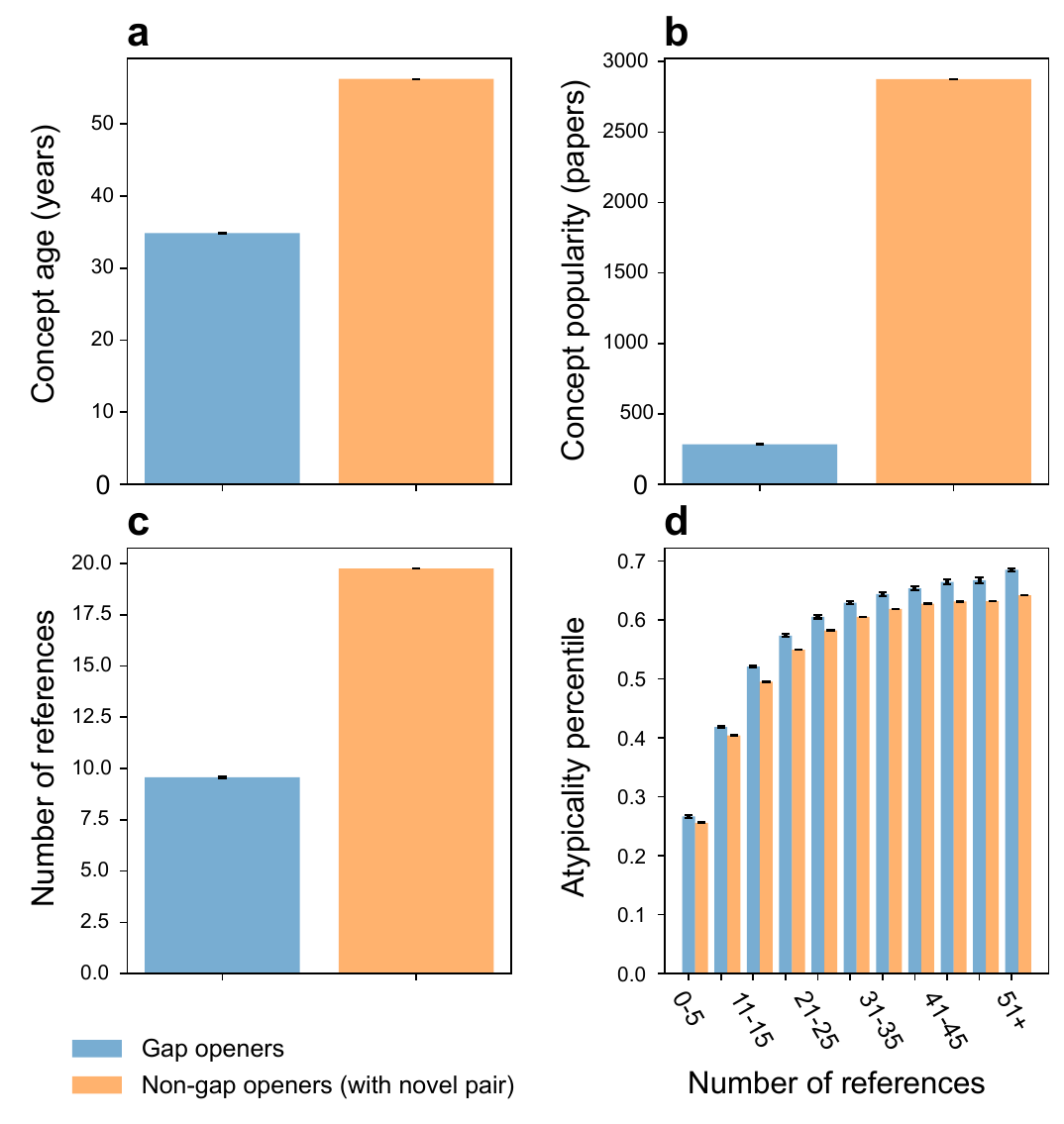}
\caption{\textbf{Characteristics of gap-opening papers: concept age, concept popularity, references, and atypicality.} Comparison of gap openers (blue) and non-gap openers (with novel pair) (orange). \textbf{(a)} Concept age, defined as the average age of the two linked level-3 concepts (n = 9,773,506). \textbf{(b)} Concept popularity, measured as the average number of prior occurrences of the two linked concepts (n = 9,773,506). \textbf{(c)} Number of references (n = 9,773,506). \textbf{(d)} A percentile across different reference counts (n = 6,616,129). Bars show mean values; error bars indicate standard errors. Non-gap openers (without novel pair) are excluded from these comparisons.}
\label{fig:3}
\end{figure}

\textbf{\emph{What distinguishes gap-opening papers?}} Beyond prevalence, gap-opening papers differ systematically from non-gap openers (with novel pair) in the types of concepts they connect and how they recombine them. Gap openers bridge newer, less established concepts, reflected in lower mean concept age and fewer prior occurrences of the concepts they link (Fig.~\ref{fig:3}a,b). Concepts linked by gap-opening papers also experience a greater proportional rise in popularity than those linked by non-gap openers, both five and ten years after publication (Fig.~\ref{fig:s2}), suggesting that gap opening helps draw attention to previously peripheral ideas. Gap-opening papers cite fewer prior works (Fig.~\ref{fig:3}c), consistent with their emergence from less densely connected areas of the literature. They also exhibit higher atypicality, as measured by the A~index \citep{uzzi2013atypical}, even after controlling for the number of references (Fig.~\ref{fig:3}d). \revision{Finally, gap-opening papers frequently use verbs signaling creation and reframing in their titles, whereas non-gap openers tend to use verbs related to demonstration, improvement, and exploitation (Fig.~\ref{fig:s3}).} Taken together, these patterns suggest that gap-opening papers are more than novel recombinations---they connect younger, less established concepts in structurally distinctive ways that reshape the knowledge landscape.

\subsection{Generative capacity of gap-opening papers}

\begin{figure}[htbp]
 \centering
 \includegraphics[width=\textwidth]{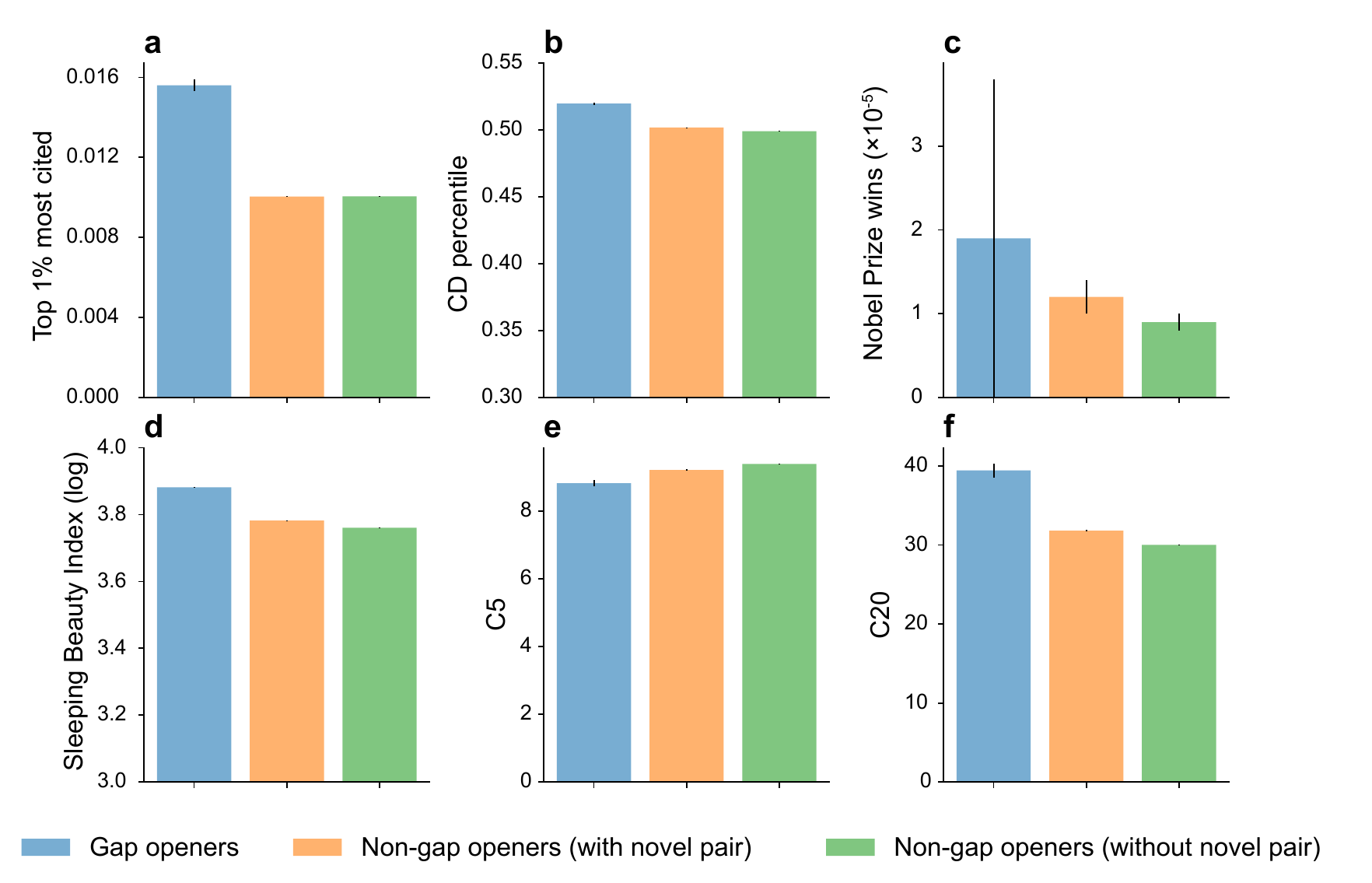}
\caption{\textbf{Generative capacity of gap-opening papers: citation impact, disruption, Nobel Prize wins, and delayed recognition.} Predicted values from regression models comparing gap openers (blue), non-gap openers (with novel pair) (orange), and non-gap openers (without novel pair) (green). \textbf{(a)} Top 1\% most cited (n = 31,983,201), papers published 1980--2020 with team sizes $\leq$20. \textbf{(b)} CD percentile (n = 21,613,648), same restrictions as (a) plus $\geq$1 reference and non-missing index values. \textbf{(c)} Nobel Prize wins (n = 8,857,237). The smaller sample reflects the rarity of Nobel-linked papers, which causes perfect or quasi-complete separation in the logistic regression; observations whose covariate combinations perfectly predict the outcome are dropped from estimation. \textbf{(d)} Sleeping Beauty (SB) index, log-transformed (n = 25,537,277), same restrictions as (a) plus non-missing index values. \textbf{(e,f)} C5 and C20, cumulative citations received within 5 and 20 years, respectively (n = 8,643,571), for papers published 1980--2001 with team sizes $\leq$20 and non-missing values for both citation measures. Error bars indicate standard errors. Full regression results are reported in Table~\ref{table:s1}; variable definitions and estimation details appear in \emph{Materials and Methods}.}
\label{fig:4}
\end{figure}

We next ask whether gap-opening papers stimulate more future research and new scientific directions. To assess this generative capacity, we examine three dimensions of how subsequent research responds to a paper---citation impact, disruption as measured by the CD percentile, and delayed recognition as measured by the Sleeping Beauty (SB) index. We estimate fixed-effects regression models that compare the three paper types across these dimensions. All models control for team size and number of references \citep{Wagner2018InternationalRC,wu2019large,Yang2024ExploringTC,wuchty2007increasing,Zeng2021FreshTA}, and---for the CD percentile and SB index---citation count \citep{Ke2015DefiningAI,wu2019large,Funk2017ADN}, and include year and field fixed effects. Fig.~\ref{fig:4} plots the predicted values from these models and Table~\ref{table:s1} reports the full regression results; variable definitions and estimation details appear in \emph{Materials and Methods}.

\textbf{\emph{Do gap-opening papers stimulate more future research?}} Gap openers are 57\% more likely than non-gap openers (without novel pair) to rank among the top 1\% most cited works among papers published in the same year and field (Fig.~\ref{fig:4}a, odds ratio [OR] = 1.57, $p<0.001$), indicating substantially greater citation impact. Non-gap openers (with novel pair), by contrast, do not differ significantly from baseline papers in their probability of ranking among the top 1\% most cited in this analysis (Fig.~\ref{fig:4}a, $p>0.05$), though they do in several replication corpora (see below), and are actually \emph{less} likely than baseline papers to appear among the top 5\% through 20\% most cited (Table~\ref{table:s2}). Gap openers also score 2.1 percentile points higher on the CD percentile (Fig.~\ref{fig:4}b, $p<0.001$), indicating that subsequent papers are more likely to cite them without also citing their predecessors. \revision{These contrasts also hold when each gap opener is compared only with observably similar papers, matched on field, year, team size, and reference count (\emph{SI Appendix}, Table~\ref{tab:matched-sampling}).}

\revision{To better understand the relationship between gap opening and disruptiveness, we conducted two additional analyses. First, atypicality and disruption have been shown to be negatively related \citep{li2025innovation, cui2022aging}. The intuition is that atypical recombination typically draws on ideas from distant domains, making the resulting work unlikely to displace any single predecessor \citep{li2025innovation}. To assess whether gap opening follows this pattern, we classify papers into four atypicality $\times$ disruption categories (atypical-and-disruptive, atypical-only, disruptive-only, and neither; see Table~\ref{table:s3} for definitions). Gap openers align most strongly with the combined atypicality-and-disruption category, consistent with the broader pattern in our results that gap opening represents a distinctively generative form of recombination---one that introduces new combinations and is more likely to displace rather than consolidate prior work. Formal Wald tests and separate margin regressions decompose this joint result (\emph{SI Appendix}, Tables~\ref{tab:multinomial}--\ref{tab:margins}).}

\revision{Second, prior work finds that work combining prior knowledge in unusual ways often receives fewer citations in the short run, as the new directions such work opens take time to attract a following \citep{wang2017bias}. To assess whether gap openers score higher on disruption even in the short run, we estimate regression models predicting the CD index and CD percentile as a function of paper type separately for each citation window and find that the gap-opener coefficients remain positive across all time horizons (Fig.~\ref{fig:s6}). Two of the gap openers highlighted previously illustrate this finding concretely, achieving both high citation impact and high disruption within five years of publication---Adleman's \citeyearpar{adleman1994} DNA computing (253 citations, CD percentile 99.1) and Marcus et al.'s \citeyearpar{marcus1993} Penn Treebank (280 citations, CD percentile 97.9). Iijima's \citeyearpar{iijima1991} carbon nanotubes shows the complementary pattern. Its disruption registered over a longer horizon (CD percentile 95.8 at ten years) even though the paper was heavily cited from the start (602 citations within five years), consistent with the delayed recognition we document below.}

For Nobel Prize-winning work, we find a positive but imprecise estimate (Fig.~\ref{fig:4}c, OR = 2.09), though the result is not statistically significant due to the small number of Nobel-linked papers and should be interpreted with caution. 

\revision{While our results highlight the generative capacity of gap-opening papers, gap opening is only one part of a broader ecology of discovery, and our classification is not a ranking of importance. If gap openers create new spaces in the knowledge landscape, one important role, among others, that non-gap openers may play is to build within spaces that already exist---extending, refining, or generalizing established lines of work. Tversky and Kahneman's \citeyearpar{tversky1992advances} ``Advances in Prospect Theory,'' for instance, is among the most cited papers in behavioral economics; our method classifies the paper as a non-gap opener (with novel pair), and as the title signals, the paper extends the authors' original 1979 prospect theory, including by refining the weighting function, rather than opening a new field. Similarly, Bollerslev's \citeyearpar{bollerslev1986generalized} GARCH model---Generalized Autoregressive Conditional Heteroskedasticity---is one of the most widely used tools in financial econometrics, yet the paper opens by describing ``a natural generalization of the ARCH process introduced in \citet{engle1982}.'' Both Tversky and Kahneman's paper and Bollerslev's introduced a first-time concept pairing that did not complete a cycle, and so both are non-gap openers (with novel pair). By contrast, the gap openers identified earlier---Adleman's DNA computation, Iijima's carbon nanotubes, Marcus et al.'s Penn Treebank, and Binnig et al.'s atomic force microscope---each opened a durable line of inquiry.}

Taken together, these results identify gap opening as a structurally distinctive and unusually generative form of recombination.

\textbf{\emph{When does this generative capacity emerge?}} The generative capacity of gap-opening papers emerges gradually. Compared to baseline, gap openers receive 6.1\% fewer citations within five years of publication but 31.4\% more within twenty years (Fig.~\ref{fig:4}e,f, both $p<0.001$). Non-gap openers (with novel pair) show a similar but much smaller pattern ($-$1.9\% at five years, $+$6.0\% at twenty years). Gap openers also score significantly higher on the Sleeping Beauty (SB) index (Fig.~\ref{fig:4}d, $\beta = 0.121$, $p<0.001$), confirming a pronounced delayed recognition effect. To pinpoint when the early citation deficit reverses, we estimate separate regressions for each citation window from 1 to 20 years. The transition from negative to positive coefficients occurs at approximately 8 years for both paper types (gap openers and non-gap openers (with novel pair); Fig.~\ref{fig:s4}), after which gap openers show consistently larger positive coefficients. 

\revision{Huang et al.'s~\citeyearpar{huang1991optical} discovery of optical coherence tomography (OCT) illustrates this pattern. Our method identifies this paper as a gap opener that linked tomographic reconstruction with superluminescent diode technology, and the paper scores in the 97th percentile on the SB index among papers in the same field. The paper appeared in \emph{Science}, yet OCT took over a decade to become the standard imaging modality in ophthalmology, and Huang received the Lasker Prize in 2023---over three decades after publication.}

These patterns suggest that gap-opening papers open intellectual territories that require time for scientific communities to explore and build upon, consistent with their greater long-term citation impact and disruption relative to the other paper types.

\subsection{Replication across databases and concept systems}

\begin{figure}[htbp]
 \centering
 \includegraphics[width=\textwidth]{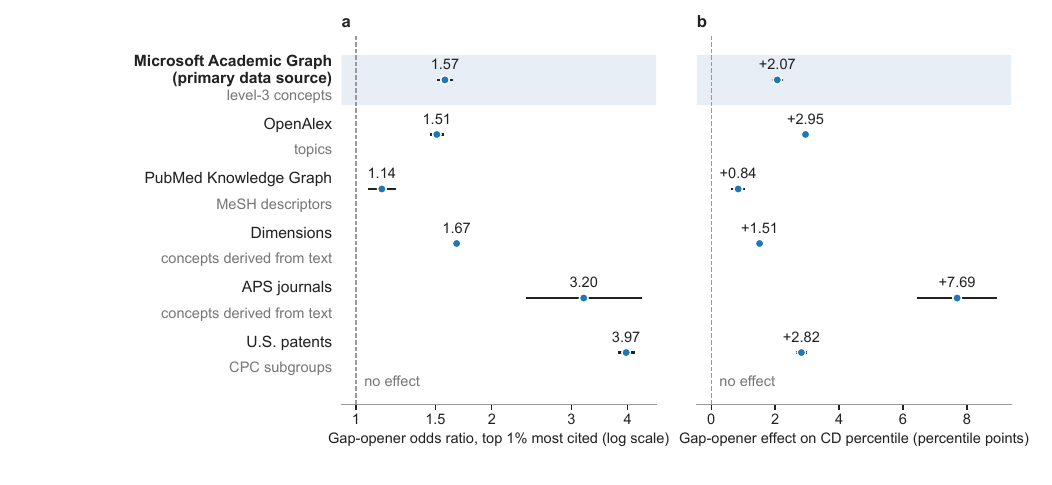}
 \caption{\textbf{Gap-opener effects on citation impact and disruption replicate across six data sources.} Gap-opener estimates for two outcomes in each data source: \textbf{(a)} odds ratio of ranking among the top 1\% most cited (log scale); \textbf{(b)} effect on CD percentile, in percentile points. The shaded top row reports the estimate from our primary data source (Microsoft Academic Graph, level-3 concepts); the remaining rows report replications in OpenAlex (topics), the PubMed Knowledge Graph (MeSH descriptors), Dimensions (concepts derived from text), APS journals (concepts derived from text), and U.S. patents (CPC subgroups). Error bars indicate 95\% confidence intervals. All twelve estimates plotted here are significant at $p<0.001$. Estimation samples for the plotted models range from roughly 420,000 observations (APS) to 101,638,624 (OpenAlex); per-model sample sizes accompany the full regressions, including additional concept vocabularies (APS PACS codes; coarser CPC subclasses), in \emph{SI Appendix}, Tables~\ref{tab:replication-openalex}--\ref{tab:replication-patents}.}
 \label{fig:replication}
\end{figure}

\revision{To assess whether these findings depend on Microsoft Academic Graph or on any particular concept vocabulary, we repeated the analysis in five additional databases---OpenAlex, the PubMed Knowledge Graph, Dimensions, American Physical Society (APS) journals, and United States patents---applying the same persistent-homology procedure in each (\emph{Materials and Methods}). The six data sources span eight concept vocabularies, from curated hierarchies (MAG level-3 concepts, OpenAlex topics, Medical Subject Headings [MeSH] descriptors) and expert classification codes (Physics and Astronomy Classification Scheme [PACS] codes, Cooperative Patent Classification [CPC] codes at two levels of granularity) to concepts extracted directly from paper text (Dimensions, APS text-derived concepts), which draw on no curated taxonomy. Estimation samples range from roughly 420,000 observations in APS journals to 101,638,624 in OpenAlex (\emph{SI Appendix}, Tables~\ref{tab:replication-openalex}--\ref{tab:replication-patents}). Fig.~\ref{fig:replication} reports the gap-opener estimates in each data source for our primary measures of citation impact and disruption (i.e., top 1\% most cited and CD percentile). In every database, gap openers are more likely than baseline papers to rank among the top 1\% most cited and score higher on the CD percentile; all twelve estimates plotted in Fig.~\ref{fig:replication} are significant at $p<0.001$ (\emph{SI Appendix}, Tables~\ref{tab:replication-openalex}--\ref{tab:replication-patents}). Gap openers were also more highly associated with these metrics than non-gap openers (with novel pair) in most instances (exceptions described below). Within MAG, the results remain robust when gap openers are identified using concepts from Levels 3 through 5 rather than Level 3 alone (\emph{SI Appendix}, Tables~\ref{tab:taxonomy-robustness}--\ref{tab:taxonomy-full}).}

\revision{The replication results are highly consistent. Gap openers have greater citation impact and disruption than non-gap openers (without novel pair) in all six data sources and all eight vocabularies. Gap openers also have larger estimates than non-gap openers (with novel pair) in 15 of the 16 outcome-by-vocabulary comparisons. Two of these differences---citation impact in APS under PACS codes and disruption in patents under CPC subgroups---are small. The sole reversal occurs for disruption in PubMed, where the estimates are $+0.8$ percentile points for gap openers and $+1.1$ for non-gap openers (with novel pair). The MAG finding that non-gap openers (with novel pair) do not exceed non-gap openers (without novel pair) in top 1\% citation impact is less general---they have positive estimates in OpenAlex, Dimensions, APS, and patents---but gap openers still have larger estimates in all of these settings except APS under PACS codes, where the estimates are similar. APS estimates also vary in magnitude across its two analyses, with stronger gap-opener associations under text-derived concepts than PACS codes for both citation impact (odds ratio 3.20 versus 1.45) and disruption ($+7.7$ versus $+4.8$ percentile points). Taken together, these results show that gap opening is associated with greater citation impact and disruption across every data source and vocabulary we examine and generally has a stronger association with both outcomes than novel concept pairing alone.}

\subsection{Team composition of gap-opening papers} 

\begin{figure}[htbp]
 \centering
 \includegraphics[width=\textwidth]{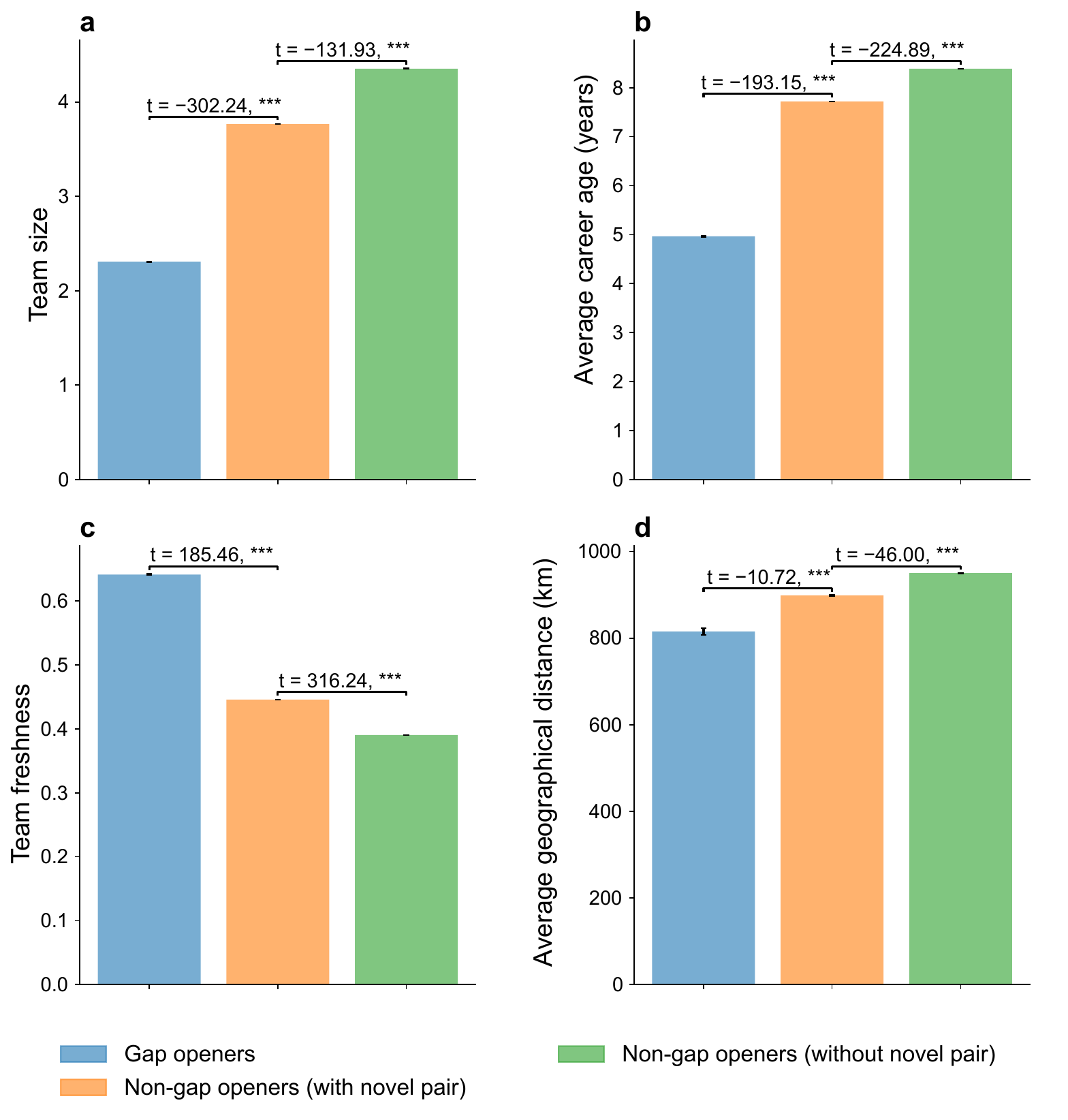}
\caption{\textbf{Collaborative conditions under which gap-opening papers emerge: team size, career age, freshness, and geographical distance.} Gap openers (blue) emerge from smaller teams with lower average career age, fresher collaborations, and greater geographical proximity than non-gap openers (with novel pair) (orange) and non-gap openers (without novel pair) (green). \textbf{(a)} Team size (n = 34,363,623). \textbf{(b)} Average career age of team members (n = 34,363,623). \textbf{(c)} Team freshness, defined as the fraction of team members with no prior collaboration with any other member of the team (n = 28,079,454), restricted to papers with team sizes 2--20. \textbf{(d)} Average geographical distance between team members' affiliations (n = 18,811,771), restricted to papers with $\geq$2 affiliations with recorded coordinates. Bars show mean values; error bars indicate standard errors. \emph{t}-values from independent-sample \emph{t}-tests are shown between paired bars (*** $p<0.001$).}
\label{fig:5}
\end{figure}

We next investigate the collaborative conditions under which gap-opening papers emerge. Gap openers are associated with smaller teams (mean = 2.31 authors), earlier-career researchers (mean career age = 4.96 years since first publication), fresher collaborations (mean = 0.64 on a 0--1 scale measuring the fraction of team members who have not previously collaborated \citep{Zeng2021FreshTA}), and greater geographical proximity (mean = 815 km between affiliations) than the other two paper types (Fig.~\ref{fig:5}, all pairwise t-tests $p<0.001$). These patterns suggest that gap-opening work emerges from specific collaborative conditions---small, early-career teams working in close proximity with unfamiliar collaborators---that differ systematically from those producing other forms of scientific recombination.

\section{Discussion}

Scientific knowledge advances through the recombination of existing theories, findings, and concepts \citep{uzzi2013atypical,hargadon1997technology,foster2015tradition,rzhetsky2015choosing, fleming2001}. Yet given the vast space of potential conceptual combinations, identifying which recombinations prove most generative remains a fundamental challenge. While researchers routinely invoke the importance of addressing ``knowledge gaps,'' no systematic method has previously existed to quantify how specific papers create these gaps within the broader landscape of scientific knowledge. Using persistent homology, we developed an approach to systematically identify papers that create topological gaps in evolving concept networks and to assess whether these papers differ in their capacity to stimulate future research.

Our findings reveal gap opening to be a particularly generative form of conceptual recombination. Gap openers are disproportionately represented among highly cited works (top 1\% to 20\% of citation counts), indicating greater citation impact than both papers that introduce novel concept pairs without opening gaps and papers that make no novel concept pairs. Papers that introduce novel concept pairs without opening gaps show either no increase or a smaller increase in the probability of ranking among the top 1\% most cited relative to baseline papers, and, under Microsoft Academic Graph's level-3 concepts,  are actually less likely to appear among moderately highly cited works (top 5\%--20\%). Gap-opening papers also tend to be more disruptive, with subsequent papers more likely to cite them without also citing their predecessors. Together, these patterns identify gap opening as a structurally distinctive and unusually generative form of recombination.

Gap-opening papers emerge from specific collaborative conditions. Teams that open gaps tend to be smaller, more geographically proximate, and composed of researchers at earlier career stages working with fresher collaborations than teams producing other forms of recombination. These teams also tend to work with younger, less established concepts. The pattern may reflect the greater cognitive flexibility of junior scientists, who are less constrained by established disciplinary frameworks \citep{wiley1998, blech2020, ollinger2008}. Consistent with this interpretation, gap-opening papers frequently employ language associated with creation and reframing in their titles, while non-gap openers more often use terms related to demonstration, improvement, or exploitation. Smaller teams and close proximity may also facilitate the intensive, unstructured collaboration needed to identify genuine knowledge gaps \citep{lin2023,brucks2022,glaeser2023}, as opposed to the more coordinated division of labor characteristic of larger teams \citep{wu2019large}.

The generative capacity of gap-opening papers may stem from their ability to simultaneously highlight what remains unknown and clarify what is known. By creating topological gaps, these papers define new territories for exploration while highlighting previously unrecognized relationships between concepts. Gaps may signal missing connections and unresolved questions, which psychological research suggests can spark curiosity and drive information seeking \citep{Golman2021, Golman2018, kedrick2023, Loewenstein1994, berlyne1960}. At the same time, the connections these papers introduce form the boundaries of topological gaps---closed loops of concepts that delineate coherent regions in the knowledge landscape. By establishing these boundaries, gap-opening papers create multiple pathways between related concepts and reduce dependency on specific connections, potentially enhancing the robustness of conceptual frameworks \citep{Fan2021,Zhou2018}. From a cognitive perspective, establishing clear conceptual boundaries may reduce the mental effort required for researchers to understand how sets of concepts relate to one another, and provide stable foundations for future work without requiring researchers to redefine fundamental relationships \citep{deJong2010}.

The delayed recognition pattern we document further illustrates the generative capacity of gap-opening papers. These works often become ``sleeping beauties'' that ultimately attract substantial scientific attention long after publication \citep{Ke2015DefiningAI}. Our regression models reveal that, compared to papers that do not introduce novel concept pairs, gap openers receive fewer citations during the first five years yet significantly more citations 20 years post-publication. Papers that introduce novel concept pairings without opening gaps exhibit similar, albeit statistically smaller, delays. This pattern may reflect the time required for scientific communities to recognize and explore the new territories opened by gap-opening papers---developing appropriate tools and redirecting attention toward unfamiliar connections \citep{dey2017}.

\revision{
\revision{The replication of the gap-opener effect across six databases and eight concept vocabularies---including vocabularies built without any curated taxonomy---indicates that the phenomenon is a general feature of evolving knowledge networks rather than an artifact of any single data source or classification system.} While we have focused on concept networks in science, the persistent homology approach we introduce applies to any network where edges accumulate over time. Research on structural holes and weak ties established that connections bridging otherwise disconnected groups in social networks create opportunities for information access and control \citep{burt1992structural,burt2004structural,ahuja2000collaboration,fleming2007collaborative,hargadon1997technology,obstfeld2005social}, but this work has primarily identified actors who successfully span such regions. Our method complements these perspectives by mapping the opportunity structure itself, locating where gaps exist independent of whether anyone has filled them. The framework enables analyses not previously possible, such as tracking when gaps emerge, how long they persist, and when subsequent connections close them. Applied to social networks, such an approach could identify brokerage opportunities before actors exploit them, enabling new empirical tests of structural holes theory and richer understanding of how network positions become advantageous.}

Our findings suggest several implications for science policy. Our approach demonstrates that computational tools are beginning to enable systematic identification of underexplored but promising areas of scientific inquiry. As such methods develop, funding agencies and research institutions could potentially use these insights to identify unexploited territories where productive research programs might be developed, informing targeted funding or strategic research investments. 

Equally important, our findings highlight critical limitations in current evaluation systems. Gap-opening papers demonstrate distinctive long-term generative capacity yet experience delayed recognition, receiving fewer citations initially before attracting substantial scientific attention, sometimes decades later. This temporal mismatch suggests that evaluation frameworks relying heavily on short-term metrics may systematically undervalue the most generative contributions to science. Research institutions and funding agencies might consider developing evaluation approaches that better account for the extended timescales over which breakthrough research demonstrates such generative capacity. 

More broadly, the declining proportion of gap-opening papers over time points to a shift in how science generates new research directions. This decline may reflect the natural maturation of knowledge networks---as concept networks become denser and publication volumes grow, opening new gaps becomes more difficult and existing gaps close more rapidly---or a genuine reduction in gap-opening activity, consistent with broader evidence of slowing scientific progress. Disentangling these explanations remains an important direction for future research.

\pagebreak

\section{Materials and Methods}\label{Methods}
\subsection{Data}
In this study, we draw on the Microsoft Academic Graph (MAG), a large-scale dataset containing over 260 million scholarly publications spanning 1800--2021. MAG provides rich metadata for each record, including publication year, venue, scientific domain, author information, and references. For our analysis, we restrict the sample to journal and conference papers published between 1900 and 2020. This time frame balances the need for a sufficiently long historical window with data completeness and reliability---coverage and metadata quality are generally poor before 1900, while papers after 2020 may be incomplete due to indexing delays.

MAG organizes scientific knowledge using a hierarchical field-of-study taxonomy with six levels (0-5), where level-0 represents broad disciplines (e.g., Computer Science, Biology), while higher levels represent increasingly specific subfields and concepts. To further ensure relevance and quality, we include only papers containing at least one level-0 field of study with a positive confidence score, and at least two level-3 fields with confidence scores above zero. The dataset spans 19 distinct level-0 fields, encompassing broad domains such as Medicine, Biology, Chemistry, Materials Science, Computer Science, Physics, Mathematics, Engineering, Psychology, Geology, Environmental Science, Sociology, Economics, Business, Political Science, Geography, History, Philosophy, and Art. After applying these criteria, our final dataset contains 34,363,623 papers.

\subsection{Concept network construction}
We construct dynamic concept networks across time for each scientific domain. In each network, nodes represent level-3 fields of study, which serve as our concepts---fine-grained research topics nested within broader level-0 fields. We construct an undirected edge between two nodes at the publication date of the first paper to which MAG assigns both of the corresponding level-3 concepts. The MAG data assign standardized concept labels to papers across the historical corpus. The labels function as analytical categories and need not reproduce the precise terminology used by authors at publication. \revision{For the illustrative cases, we use historically appropriate descriptions of the ideas represented by MAG's standardized concept labels, based on the underlying publications. Because a gap can be bounded by more than one cycle, the illustrations describe a cycle that makes the historical connections clear, whereas the gap-size analyses use the minimal bounding cycle.} The resulting network therefore measures the first co-assignment of concepts within the MAG classification rather than the first use of particular terms.

\revision{We restrict our analysis to level-3 fields based on three methodological considerations. First, doing so helps to minimize the effect of potential taxonomic artifacts. The MAG taxonomy is hierarchical. Simultaneously analyzing multiple levels could potentially introduce edges with inconsistent meanings, with some indicating genuine conceptual co-occurrence, others simply reflecting the hierarchical structure of the taxonomy. Second, level-3 concepts offer the optimal balance of granularity and coverage; as shown in Fig.~\ref{fig:s5}, they are the most numerous in the hierarchy, capturing specific recombinations that coarser levels (0--2) would obscure. Third, level 3 maximizes signal-to-noise ratio. While levels 4--5 are more granular, they exhibit higher sparsity (i.e., fewer connections between concepts) and potentially lower confidence in automated indexing. Our dataset includes 100,650 unique level-3 concepts in total. On average, each level-0 field encompasses approximately 21,490 level-3 concepts, reflecting a diverse landscape of scientific research.} \revision{We assess whether the findings depend on this choice or on the MAG taxonomy more broadly by repeating the MAG analysis with concepts from Levels 3--5 and replicating the main results across five additional databases and eight concept vocabularies, including several derived directly from text (see ``Robustness tests, gap-level analyses, and replication'' below).}

We then apply persistent homology (discussed below) to each dynamic concept network to identify birth pairs---pairs of concepts whose connection marks the emergence of a topological feature. Papers that first establish links between birth pairs are defined as gap openers. Papers that link two concepts for the first time, but which do not constitute birth pairs, are categorized as non-gap openers (with novel pair). All remaining papers are considered to involve no novel concept pairings, and are categorized as non-gap openers (without novel pair).

To assess the significance of the observed topological structures and control for the exponential growth of science, we construct a null model. Specifically, we preserve the number of level-3 concepts associated with each paper, but randomly assign concept labels to each paper. This procedure retains the structural sparsity and concept richness of the empirical data, while removing any meaningful co-occurrence patterns. We then apply the same topological analysis to the randomized networks. By comparing the empirical and random models, we can determine whether the observed birth pairs---and the emergence of topological features---reflect meaningful structural patterns rather than artifacts of network growth.

\subsection{Persistent homology}
The following section provides the mathematical foundation for our approach to identifying knowledge gaps using persistent homology\footnote{Our presentation of persistent homology applied to knowledge networks draws on an earlier paper by our team \citep{gebhart2020emergence}, but substitutes the co-occurrence filtration for a temporal filtration to identify knowledge gaps.}. Readers primarily interested in the empirical findings may skip to the next section, as the key intuition---that persistent homology detects when new concept combinations create structural holes in networks---has been established in the main text.

Assume we are presented with a time-varying graph $G = (V,E,\tau)$ where $V$ is the set of vertices, $E$ the set of edges, and $\tau: E \rightarrow [0, t_{\text{max}}]$ a function which tracks the inclusion of edges into the graph over time. In other words, $\tau$ tracks the time at which two nodes are first connected in the graph. 
Here, $t_{\text{max}} = \max_{e \in E}\tau(e) \in \mathbb{R}^+$ is a maximal time after which new edges cease inclusion into the graph. 
Note that $\tau$ can be viewed equivalently as a weighting on the edges of $G$.
We assume that once edges are added to the graph they remain in the graph indefinitely, reflecting the accumulative nature of scientific knowledge in that once a connection between concepts is made, that connection persists in the body of scientific knowledge through time. 

Therefore, the time-varying structure of the graph results from the construction of $G = G_{t_\text{max}}$ through time as measured by $\tau$. 
For convenience, we will denote the subgraph of $G$ consisting of edges added up to time $t$ by $G_t = (V, \{e \in E \mid \tau(e) \leq t\}, \tau)$.
Using this notation, let $t_1 = \min_{e \in E}\tau(e)$ and define the graph $G_1 \subset G$ as the graph consisting of a single edge corresponding to the first edge to appear in the temporal construction of the graph (it is also a subgraph of $G$). 
Because our graph has a finite edge set $E$, and $\tau$ is bounded by $t_{\text{max}}$, we can discretize $G$ as a sequence of subgraphs $\emptyset = G_{t_0} \subset G_{t_1} \subset G_{t_2} \dots G_{t_i} \dots \subset G_{t_M} = G$ where $M$ is the number of distinct values of $\tau$ and $G_{t_M} = G_{t_{\text{max}}}$.
Note that the number of subgraphs in this sequence $M \leq |E|$ as $\tau$ may assign multiple edges to the same value $t_i$, resulting in a ``tie'' with respect to the ordering of edge inclusion, resulting in the incorporation of all tied edges when moving from $G_{t_{i-1}}$ to $G_{{t_i}}$.

Given the above decomposition of $G$ as a sequence of subgraphs through the monotonic ordering on edges induced by $\tau$, each inclusion of $G_{t_i}$ into $G_{t_{i+1}}$ ($G_{t_i} \xhookrightarrow{} G_{t_{i+1}}$) is the inclusion of the subgraph $G_{t_i}$ into the larger graph $G_{t_{i+1}}$ which includes all of $G_{t_i}$ plus additional edges $\{e \in E \mid \tau(e) = t_{i+1} \}$. 

This inclusion relationship induced by $\tau$ is called a \emph{filtration} of $G$. 
At each step $t_i$ in this decomposition of the graph, we compute the \emph{clique complex} (also known as the \emph{flag complex} for networks) $K(G_{t_i})$ of $G_{t_i}$ by treating each $k$-clique of $G_{t_i}$ as a $(k-1)$-dimensional \emph{simplex}. 
In other words, we ``fill in'' all cliques of the graph at each step $t_i$ such that $0$-simplices correspond to nodes, $1$-simplices to edges, $2$-simplices to $3$-cliques (triangles), $3$-simplices to $4$-cliques (tetrahedra), and so on. 
Each $k$-simplex $\sigma = (v_0, v_1, \dots, v_k)$ may be realized geometrically as the convex hull of $k+1$ affinely-positioned nodes. 
As a simplicial complex, $K(G_{t_i})$ is closed under taking subsets of $V$ so that any subset of a simplex must also be a simplex. 
The filtration of $G$ extends to a filtration of simplicial complexes on $G$ such that $\emptyset = K(G_{t_0}) \xhookrightarrow{} K(G_{t_1}) \xhookrightarrow{} \dots \xhookrightarrow{} K(G_{t_M}) = K(G)$. 
For brevity, we hereafter write $K(G_{t_i})$ as $K_{t_i}$ unless otherwise specified.
See Fig.~\ref{fig:1} for an illustration of the construction of a simplicial complex from a time-varying graph.

To track the topological features of $K_{t_i}$, we introduce the \emph{chain group} $C_k(K_{t_i})$, a vector space with basis elements the $k$-simplices $\sigma \in K_{t_i}$. 
Elements of the space $\sum_{\dim \sigma_j = k} c_j \sigma_j \in C_k(K_{t_i})$ are linear combinations of these basis elements $\sigma_j$ and are referred to as \emph{$k$-chains}. We are free to choose any abelian group as the coefficient field, with different choices offering different interpretations of the corresponding homology. 
We use $\mathbb{Z}_2$ for computational simplicity and ease of interpretation, since elements are represented as either present or absent over this field. 

To measure structure in particular dimensions, we need to know how cells in higher dimensions map onto lower-dimensional cells as the filtration proceeds. 
Given an (arbitrary but consistent) orientation of simplices, we define the \emph{boundary operator} $\partial_{k}: C_{k}(K_{t_i}) \rightarrow C_{k-1}(K_{t_i})$ as
\begin{equation*}
 \partial_{k}(\sigma) = \sum\limits_{j=0}^k (-1)^j (v_0, v_1, \dots, v_{j-1}, v_{j+1}, \dots, v_k).
\end{equation*}

Applying the boundary operator twice yields zero, $\partial_k \circ \partial_{k+1} = 0$. 
Cycles in dimension $k$ correspond precisely to the elements of $C_k(K_{t_i})$ that are mapped to zero by $\partial_k$---the cycles of $C_k(K_{t_i})$ are elements of $\ker{\partial_k}$. 
The image of the $(k+1)$-dimensional boundary, $\im \partial_{k+1}$, comprises the $k$-boundaries. 
The upper boundary map $\partial_{k+1}$ takes the interior of a $(k+1)$-dimensional simplex to its boundary. Therefore, $\im \partial_{k+1} \subseteq \ker\partial_k$ and $\partial_k \circ \partial_{k+1} = 0$ as noted above. 

Elements of $\ker \partial_k$ that are not in the image of $\partial_{k+1}$ form the \emph{$k$-dimensional holes} of the simplicial complex, in that they are collections $k$-cycles which are not the result of the disassembly of $k+1$-dimensional simplices by $\partial_{k+1}$.
We would like to count the number of $k$-dimensional holes within the simplicial complex, but there may be many cycles which, by the definition of $\partial_k$, form a boundary of this $k$-dimensional hole. 
Therefore, we must associate all cycles enclosing each unique $k$-dimensional hole. 
In other words, we form an equivalence class of the cycles by associating any two cycles $x,y \in \ker\partial_k$ if $x - y \in \im \partial_{k+1}$. 
The \emph{homology group in dimension $k$} $H_k(K(G)) = \ker\partial_k / \im\partial_{k+1}$ of simplicial complex $K(G)$ is precisely the group formed by these equivalence classes of cycles. 

Similar to how the inclusion relationship induced by the temporal filtration on the graph, $G_{t_i} \subset G_{t_{i+1}}$ extends to a filtration on the associated simplicial complexes $K(G_{t_i}) \xhookrightarrow{} K(G_{t_{i+1}})$, this inclusion relationship over the simplicial structure extends again to the chain groups by mapping basis elements of $C_k(K_{t_i})$ to basis elements of $C_k(K_{t_{i+1}})$ such that $C_k(K_{t_i}) \xhookrightarrow{} C_k(K_{t_{i+1}})$. 
Note that the homology of $K(G)$ is defined solely in terms of the chain groups. 
These induced maps imply that, for proper choice of basis, we can map cycles to cycles through the induced map $H_k(K_{t_i}) \rightarrow H_k(K_{t_{i+1}})$. 
Through this mapping on homology groups, we can track holes across the entire filtration of $G$. This insight underpins the persistent homology algorithm---\emph{persistent homology in dimension $k$} is the computation of a collection of ranks induced by the inclusion maps of a filtration on homology and consists of ranks of induced homology maps $\beta_k^{t_i, t_j} = \rank H_k\left(K(G_{t_i}) \xhookrightarrow{} K(G_{t_j})\right)$ for all $i,j$ such that $1 \leq i \leq j \leq M$. 
A simplex $\sigma_i$ is called a \emph{birth simplex} if and only if its boundary is a linear combination of boundaries in $K(G_{t_{i-1}})$. 
Conversely, $\sigma_j$ is a \emph{death simplex} if and only if its boundary is not a linear combination of boundaries in $K(G_{t_{j-1}})$. 
While methods for efficient calculation of persistent homology can become rather involved, at its core this computation takes the form of a row reduction operation similar in spirit to Gaussian elimination~\citep{otter2017roadmap} given $\partial_{\bullet}$. 

For each $i$, the inclusion of a $k$-dimensional simplex $\sigma_i$ to $K(G_{t_{i-1}})$ either creates a $k$-dimensional homology class or destroys a $(k-1)$-dimensional class. 
Each death $k$-simplex $\sigma_j$ is paired to a unique birth $(k-1)$-simplex $\sigma_i$, $i < j$ to form a \emph{persistence pair} which encodes the notion that the addition of simplex $\sigma_i$ creates a homology class that is then destroyed by $\sigma_j$. 
Birth simplices in dimension $k$ not paired with a corresponding death simplex are called \emph{essential simplices} and correspond to homology generators of $H_k(K(G))$--the invariants of the concept network at time $t_{\text{max}}$ at the end of the filtration.

\subsubsection{Relationship to ``structural holes''} \revision{Our use of persistent homology to detect knowledge gaps shares a conceptual affinity with Burt's notion of structural holes \cite{burt1992structural,burt2004structural}, as both approaches focus on where connections are absent or weak in networks. However, the two methods differ in what they measure. Burt's constraint measure characterizes the extent to which an individual node's contacts are interconnected, identifying actors who occupy advantageous brokerage positions (as indicated by low constraint). Persistent homology, by contrast, detects topological features of the network as a whole, identifying when cycles (closed loops of connections) form and dissolve as the network evolves. This distinction is consequential for our purposes. Constraint helps to identify specific actors who span holes; persistent homology identifies where holes exist and when they are created, independent of any particular node's position.}

\subsection{Generative capacity measures}

\subsubsection{Citation impact}
We measure citation impact using both threshold-based and count-based approaches. For threshold-based measures, we classify papers as falling within the top 1\%, 5\%, 10\%, 15\%, or 20\% most cited works published in the same year and level-0 field, based on cumulative citations through 2021, the last year of complete citation coverage. These binary indicators capture whether a paper achieves exceptional citation impact relative to disciplinary and temporal peers. For count-based measures, we use cumulative citation counts over windows of $W$ years after publication, considering citations from papers published in the interval $(y_f,\, y_f + W]$. We report results for $W = 5$ (C5) and $W = 20$ (C20); the short window captures early reception, while the long window captures sustained impact.

\subsubsection{Disruption}
Some scientific contributions make prior work less central to what follows, while others consolidate existing knowledge by building upon and reinforcing established foundations \citep{Funk2017ADN,wu2019large}. Because gap-opening papers may stimulate new directions rather than extend existing ones, disruption provides a key indicator of their generative capacity. To measure where a paper falls along this spectrum, we use the CD index \citep{Funk2017ADN}. For a focal paper $f$, denote by $B_f$ its references (backward citations), by $C_f$ the papers that cite the focal paper (forward citations), and by $C_B$ the papers that cite any $i \in B_f$. We partition the papers that cite either $f$ or any member of $B_f$ into three disjoint sets: $I_f = C_f \setminus C_B$ (papers that cite $f$ but not the references), $J_f = C_f \cap C_B$ (papers that cite both $f$ and the references), and $K_f = C_B \setminus C_f$ (papers that cite the references but not $f$). Writing $n_I = |I_f|$, $n_J = |J_f|$, and $n_K = |K_f|$, the CD index is defined as:

\[
\text{CD index} = \frac{n_I - n_J}{n_I + n_J + n_K}
\]

We also write $n_C = |C_f| = n_I + n_J$ for the number of forward citations, $n_B = |B_f|$ for the number of references, and $y_f$ for the publication year of $f$. We compute the index over a forward citation window of $W$ years, considering citations from papers published in the interval $(y_f,\, y_f + W]$.

The index ranges from $-1$ to $1$, with higher values indicating that citing works tend to reference the focal paper without its predecessors---a signature of disruptive influence---and lower values indicating consolidating influence. 

We restrict computation of the CD index to papers with at least one reference and at least one forward citation. For our main analyses, we convert the CD index to within-year percentile ranks (CD percentile), following \citet{wu2019large}, to aid interpretation of effect sizes and to facilitate comparison with other percentile-scaled measures such as the A~percentile; results are robust to using the raw CD index (Table~\ref{table:s4}, Fig.~\ref{fig:s6}). Throughout, we use `disruption' to refer to the underlying concept, `CD index' for the raw measure, and `CD percentile' for the percentile-transformed version. Unless otherwise noted, we compute the CD index using every forward citation recorded in the data, while the estimation samples are restricted to papers published 1980--2020; window-specific analyses (e.g., Table~\ref{table:s4}, Figs.~\ref{fig:s4} and~\ref{fig:s6}) state the relevant $W$ explicitly.

\subsubsection{Delayed recognition}
Scientific papers that open new intellectual territories may require time for communities to recognize and build upon, resulting in citation trajectories characterized by long dormant periods followed by sudden surges of attention \citep{Ke2015DefiningAI,van2004sleeping}. To measure this phenomenon, we use the Sleeping Beauty (SB) index, which quantifies the deviation between a paper's actual citation trajectory and a reference line defined by three parameters---the publication year, the peak annual citation count within an observation window, and the time $t_m$ at which the peak occurs \citep{Ke2015DefiningAI}.

Let $c_t$ denote the number of citations received in the $t$-th year after publication, where $t \in [0, T]$ represents the paper's age. Let $c_{t_m}$ be the maximum number of yearly citations received within the interval, occurring at time $t_m \in [0, T]$. The linear reference trajectory $l_t$ connecting the citation values at $t = 0$ and $t = t_m$ is defined as:

\[
l_t = \frac{c_{t_m} - c_0}{t_m} \cdot t + c_0
\]

where $c_0$ is the number of citations in the year of publication and $\frac{c_{t_m} - c_0}{t_m}$ represents the slope of the reference line.
For each $t \leq t_m$, we compute the normalized difference between the reference line and actual citations:

\[
\frac{l_t - c_t}{\max\{1, c_t\}}
\]

Summing from $t = 0$ to $t = t_m$, the SB index is given by:

\[
\text{SB index} = \sum_{t = 0}^{t_m} \frac{l_t - c_t}{\max\{1, c_t\}}
\]

We restrict computation of the SB index to papers with at least one citation. By definition, the SB index equals zero if $t_m = 0$ or if the citation trajectory exactly follows the linear reference (i.e., $c_t = l_t$ for all $t$). Positive values indicate delayed recognition, where citations accumulate slowly before surging near the peak. Throughout, we use `delayed recognition' to refer to the underlying concept and `SB index' for the measure. Since the SB index can assume negative values and exhibits a heavy right skew, we employ a logarithmic transformation and use the resulting measure, defined as:

\[
\text{SB index (log)} = \log_2\big(\text{SB index} + (|\min(\text{SB index})| + 1)\big)
\]

in all regression analyses. Unless otherwise noted, we compute the SB index using every citation recorded in the data.

\subsubsection{Nobel Prize}
We construct a binary indicator coded as 1 for papers identified as Nobel Prize-winning works and 0 otherwise, drawing on the dataset of Nobel laureate publication records compiled by \citet{li2019dataset}, which provides MAG paper identifiers that allow direct linkage to our dataset.

\subsection{Paper and team characteristics}

\subsubsection{Atypicality}

Our study distinguishes between two constructs that capture different aspects of a paper's contribution. The first---whether a paper introduces a \emph{novel concept pair}---is a structural property of the concept network, defined by whether a paper first connects two previously unlinked level-3 concepts. This property is integral to our paper typology, distinguishing gap openers and non-gap openers (with novel pair) from baseline papers. The second---\emph{atypicality}---captures how unusual the journal sources are that a paper draws upon \citep{uzzi2013atypical}. Atypicality is one of the most widely used measures in the science of science literature for capturing how unusually a paper combines prior work \citep{uzzi2013atypical,wang2017bias,foster2015tradition}, and we employ the measure here as an outcome to characterize and compare the three paper types. A paper may introduce a novel concept pair without citing atypical journal combinations, and vice versa; the two constructs are complementary but distinct.

To measure atypicality, we follow \citet{uzzi2013atypical} and \citet{li2025innovation}. For each pair of journals $(m, n)$ co-cited in a paper's reference list, a $z$-score is computed as
\[
z_{mn} = \frac{\mathrm{obs}_{mn} - \mathrm{exp}_{mn}}{\sigma_{mn}}
\]
where $\mathrm{obs}_{mn}$ is the observed co-citation frequency of journals $m$ and $n$ across all papers published in that year, $\mathrm{exp}_{mn}$ is the expected frequency under a null model that randomly reassigns journal labels to references, and $\sigma_{mn}$ is the corresponding standard deviation. Lower $z$-scores indicate more atypical pairings. Following \citet{uzzi2013atypical}, we take the 10th percentile of the $z$-score distribution across all journal pairs in a paper's reference list as the paper-level atypicality score. We then reverse-code by negating this value so that higher scores indicate greater atypicality. Because the reverse-coded measure is no longer on the original scale of \citet{uzzi2013atypical}, we refer to it by a distinct name, the A~index \citep{li2025recombination,li2025innovation}, whose form also parallels the paper's other outcome indices, the CD index and the SB index:

\[
\text{A index} = 
\begin{cases}
-\log_2(P_{10} + 1), & \text{if } P_{10} > 0 \\
\log_2(-P_{10} + 1), & \text{if } P_{10} \leq 0
\end{cases}
\]

where $P_{10}$ denotes the 10th percentile of the $z$-score distribution across all journal pairs in the reference list. 

We restrict computation of the A~index to papers with at least two references. For our main analyses, we convert the A~index to within-year percentile ranks (A~percentile) to aid interpretation of effect sizes and to facilitate comparison across percentile-scaled measures, so that higher percentiles correspond to greater atypicality. Throughout, we use `atypicality' to refer to the underlying concept, `A~index' for the reverse-coded raw measure, and `A~percentile' for the percentile-transformed version.

\subsubsection{Paper attributes}
We characterize gap-opening papers using several attributes of the concepts they link and the references they cite. \emph{Concept age} is the average age of the two linked level-3 concepts, where the age of a concept is defined as the number of years since its first appearance in the dataset. \emph{Concept popularity} is the average number of papers assigned each of the two linked level-3 concepts prior to the focal paper's publication year. \emph{Number of references} is the count of cited works in the focal paper's reference list. These measures are reported as descriptive comparisons between gap openers and non-gap openers (with novel pair) (Fig.~\ref{fig:3}). We also track how concept popularity evolves after publication by measuring the average occurrences of the two linked concepts five and ten years after the focal paper's publication year (Fig.~\ref{fig:s2}).

\subsubsection{Team attributes}
We characterize the collaborative conditions under which gap-opening papers emerge using four measures. \emph{Team size} is the number of authors on a paper. \emph{Career age} is the average number of years since each team member's first publication in the dataset. \emph{Team freshness} is defined as the fraction of authors in a paper who have no prior co-authorship history with any of the other authors on that paper \citep{Zeng2021FreshTA}; the measure is undefined for solo-authored papers, so it is computed for papers with team sizes 2--20 (the manuscript-wide cap). \emph{Geographical distance} is the average distance (in km) between the affiliations of team members; this measure is restricted to papers with at least two affiliations with recorded coordinates. These measures are reported as descriptive comparisons across the three paper types using independent-sample $t$-tests (Fig.~\ref{fig:5}).

\subsubsection{Title verb analysis}

To characterize linguistic differences between gap openers and non-gap openers (with novel pair), we analyze the frequency of verbs in paper titles. Using spaCy, we lemmatize all tokens in English-language titles and extract verbs, yielding 12,483 unique verb lemmas across the two groups. Verbs were retained only if they appeared in the titles of both groups, ensuring that the frequency ratio $r$ is well-defined for all verbs analyzed. Papers classified as non-gap openers without any novel pair were excluded. The final sample includes 274,230 gap openers and 9,144,020 non-gap openers (with novel pair) with non-missing English-language titles.

For each verb, we compute a relative frequency ratio $r$:
\[
r = \frac{f_{\text{gap}}(\text{verb})}{f_{\text{non-gap}}(\text{verb})}
\]

where $f_{\text{gap}}(\text{verb})$ and $f_{\text{non-gap}}(\text{verb})$ denote the proportion of titles containing a given verb in the gap-opener and non-gap-opener (with novel pair) groups, respectively. Values of $r > 1$ indicate verbs overrepresented among gap openers; values of $r < 1$ indicate verbs overrepresented among non-gap openers (with novel pair). No statistical tests were applied to the frequency ratios; the results are intended to illustrate qualitative differences in verb usage between the two groups. Fig.~\ref{fig:s3} displays a representative selection of verbs for each group.

\subsection{Statistical analysis}

To systematically assess the generative capacity of gap-opening papers, we estimate fixed-effects regression models comparing our three paper categories---gap openers, non-gap openers (with novel pair), and non-gap openers (without novel pair). In each model, non-gap openers (without novel pair) serve as the baseline category. We employ different model specifications appropriate to the distribution of each dependent variable. Specifically, we use logistic regression for binary outcomes (top 1\% most cited, Nobel Prize wins), Ordinary Least Squares (OLS) for continuous measures (CD percentile, SB index), and Poisson regression for count data (C5 and C20 citation counts). \revision{The primary outcome models use one observation per paper and robust standard errors. The pooled citation-window models use one observation per paper per window, with standard errors clustered by paper. In the replication corpora, records assigned to multiple fields or CPC sections contribute one observation per assignment, with standard errors clustered by paper or patent to account for repeated observations; APS papers have a single field assignment and use robust standard errors. Each table note states the unit of analysis and standard-error treatment.} For logistic regression models, we report log-odds coefficients; exponentiated coefficients yield odds ratios (e.g., a coefficient of 0.454 corresponds to OR = 1.57). For Poisson regression models, exponentiated coefficients yield incidence rate ratios representing multiplicative effects on citation counts (e.g., a coefficient of $-$0.063 corresponds to a 6.1\% decrease).

All models include fixed effects for publication year and scientific field (level-0) to control for systematic differences across fields and time periods. We further control for team size and the log of the number of references, as prior research indicates these factors influence citation impact and disruption \citep{Wagner2018InternationalRC,wu2019large,Yang2024ExploringTC,wuchty2007increasing,Zeng2021FreshTA}. For models where the dependent variable is derived from citation data---specifically, the CD percentile and SB index---we also control for total citation count (log-transformed) to isolate variation in the shape of citation patterns from variation in citation volume \citep{Ke2015DefiningAI,Funk2017ADN}.

Table~\ref{table:s1} reports the main results across six outcome measures; Fig.~\ref{fig:4} plots the corresponding predicted values. The sample is restricted to papers published 1980--2020 with team sizes $\leq$20; the two restrictions jointly retain 93\% of the corpus. The publication window reflects the sparser coverage of reference and author metadata in MAG's earlier decades and allows post-publication citation accumulation at the recent end; the team-size cap excludes the extreme tail of hyperauthored papers, which form under a qualitatively different collaborative regime and whose authorship practices complicate the attribution of individual contributions \citep{cronin2001hyperauthorship,milojevic2014principles}, and follows prior studies of scientific collaboration that impose the same 20-author limit \citep{azoulay2011incentives,bikard2015exploring,lu2020contributorship}. Observation counts vary across columns because each model further requires non-missing values for the relevant dependent variable and, where applicable, non-missing citation counts used as controls. For example, we restrict the sample to papers with at least one reference and one citation for the CD percentile. The C5 and C20 models share the same sample (n = 8,643,571) because we require non-missing values for both measures simultaneously, which restricts the sample to papers published 1980--2001 to ensure 20 years of post-publication citation data. Table~\ref{table:s2} extends the citation impact analysis to additional thresholds (top 5\%, 10\%, 15\%, and 20\% most cited), using the same sample and specification as the top 1\% model in Table~\ref{table:s1}.

Table~\ref{table:s3} reports an unordered multinomial logistic regression that classifies papers into four joint atypicality--disruption categories based on whether a paper is atypical ($\text{A index} > 0$) and disruptive ($\text{CD index} > 0$). The model estimates the probability of falling into each category---atypical-and-disruptive, atypical-only, disruptive-only, and neither---as a function of paper type, with the neither category serving as the baseline outcome. The sample is restricted to papers with non-missing values for both the A~index and CD~index, yielding a smaller sample than the CD percentile model in Table~\ref{table:s1}.

To determine whether the joint atypicality--disruption association is reducible to either dimension, we estimate a second multinomial model that adds citation count to the controls and use Wald tests to compare the atypical-and-disruptive coefficient with the atypical-only and disruptive-only coefficients. We also estimate the two margins separately using logistic regressions for binary atypicality and disruptiveness and OLS regressions for A percentile and CD percentile (\emph{SI Appendix}, Tables~\ref{tab:multinomial}--\ref{tab:margins}).

Table~\ref{table:s4} reports pooled regressions across citation windows from 1 to 20 years, stacking observations at the paper--window level to estimate average effects across horizons. Citation window is included as an additional covariate. The sample includes 3,312,397 unique papers, each contributing 20 observations (one per window). Standard errors are clustered at the paper level to account for within-paper dependence across windows. These models complement the window-by-window estimates plotted in Figs.~\ref{fig:s4} (citations only, n = 8,643,571) and~\ref{fig:s6} (citations and disruption, n = 3,312,397) by providing a single summary coefficient for each paper type. The window-by-window estimates in Figs.~\ref{fig:s4} and~\ref{fig:s6} are obtained from separate regressions for each citation window, using the same control variables and model specifications but omitting the citation window covariate, which instead serves as the stratification variable.

\pagebreak

\subsection{Robustness tests, gap-level analyses, and replication}

We conduct three sets of analyses beyond the primary specifications. We first assess whether the estimated gap-opener associations depend on concept granularity, the distinction between introducing a novel concept pair and occupying a gap-opening position, observed paper- and concept-level characteristics, or the construction of the comparison sample. We then examine how two properties of gaps---their duration and size---relate to the reception of opening papers and the subsequent development of the gaps. Finally, we replicate the primary citation-impact and disruption models across additional databases and concept systems.

\subsubsection{Robustness to measurement and model specification}

To assess sensitivity to concept granularity, we re-identify gap openers using all MAG concepts from Levels 3 through 5 rather than Level 3 alone. We exclude Levels 0 through 2 because they represent broad disciplinary categories rather than specific concepts. The top 1\% and CD-percentile estimates remain unchanged in sign and significance, although several other estimates shift: the 5-year-citation coefficient turns positive, the Nobel Prize coefficient becomes significant, and non-gap openers (with novel pair) show a small positive top 1\% coefficient that is absent under Level-3 concepts, against a gap-opener coefficient more than five times larger. These shifts reflect the broader vocabulary admitting papers outside the main corpus and narrowing the baseline group, so that the citation-count and top-percentile coefficients rise (\emph{SI Appendix}, Tables~\ref{tab:taxonomy-robustness}--\ref{tab:taxonomy-full}).

To separate the contribution of introducing a novel concept pair from that of occupying a gap-opening position, we estimate two alternative specifications. The first uses a nested parameterization with one indicator for introducing any novel pair and a second indicator for whether that pair opens a gap. The second restricts the sample to papers that introduce novel concept pairs and compares gap openers only with non-gap openers (with novel pair). Both specifications identify an association between gap-opening position and the primary outcomes independent of novel-pair status (\emph{SI Appendix}, Tables~\ref{tab:nested-model}--\ref{tab:within-sample}).

We assess alternative paper- and concept-level explanations using two additional control specifications. The first adds the CD index, atypicality, team mean age, team freshness, and geographical distance to the citation models. The second adds concept age, prior concept popularity, and atypicality. The positive associations with long-term citation impact persist under both specifications and, under the concept-level controls, strengthen. Under the fuller paper-level controls, the 5-year-citation coefficient turns positive while the long-term association remains (\emph{SI Appendix}, Tables~\ref{tab:net-effect} and~\ref{tab:concept-quality}).

Finally, we use coarsened exact matching to compare each gap opener with up to five control papers in the same field and publication year and in the same team-size and reference-count bins, as well as the same prior-citation bin where the corresponding specification includes citations as a control. We estimate models with matched-set fixed effects for two comparison samples: all eligible control papers and, more restrictively, non-gap openers (with novel pair). The analysis covers the outcomes in Fig.~\ref{fig:4} other than Nobel Prize wins, which are too rare for estimation within matched sets; the top 1\% outcome is estimated using a linear probability model. Gap openers remain more disruptive and more highly cited in the long run in both matched comparisons (\emph{SI Appendix}, Table~\ref{tab:matched-sampling}).

\subsubsection{Gap-level analyses}

We examine two properties of the gaps themselves. Gap duration is the elapsed time between the paper that opens a gap and the first paper whose new concept pairing completes the simplices that fill its interior. For papers that open multiple gaps, we average duration across those gaps. Among gap-opening papers whose gaps subsequently close, we regress citation and disruption outcomes on log gap duration, controlling for team size, references, the A index, and year and field fixed effects. Papers whose gaps close more quickly have greater citation impact and disruption than those whose gaps persist longer (\emph{SI Appendix}, Table~\ref{tab:gap-duration}). We interpret this relationship as an association rather than a mechanism because the subsequent research attention that generates citations and disruption may also hasten gap closure.

We measure gap size by the length of the shortest cycle bounding the gap. At the gap level, we regress three measures of subsequent development on log gap size: the number of follow-on papers per possible pair of boundary concepts, the length of time the gap remains open, and the number of concept triangles in the filling structure identified when the gap closes. Follow-on papers are papers published after the gap opens that use at least two concepts from its boundary. In the raw counts, four-concept gaps are followed over the remainder of MAG's observation period by an average of 139 such papers, compared with 6,100 for gaps bounded by cycles of six or more links. Larger gaps attract more follow-on work, remain open longer, and have larger filling structures when they close, although the opening paper's own citation impact does not increase with gap size (\emph{SI Appendix}, Table~\ref{tab:gap-size}).

\subsubsection{Replication across databases and concept systems}

We replicate the analysis in five additional databases---OpenAlex, the PubMed Knowledge Graph, Dimensions, APS journals, and U.S. patents---which, together with MAG, span six data sources and eight concept vocabularies. Within each corpus, we construct concept networks, identify gap openers using the same persistent-homology procedure, and estimate the top 1\% most-cited and CD-percentile specifications separately. Persistent homology is implemented using the GUDHI \citep{maria2014gudhi} and OAT \citep{oat2026} libraries. Corpus-specific data sources, inclusion criteria, concept definitions, network-construction procedures, field assignments, units of analysis, and standard-error treatments are documented in \emph{SI Appendix}, section ``Replication data sources and network construction.'' Full regression results appear in \emph{SI Appendix}, Tables~\ref{tab:replication-openalex}--\ref{tab:replication-patents}.

\section{Use of AI and generative AI software}
The authors used Anthropic Claude Opus 5 and OpenAI GPT-5 to improve the language, style, and readability of the manuscript and to assist in reviewing analysis code and cross-checking reported results. The authors independently verified all analyses and results and take full responsibility for the content of the manuscript.

\section{Data availability}
Data associated with this study are freely available in a public repository at \url{https://doi.org/10.5281/zenodo.15694240}.

\section{Code availability}
Open-source code related to this study is available at \url{https://doi.org/10.5281/zenodo.15694240}.

\clearpage

\bibliographystyle{plainnat}
\bibliography{bibliography}

\clearpage

\clearpage

\clearpage

\clearpage

\clearpage

\clearpage

\clearpage
\def\SIinmain{}  
\ifdefined\SIinmain\else
\documentclass{article}

\usepackage[utf8]{inputenc}
\usepackage[T1]{fontenc}

\usepackage{mathpazo} 

\usepackage[margin=1.0in]{geometry}
\usepackage[bottom]{footmisc}

\usepackage{graphicx}
\usepackage[skip=0pt, labelfont=bf, font=footnotesize]{caption}
\usepackage[capposition=top]{floatrow}
\floatsetup[table]{captionskip=0pt}
\usepackage{tikz}

\usepackage{booktabs}
\usepackage{longtable}
\usepackage{tabularx}
\usepackage{threeparttable}
\usepackage{threeparttablex}
\usepackage{makecell}
\usepackage{etoolbox}
\BeforeBeginEnvironment{tabular}{\small}

\newcounter{magicrownumbers}
\newcommand\rownumber{\stepcounter{magicrownumbers}\arabic{magicrownumbers}. }

\usepackage{pdflscape} 

\usepackage{authblk}

\usepackage{xcolor}
\usepackage{nameref}



\usepackage{booktabs} 
\usepackage{nicefrac}  
\usepackage{microtype}
\usepackage[utf8]{inputenc}

\usepackage{wrapfig}
\usepackage{multirow}

\usepackage{amssymb}
\usepackage{amsthm}
\usepackage{bbm}
\usepackage{bbm}
\usepackage{array}
\usepackage[colorlinks=true,urlcolor=purple,
linkcolor=purple,citecolor=purple]{hyperref}
\usepackage{amsmath}
\usepackage{mathtools}
\usepackage{float}

\usepackage{lscape}
\usepackage{graphicx}

\usepackage{booktabs}
\usepackage{longtable}
\usepackage{threeparttable}
\usepackage{threeparttablex}

\usepackage{mathrsfs}
\usepackage{graphicx}
\usepackage{color}
\usepackage[numbers,sort&compress]{natbib}

\usepackage{algorithm}
\usepackage{algpseudocode}
\usepackage{thmtools,thm-restate} 

\makeatletter
\def\paragraph{\@startsection{paragraph}{4}%
  \z@\z@{-\fontdimen2\font}%
  {\normalfont\bfseries}}
\makeatother

\usepackage{tikz}
\usetikzlibrary{decorations.markings, backgrounds, calc, positioning, shapes, shadows, arrows, fit, automata}
\usepackage{tikz-cd}
\usetikzlibrary{arrows.meta}
\tikzset{>={Latex}}

\newcommand{\revision}[1]{\textcolor{black}{#1}}

\makeatletter
\setlength{\@fptop}{0pt}
\makeatother

\newcommand{\equalcontrib}{\textsuperscript{\textdagger}}

\title{Opening Knowledge Gaps Drives Scientific Progress
\thanks{We thank the National Science Foundation (grant nos. 1829168, 1932596, 2318172, and 2449660 to R.J.F.; grant no. 2449659 to K.K.), the Alfred P. Sloan Foundation (grant no. G-2024-25123 to R.J.F. and T.G.), and the National Natural Science Foundation of China (grant no. 72004177 to Y.W.) for financial support of work related to this project. R.J.F.\ is also grateful for many helpful conversations related to this work with Lori Ziegelmeier, Jason Owen-Smith, Tanisha Dodla, Nadezhda Dominguez Salinas, Gavin Engelstad, Jingyi Guan, Floyd Liu, Lucia Luo, Frances McConnell, Tam Nguyen, Adam Schroder, and Ethan Scheelk.}}

\author[1]{Kara Kedrick\equalcontrib}
\author[2,3]{Wenlong Yang\equalcontrib}
\author[4]{Thomas Gebhart}
\author[2]{Yang Wang\thanks{Corresponding author: \texttt{yang.wang@xjtu.edu.cn}}}
\author[5]{Russell J. Funk\thanks{Corresponding author: \texttt{rfunk@umn.edu}}}

\affil[1]{Institute for Complex Systems Dynamics, Carnegie Mellon University}
\affil[2]{School of Public Policy and Administration, Xi'an Jiaotong University}
\affil[3]{College of Public Administration, Huazhong University of Science and Technology}
\affil[4]{Computer Science and Engineering, University of Minnesota}
\affil[5]{Carlson School of Management, University of Minnesota}

\date{}

\usepackage{xr}
\externaldocument{main} 

\usepackage{adjustbox}

\begin{document}
\fi

\ifdefined\SIinmain\else\pagenumbering{arabic}\fi
\beginsupplement

\begin{center}
\vspace*{1cm}

{\LARGE \textbf{SI Appendix for}}

\vspace{0.8cm}

{\Large \textbf{Opening Knowledge Gaps Drives Scientific Progress}}

\vspace{1cm}

Kara Kedrick\textsuperscript{\textdagger},
Wenlong Yang\textsuperscript{\textdagger},
Thomas Gebhart,
Yang Wang\textsuperscript{*},
Russell J. Funk\textsuperscript{*}

\vspace{0.6cm}

\textsuperscript{\textdagger}These authors contributed equally.

\vspace{0.3cm}

\textsuperscript{*}Corresponding authors:\\
Yang Wang, \texttt{yang.wang@xjtu.edu.cn}\\
Russell J. Funk, \texttt{rfunk@umn.edu}

\vspace{1cm}

\textbf{This PDF file includes:}

\vspace{0.3cm}

\begin{flushleft}
\hspace{3cm}Supplementary text\\
\hspace{3cm}Figures~\ref{fig:s1} to ~\ref{fig:s6}\\
\hspace{3cm}Tables~\ref{table:s1} to~\ref{tab:concept-quality}\\
\end{flushleft}

\end{center}

\clearpage

\section*{SI Appendix}

\subsection*{Index of tables and figures}
\begingroup
\setlength{\parindent}{0pt}
\setlength{\parskip}{1.5pt}

{\bfseries Supplementary analyses supporting the main results}\par\nobreak
\begingroup\leftskip=1.7em
\hangindent=1.8em Cross-field differences in gap-opener prevalence.\par
\hangindent=1.8em Additional analyses supporting the main results.\par
\endgroup
\addvspace{7pt}

{\bfseries Tables}\par\nobreak
\begingroup\leftskip=1.7em
\hangindent=1.8em Table~\ref{table:s1}. Generative capacity of gap-opening papers across citation impact, disruption, Nobel prizes, and delayed recognition.\par
\hangindent=1.8em Table~\ref{table:s2}. Citation impact of gap-opening papers at the top 5\%, 10\%, 15\%, and 20\% thresholds.\par
\hangindent=1.8em Table~\ref{table:s3}. Joint atypicality--disruption profiles of the three paper types.\par
\hangindent=1.8em Table~\ref{table:s4}. Citation and disruption trajectories of gap-opening papers pooled across citation windows.\par
\endgroup
\addvspace{7pt}

{\bfseries Figures}\par\nobreak
\begingroup\leftskip=1.7em
\hangindent=1.8em Figure~\ref{fig:s1}. Cross-field variation in gap-opening papers: publication volume, gap-closure duration, concept counts, and citations.\par
\hangindent=1.8em Figure~\ref{fig:s2}. Concept popularity of gap-opening papers: occurrences five and ten years after publication.\par
\hangindent=1.8em Figure~\ref{fig:s3}. Linguistic characteristics of gap-opening papers: verb frequency differences between gap openers and non-gap openers (with novel pair).\par
\hangindent=1.8em Figure~\ref{fig:s4}. Citation trajectories of gap-opening papers: estimated coefficients across citation windows (1--20 years).\par
\hangindent=1.8em Figure~\ref{fig:s5}. Microsoft Academic Graph taxonomy: number of concepts at each hierarchical level.\par
\hangindent=1.8em Figure~\ref{fig:s6}. Citation and disruption trajectories of gap-opening papers: estimated coefficients across citation windows (1--20 years).\par
\endgroup
\addvspace{7pt}
{\bfseries Robustness tests, gap-level analyses, and replication}\par\nobreak
\begingroup\leftskip=1.7em
\hangindent=1.8em Table~\ref{tab:taxonomy-robustness}. Robustness of gap-opener effects across Microsoft Academic Graph (MAG) taxonomy levels.\par
\hangindent=1.8em Table~\ref{tab:taxonomy-full}. Gap-opener effects across all outcomes using MAG Levels 3--5 concepts.\par
\hangindent=1.8em Table~\ref{tab:replication-openalex}. Replication of gap-opener effects in OpenAlex.\par
\hangindent=1.8em Table~\ref{tab:replication-pubmed}. Replication of gap-opener effects in the PubMed Knowledge Graph.\par
\hangindent=1.8em Table~\ref{tab:replication-dimensions}. Replication of gap-opener effects in Dimensions.\par
\hangindent=1.8em Table~\ref{tab:replication-aps}. Replication of gap-opener effects in APS journals under text-derived concepts and PACS codes.\par
\hangindent=1.8em Table~\ref{tab:replication-patents}. Replication of gap-opener effects in U.S. patents under CPC codes at two levels of granularity.\par
\hangindent=1.8em Table~\ref{tab:nested-model}. Decomposition of novel-pair and gap-opening contributions.\par
\hangindent=1.8em Table~\ref{tab:within-sample}. Gap-opener effect among papers with novel concept pairs.\par
\hangindent=1.8em Table~\ref{tab:gap-duration}. Effect of gap duration on citation impact and disruption.\par
\hangindent=1.8em Table~\ref{tab:net-effect}. Gap-opener effect after controlling for established impact predictors.\par
\hangindent=1.8em Table~\ref{tab:matched-sampling}. Gap-opener effects under coarsened exact matching.\par
\hangindent=1.8em Table~\ref{tab:multinomial}. Joint atypicality--disruption profiles after controlling for citations.\par
\hangindent=1.8em Table~\ref{tab:wald}. Wald comparisons of the joint atypicality--disruption coefficient with each single-margin coefficient.\par
\hangindent=1.8em Table~\ref{tab:margins}. Separate contributions of atypicality and disruption to the gap-opener effect.\par
\hangindent=1.8em Table~\ref{tab:gap-size}. Generative capacity of knowledge gaps by gap size.\par
\hangindent=1.8em Table~\ref{tab:concept-quality}. Gap-opener effect after controlling for concept-level characteristics.\par
\endgroup
\addvspace{7pt}

{\bfseries Replication data sources and network construction}\par\nobreak
\begingroup\leftskip=1.7em
\hangindent=1.8em Shared analytical procedure.\par
\hangindent=1.8em OpenAlex.\par
\hangindent=1.8em PubMed Knowledge Graph.\par
\hangindent=1.8em Dimensions.\par
\hangindent=1.8em American Physical Society journals.\par
\hangindent=1.8em U.S. patents.\par
\endgroup
\endgroup

\clearpage

\section*{Supplementary analyses supporting the main results}

\revision{This section provides additional analyses supporting the descriptive and outcome patterns reported in the main text. It first examines why gap-opener prevalence varies across fields, then reports supplementary analyses of concept characteristics, linguistic patterns, the MAG taxonomy, citation and disruption trajectories, and the regression models underlying the main results.}

\subsection*{Cross-field differences in gap-opener prevalence}

\begin{figure}[htbp]
 \centering
 \includegraphics[width=\textwidth,height=0.85\textheight,keepaspectratio]{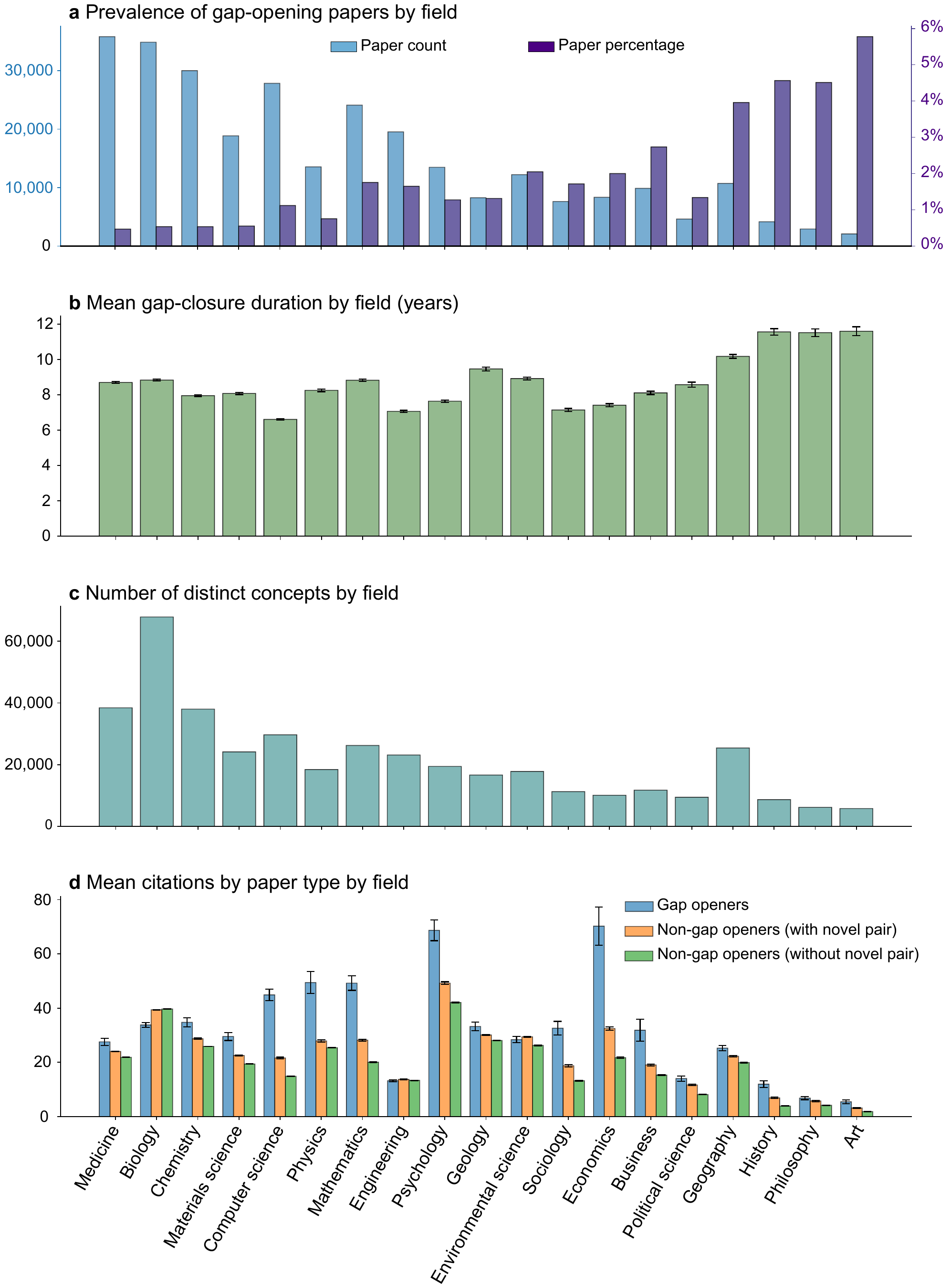}
\caption{\textbf{Cross-field variation in gap-opening papers: publication volume, gap-closure duration, concept counts, and citations.} 
\textbf{(a)} Absolute number of gap-opening papers (left axis) and their fraction among all papers in the field (right axis; n = 289,041). \textbf{(b)} Mean gap-closure duration (in years) by field, among gap openers whose gaps have closed (n = 289,012). \textbf{(c)} Number of distinct level-3 concepts in each field. \textbf{(d)} Mean citations across fields for three paper types: gap openers, non-gap openers (with novel pair), and non-gap openers (without novel pair) (n = 34,363,623). Error bars in panels (b) and (d) indicate standard errors. Fields are ordered consistently across panels. These results provide detail underlying the cross-field patterns shown in Fig.~\ref{fig:2}b.}
\label{fig:s1}
\end{figure}

\begin{figure}[htbp]
 \centering
 \includegraphics[width=\textwidth]{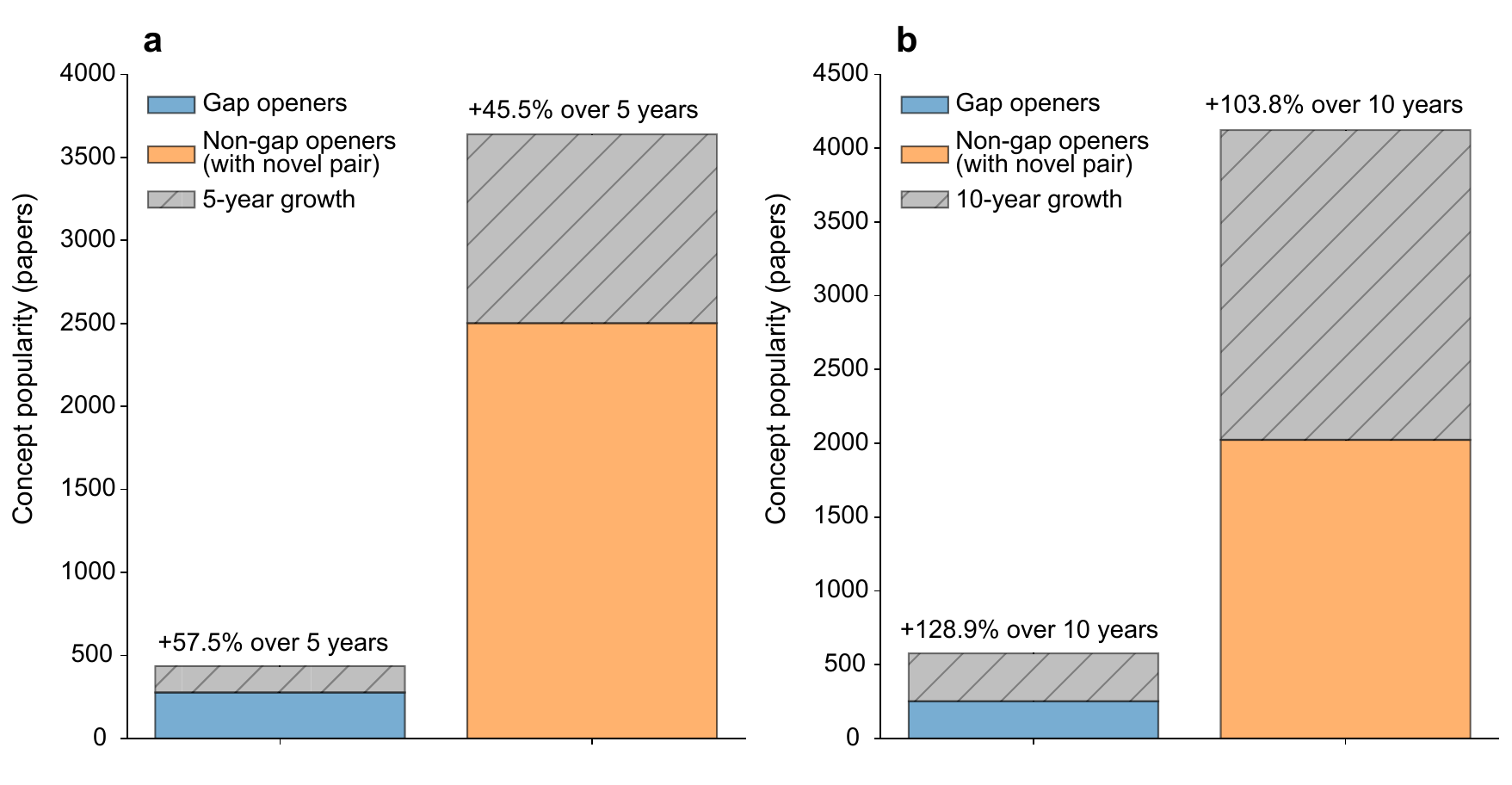}
\caption{\textbf{Concept popularity of gap-opening papers: occurrences five and ten years after publication.} Comparison of gap openers (blue) and non-gap openers (with novel pair) (orange). \textbf{(a)} Average occurrences of the two linked concepts five years after publication, for papers published up to and including 2016 (n = 8,528,430). \textbf{(b)} Average occurrences ten years after publication, for papers published up to and including 2011 (n = 6,785,227). Gray segments represent the increase in concept popularity over the respective time windows. Error bars indicate standard errors.}
\label{fig:s2}
\end{figure}

\begin{figure}[htbp]
 \centering
 \includegraphics[width=\textwidth]{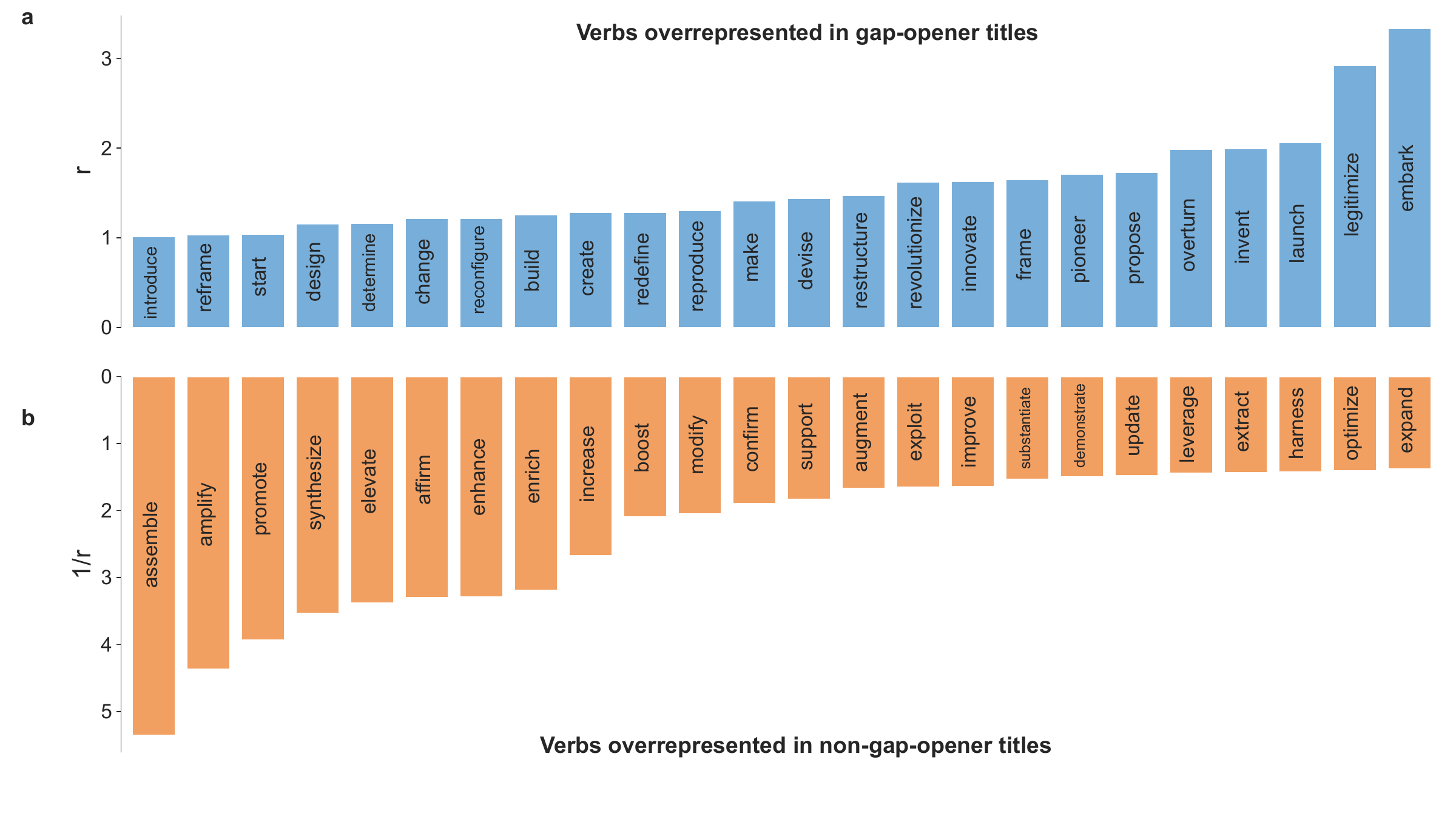}
\caption{\textbf{Linguistic characteristics of gap-opening papers: verb frequency differences between gap openers and non-gap openers (with novel pair).} For verbs appearing in the titles of both groups, we compute the relative frequency ratio $r$, defined as the proportion of titles containing a given verb in the gap-opener group divided by the corresponding proportion in the non-gap-opener (with novel pair) group. The comparison includes 274,230 gap-opening papers with titles and 9,144,020 non-gap openers (with novel pair) with titles. \textbf{(a)} Verbs with $r > 1$. \textbf{(b)} Verbs with $r < 1$, displayed as $1/r$ for visual comparability. Gap-opening papers more frequently use verbs signaling creation (e.g., `build', `make', `invent') and reframing (e.g., `redefine', `restructure', `overturn'). In contrast, non-gap openers (with novel pair) more often use verbs signaling demonstration (e.g., `affirm', `confirm', `support', `demonstrate'), improvement (e.g., `amplify', `promote', `enhance', `improve'), and exploitation (e.g., `expand', `leverage', `extract', `harness').} \label{fig:s3}
\end{figure}

\begin{figure}[htbp]
 \centering
 \includegraphics[width=\textwidth]{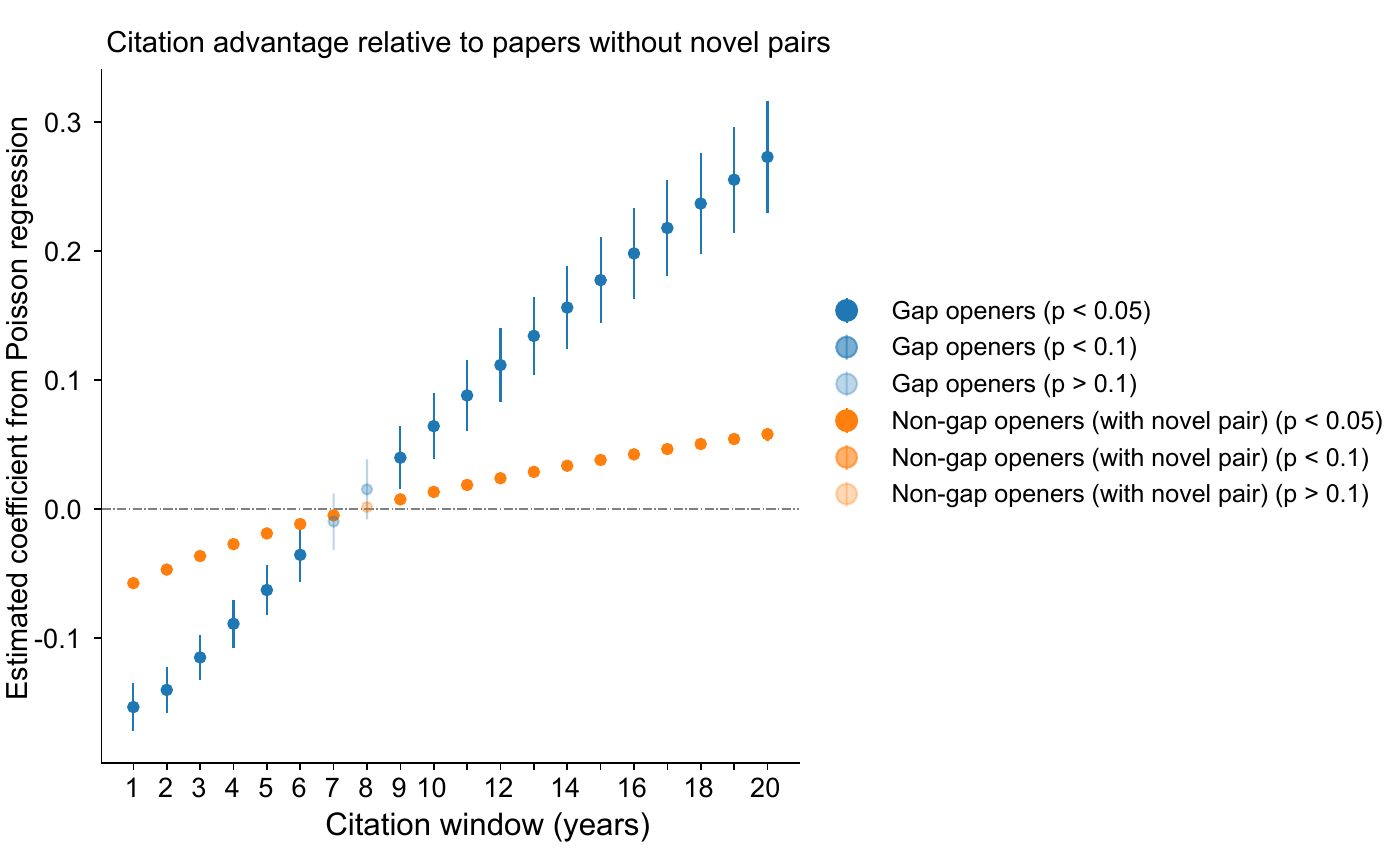}

\caption{\textbf{Citation trajectories of gap-opening papers: estimated coefficients across citation windows (1--20 years).} We estimate Poisson regressions for citation counts at each window from 1 to 20 years after publication. Each point shows the estimated coefficient for gap openers (blue) and non-gap openers (with novel pair) (orange) relative to the baseline group, non-gap openers (without novel pair). The sample includes 8,643,571 papers published 1980--2001 with team sizes $\leq$20 and complete citation data for all windows (C1--C20). Error bars indicate 95\% confidence intervals. Fig.~\ref{fig:s6} replicates the citation analysis on the more restrictive sample that also requires complete CD index data (n = 3,312,397) and extends it to disruption measures; Table~\ref{table:s4} reports pooled estimates across all windows for that restrictive sample.}
\label{fig:s4}
\end{figure}

\begin{figure}[htbp]
 \centering
 \includegraphics[width=\textwidth]{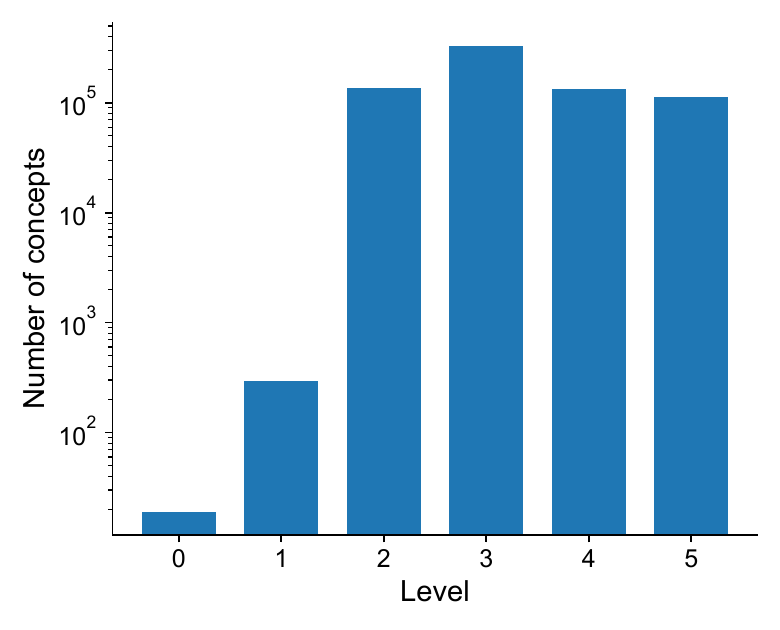}
\caption{\textbf{Microsoft Academic Graph taxonomy: number of concepts at each hierarchical level.} Number of distinct concepts at each level (0--5) of the full Microsoft Academic Graph taxonomy. Of the level-3 concepts, 100,650 appear in our corpus. The vertical axis is plotted on a logarithmic scale.} 
\label{fig:s5}
\end{figure}

\begin{figure}[htbp]
 \centering
 \includegraphics[width=\textwidth]{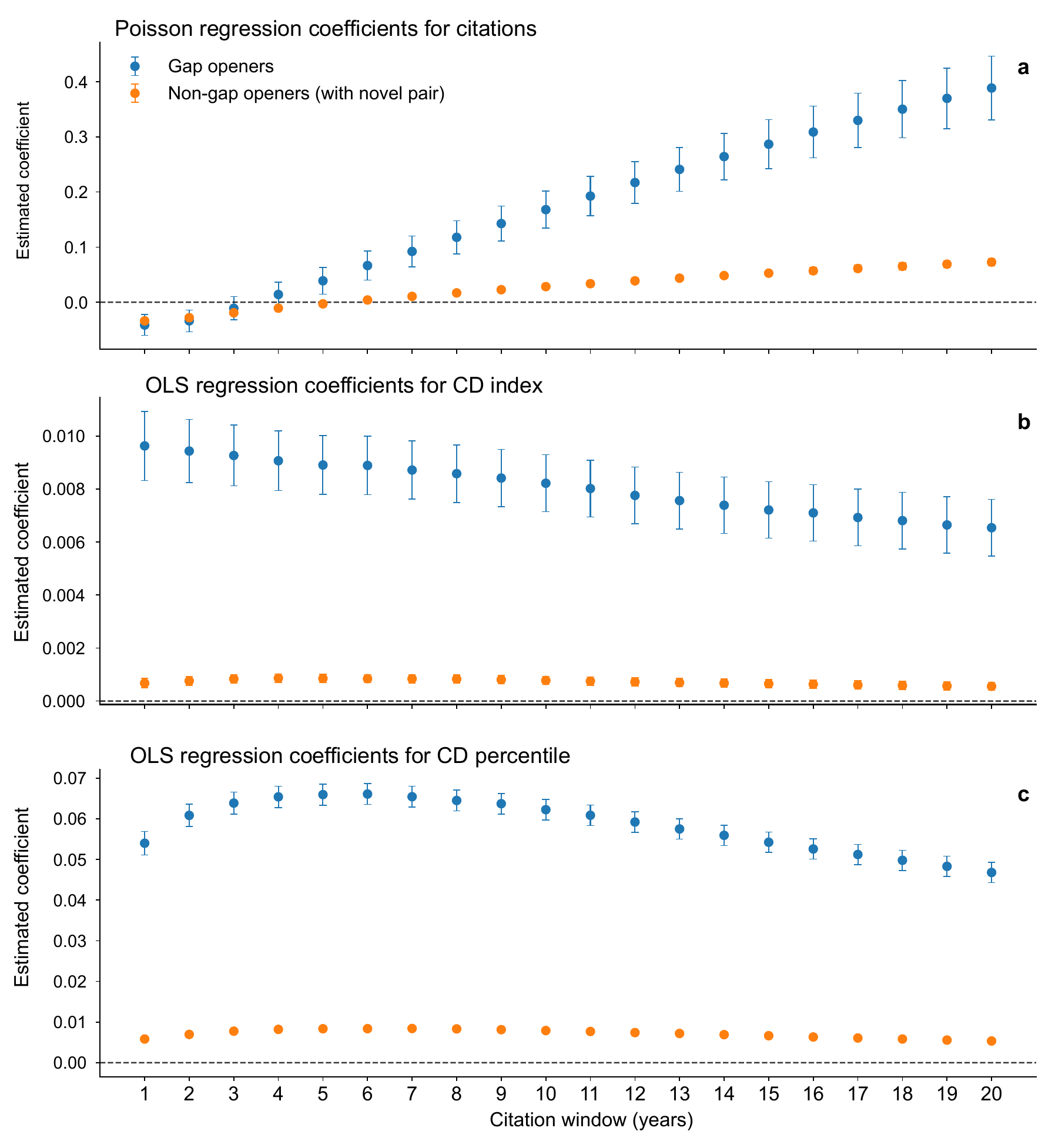}
\caption{\textbf{Citation and disruption trajectories of gap-opening papers: estimated coefficients across citation windows (1--20 years).} We estimate Poisson regressions for citation counts and OLS regressions for the CD index and CD percentile at each window from 1 to 20 years after publication. Each point reports the coefficient for gap openers (blue) and non-gap openers (with novel pair) (orange) relative to the baseline group, non-gap openers (without novel pair). The sample includes 3,312,397 papers published 1980--2001 with team sizes $\leq$20 and complete data on citations and the CD index for all windows (C1--C20 and CD1--CD20). Error bars indicate 95\% confidence intervals. The crossover of the citation coefficients from negative to positive occurs earlier in this more restrictive sample (around year 4 for gap openers) than in Fig.~\ref{fig:s4} (around year 8). Table~\ref{table:s4} reports the corresponding pooled estimates across all windows. The citation trajectory in panel (a) replicates Fig.~\ref{fig:s4} on this more restrictive sample; panels (b) and (c) extend the analysis to the CD index and CD percentile.}
\label{fig:s6}
\end{figure}

\revision{The proportion of gap-opening papers varies substantially across fields, from 0.47\% in Medicine to 5.78\% in Art. This analysis evaluates whether that variation reflects publication volume, gap-closure speed, the number of available concepts, or differences in the relationship between gap opening and citation impact across fields (Fig.~\ref{fig:s1}).}

\revision{\textbf{\emph{Publication volume.}} Much of the cross-field variation in proportions reflects differences in the size of the denominator. Fast-moving fields generate many gap openers in absolute terms---Medicine (35,840), Biology (34,891), and Computer Science (27,847)---but these counts are drawn from much larger publication volumes (e.g., Medicine 7,645,783 papers; Biology 6,517,210; Computer Science 2,487,978), yielding small proportions (Fig.~\ref{fig:s1}a). Slower-moving fields have far smaller corpora (Art 36,120; History 91,215; Philosophy 64,908) and correspondingly fewer gap openers (Art 2,086; History 4,164; Philosophy 2,928), yet much higher proportions.}

\revision{\textbf{\emph{Gap-closure speed.}} Publication volume alone does not fully account for the pattern. We find that gaps also close substantially faster in fast-moving fields (Fig.~\ref{fig:s1}b). The average gap duration is 6.61 years in Computer Science, 8.70 years in Medicine, and 8.83 years in Biology, compared with 11.56 years in History, 11.51 years in Philosophy, and 11.60 years in Art. In fast-moving fields, newly introduced conceptual recombinations are followed up and consolidated more quickly, so fewer gaps remain open for long. In slower-moving fields, gaps persist much longer before the community closes them. Gap duration therefore tracks the pace at which each field consolidates new recombinations, complementing the publication-volume account of the prevalence differences.}

\revision{\textbf{\emph{Concept counts.}} An alternative explanation is that slower-moving fields have more concepts, creating more combinatorial opportunities for gap formation. The data do not support this account. The number of distinct level-3 concepts is much larger in Medicine (38,432), Biology (67,858), and Computer Science (29,671) than in History (8,663), Philosophy (6,184), and Art (5,797) (Fig.~\ref{fig:s1}c).}

\revision{\textbf{\emph{Citation impact.}} Gap openers received more citations than non-gap openers (with novel pair) and non-gap openers (without novel pair) in most fields, with particularly strong separation in some fast-moving fields (e.g., Computer Science: 44.9 vs 21.6 and 14.8 mean citations) and the same pattern in slower-moving fields as well (e.g., Art: 5.4 vs 3.1 and 1.8; Fig.~\ref{fig:s1}d).}

\medskip

\revision{Taken together, these results point to a consistent explanation: the cross-field differences in gap-opener prevalence reflect publication volume and gap-closure speed rather than differences in concept counts or the meaningfulness of gap opening itself. These patterns align with prior work on how knowledge develops across fields \citep{biglan1973,kolb1981,becher1994significance}. Fields in the physical and biological sciences, characterized by more unified conceptual frameworks and cumulative progress \citep{popper1963,becher1994significance,cole1995}, tend to fill gaps quickly, reducing their observable prevalence. Fields in the social sciences and humanities, characterized by greater pluralism and weaker consensus \citep{cole1994,cole2006,becher1994significance}, close gaps more slowly, allowing them to persist longer. Importantly, gap opening is associated with higher citation impact across fields, suggesting that the creation of topological gaps marks generative work across fields of differing pace and scale. The ordering is not uniform in every field---in Biology, Engineering, and Environmental science the three groups are close together or slightly reordered---but gap openers lead in the large majority.}

\clearpage

\subsection*{Additional analyses supporting the main results}

\revision{The remaining figures and tables provide additional evidence on the characteristics of gap-opening papers, the MAG taxonomy, citation and disruption trajectories, and the regression models underlying the main-text results.}

\clearpage

\clearpage

\clearpage

\clearpage

\clearpage

\begin{table}[htbp]
\centering
\caption{Generative capacity of gap-opening papers across citation impact, disruption, Nobel prizes, and delayed recognition.}\label{table:s1}

\begin{adjustbox}{max width=\textwidth,center}
\begin{threeparttable}
\begin{tabular}{lcccccc}
\toprule

 & (1) & (2) & (3) & (4) & (5) & (6) \\

 & \makecell{Top 1\%\\most cited} & \makecell{CD\\percentile} & \makecell{Nobel\\prize wins} & \makecell{SB index\\(log)} & C5 & C20 \\

 & \emph{Logistic} & \emph{OLS} & \emph{Logistic} & \emph{OLS} & \emph{Poisson} & \emph{Poisson} \\

\midrule

Gap openers (with novel pair) & 0.454*** & 0.021*** & 0.736 & 0.121*** & $-$0.063*** & 0.273*** \\
 & (0.0206) & (0.0009) & (1.0266) & (0.0017) & (0.0099) & (0.0220) \\[4pt]

Non-gap openers (with novel pair) & $-$0.001 & 0.003*** & 0.307 & 0.022*** & $-$0.019*** & 0.058*** \\
 & (0.0041) & (0.0001) & (0.2251) & (0.0002) & (0.0017) & (0.0026) \\[4pt]

Non-gap openers (without novel pair) & Baseline & Baseline & Baseline & Baseline & Baseline & Baseline \\[6pt]

Team size & 0.117*** & $-$0.002*** & 0.218*** & $-$0.008*** & 0.099*** & 0.076*** \\
 & (0.0006) & (0.0000) & (0.0248) & (0.0000) & (0.0004) & (0.0005) \\[4pt]

Number of references (log) & 0.741*** & $-$0.040*** & 0.159** & $-$0.011*** & 0.242*** & 0.254*** \\
 & (0.0023) & (0.0001) & (0.0600) & (0.0001) & (0.0005) & (0.0008) \\[4pt]

Citations (log) & -- & $-$0.041*** & -- & 0.001*** & -- & -- \\
 & & (0.0000) & & (0.0000) & & \\

\midrule

Year fixed effects & \checkmark & \checkmark & \checkmark & \checkmark & \checkmark & \checkmark \\
Field fixed effects & \checkmark & \checkmark & \checkmark & \checkmark & \checkmark & \checkmark \\

\midrule

R$^2$ & 0.106 & 0.146 & 0.076 & 0.264 & 0.197 & 0.183 \\
Observations & 31,983,201 & 21,613,648 & 8,857,237 & 25,537,277 & 8,643,571 & 8,643,571 \\

\bottomrule
\end{tabular}

\begin{tablenotes}
\footnotesize
\item \emph{Notes:} This table compares gap-opening papers, papers that introduce novel concept pairs without opening gaps, and papers that introduce no novel concept pairs across measures of citation impact, disruption, Nobel Prize wins, and delayed recognition. The dependent variables are as follows: Top 1\% most cited is a binary indicator for papers ranking in the top 1\% of citations within the same year and field; CD percentile is the within-year percentile rank of the CD index, which measures whether a paper's influence displaces or consolidates prior work; Nobel Prize wins is a binary indicator for papers identified as Nobel Prize-winning work; SB index is the Sleeping Beauty Index, which measures delayed recognition (log-transformed); C5 and C20 are cumulative citation counts received within 5 and 20 years of publication, respectively. The unit of analysis is the paper; each paper contributes one observation. All models are restricted to papers published 1980--2020 with team sizes $\leq$20, except the C5 and C20 models, which cover 1980--2001; the two restrictions jointly retain 93\% of the corpus, and the team-size cap excludes the extreme tail of hyperauthored papers (see \emph{Materials and Methods} for justification of both restrictions). Sample sizes vary across models due to additional requirements: the CD percentile model requires $\geq$1 reference and non-missing index values; the SB index model requires non-missing index values; and the C5/C20 models require non-missing values for both citation measures simultaneously to ensure comparability across citation windows, which effectively restricts the sample to papers published early enough to have at least 20 years of post-publication citation data. The Nobel Prize model has a smaller sample because the rarity of Nobel-linked papers causes perfect or quasi-complete separation in the logistic regression; observations whose covariate combinations perfectly predict the outcome are dropped from estimation. See \emph{Materials and Methods} for variable definitions and estimation details. Fig.~\ref{fig:4} plots the predicted values from these models. Heteroskedasticity-robust standard errors in parentheses. For logistic and Poisson models, R$^2$ denotes pseudo-R$^2$.
\item \checkmark\ indicates the variable or fixed effect is included.
\item *$p<0.05$, **$p<0.01$, ***$p<0.001$.
\end{tablenotes}
\end{threeparttable}
\end{adjustbox}
\end{table}

\clearpage

\begin{table}[htbp]
\centering
\caption{Citation impact of gap-opening papers at the top 5\%, 10\%, 15\%, and 20\% thresholds.}\label{table:s2}
\begin{threeparttable}
\begin{tabular}{lcccc}
\toprule

 & (1) & (2) & (3) & (4) \\

 & \makecell{Top 5\%\\most cited} & \makecell{Top 10\%\\most cited} & \makecell{Top 15\%\\most cited} & \makecell{Top 20\%\\most cited} \\

 & \emph{Logistic} & \emph{Logistic} & \emph{Logistic} & \emph{Logistic} \\

\midrule

Gap openers (with novel pair) & 0.259*** & 0.184*** & 0.147*** & 0.127*** \\
 & (0.0106) & (0.0081) & (0.0070) & (0.0064) \\[4pt]

Non-gap openers (with novel pair) & $-$0.013*** & $-$0.009*** & $-$0.007*** & $-$0.005*** \\
 & (0.0019) & (0.0014) & (0.0012) & (0.0011) \\[4pt]

Non-gap openers (without novel pair) & Baseline & Baseline & Baseline & Baseline \\[6pt]

Team size & 0.103*** & 0.100*** & 0.100*** & 0.101*** \\
 & (0.0003) & (0.0002) & (0.0002) & (0.0002) \\[4pt]

Number of references (log) & 0.612*** & 0.567*** & 0.543*** & 0.531*** \\
 & (0.0009) & (0.0006) & (0.0005) & (0.0004) \\

\midrule

Year fixed effects & \checkmark & \checkmark & \checkmark & \checkmark \\
Field fixed effects & \checkmark & \checkmark & \checkmark & \checkmark \\

\midrule

Pseudo-R$^2$ & 0.112 & 0.120 & 0.127 & 0.135 \\
Observations & 31,983,201 & 31,983,201 & 31,983,201 & 31,983,201 \\

\bottomrule
\end{tabular}

\begin{tablenotes}
\footnotesize
\item \emph{Notes:} This table compares gap-opening papers, papers that introduce novel concept pairs without opening gaps, and papers that introduce no novel concept pairs in their likelihood of ranking among the top 5\%, 10\%, 15\%, and 20\% most cited papers published in the same year and field. The models extend the Top 1\% most-cited specification in Table~\ref{table:s1} to four additional citation thresholds. Each dependent variable is a binary indicator for whether a paper ranks in the top $k$\% of citations within the same year and field, where $k \in \{5, 10, 15, 20\}$. The unit of analysis is the paper; each paper contributes one observation. The sample and specification are identical to the Top 1\% model in Table~\ref{table:s1}: papers published 1980--2020 with team sizes $\leq$20. See \emph{Materials and Methods} for variable definitions and estimation details. Heteroskedasticity-robust standard errors in parentheses.
\item \checkmark\ indicates the variable or fixed effect is included.
\item *$p<0.05$, **$p<0.01$, ***$p<0.001$.
\end{tablenotes}

\end{threeparttable}
\end{table}

\clearpage

\begin{table}[htbp]
\centering
\caption{Joint atypicality--disruption profiles of the three paper types.}\label{table:s3}
\begin{threeparttable}
\begin{tabular}{lcccc}
\toprule

 & (1) & (2) & (3) & (4) \\

 & \makecell{Atypical=1\\ Disruptive=1} & \makecell{Atypical=1\\ Disruptive=0} & \makecell{Atypical=0\\ Disruptive=1} & \makecell{Atypical=0\\ Disruptive=0\\(baseline outcome)} \\

\midrule

Gap openers (with novel pair) & 0.796*** & 0.697*** & 0.175*** & -- \\
 & (0.0107) & (0.0085) & (0.0098) & \\[4pt]

Non-gap openers (with novel pair) & 0.469*** & 0.486*** & 0.055*** & -- \\
 & (0.0017) & (0.0012) & (0.0017) & \\[4pt]

Non-gap openers (without novel pair) & Baseline & Baseline & Baseline & -- \\[6pt]

Team size & $-$0.008*** & 0.036*** & $-$0.038*** & -- \\
 & (0.0003) & (0.0002) & (0.0003) & \\[4pt]

Number of references (log) & $-$0.106*** & 0.483*** & $-$0.785*** & -- \\
 & (0.0009) & (0.0006) & (0.0009) & \\

\midrule

Year fixed effects & \checkmark & \checkmark & \checkmark & -- \\
Field fixed effects & \checkmark & \checkmark & \checkmark & -- \\

\midrule

Pseudo-R$^2$ & \multicolumn{4}{c}{0.097} \\
Observations & \multicolumn{4}{c}{20,880,896} \\

\bottomrule
\end{tabular}

\begin{tablenotes}
\footnotesize
\item \emph{Notes:} This table compares gap-opening papers, papers that introduce novel concept pairs without opening gaps, and papers that introduce no novel concept pairs across four outcomes---being both atypical and disruptive, being atypical only, being disruptive only, or being neither atypical nor disruptive. The estimates come from an unordered multinomial logistic regression. Papers are classified as atypical if they exhibit a positive A~index ($\text{A index} > 0$, indicating that the most atypical journal pairing in the reference list is rarer than expected by chance) and as disruptive if they exhibit a positive CD index ($\text{CD index} > 0$). The ``neither'' category serves as the baseline outcome; coefficients in columns (1)--(3) are estimated relative to this category. The unit of analysis is the paper; each paper contributes one observation. The sample is restricted to papers with non-missing values for both the A~index and CD~index, yielding a smaller sample than the CD percentile model in Table~\ref{table:s1}. See \emph{Materials and Methods} for variable definitions and estimation details. Robust standard errors in parentheses.
\item \checkmark\ indicates the variable or fixed effect is included.
\item *$p<0.05$, **$p<0.01$, ***$p<0.001$.
\end{tablenotes}

\end{threeparttable}
\end{table}

\clearpage

\begin{table}[htbp]
\centering
\caption{Citation and disruption trajectories of gap-opening papers pooled across citation windows.}\label{table:s4}
\begin{threeparttable}
\begin{tabular}{lccc}
\toprule

 & (1) & (2) & (3) \\

 & Citations & \makecell{CD\\index} & \makecell{CD\\percentile} \\

 & \emph{Poisson} & \emph{OLS} & \emph{OLS} \\

\midrule

Gap openers (with novel pair) & 0.251*** & 0.008*** & 0.060*** \\
 & (0.0213) & (0.0005) & (0.0012) \\[4pt]

Non-gap openers (with novel pair) & 0.042*** & 0.001*** & 0.008*** \\
 & (0.0027) & (0.0001) & (0.0003) \\[4pt]

Non-gap openers (without novel pair) & Baseline & Baseline & Baseline \\[6pt]

Team size & 0.065*** & $-$0.001*** & $-$0.006*** \\
 & (0.0006) & (0.0000) & (0.0001) \\[4pt]

Number of references (log) & 0.347*** & $-$0.033*** & $-$0.123*** \\
 & (0.0016) & (0.0001) & (0.0001) \\[4pt]

Citation window & 0.089*** & $-$0.001*** & $-$0.009*** \\
 & (0.0001) & (0.0000) & (0.0000) \\[4pt]

Citations (log) & -- & 0.009*** & 0.061*** \\
 & & (0.0000) & (0.0001) \\

\midrule

Year fixed effects & \checkmark & \checkmark & \checkmark \\
Field fixed effects & \checkmark & \checkmark & \checkmark \\

\midrule

R$^2$ & 0.257 & 0.228 & 0.324 \\
Observations & 66,247,940 & 66,247,940 & 66,247,940 \\

\bottomrule
\end{tabular}

\begin{tablenotes}
\footnotesize
\item \emph{Notes:} This table compares gap-opening papers, papers that introduce novel concept pairs without opening gaps, and papers that introduce no novel concept pairs in citation counts, the CD index, and the CD percentile across citation windows from 1 to 20 years after publication. These models pool observations across citation windows from 1 to 20 years after publication, stacking the data at the paper--window level to estimate average effects across horizons. The unit of analysis is the paper--window pair; each paper contributes one observation per citation window. The citation window variable controls for the length of the observation window. The large observation count reflects this stacked structure (each paper contributes 20 observations, one per window). The sample includes papers published 1980--2001 with team sizes $\leq$20 and complete citation and CD index data for all windows (C1--C20), yielding 3,312,397 unique papers (66,247,940 paper--window observations); the earlier end date ensures at least 20 years of post-publication data within the observation period. These pooled estimates complement the window-by-window coefficients plotted in Figs.~\ref{fig:s4} and~\ref{fig:s6}. See \emph{Materials and Methods} for variable definitions and estimation details. Standard errors in parentheses are clustered by paper. For the Poisson model, R$^2$ denotes pseudo-R$^2$.
\item \checkmark\ indicates the variable or fixed effect is included.
\item *$p<0.05$, **$p<0.01$, ***$p<0.001$.
\end{tablenotes}

\end{threeparttable}
\end{table}

\clearpage
\section*{Robustness tests, gap-level analyses, and replication}

\revision{The analyses below evaluate whether the main findings depend on the concept taxonomy, data source, model specification, comparison-group construction, or established predictors of scientific impact. They also examine whether gap duration and gap size are associated with the reception of gap-opening papers and the subsequent research those gaps organize.}

\subsection*{Robustness to concept-taxonomy level}
We first test whether the findings depend on the level of the Microsoft Academic Graph (MAG) concept taxonomy used to define concepts. We re-identify gap openers using all MAG concepts from Levels 3 through 5 rather than Level 3 alone. Levels 0 through 2 are excluded because they denote highly aggregated disciplinary labels---for example, Level 0 includes Computer Science and Biology---rather than specific concepts. The top 1\% most cited and CD percentile outcomes for gap openers retain their sign and significance under this broader concept set (Table~\ref{tab:taxonomy-robustness}). Several estimates change (Table~\ref{tab:taxonomy-full}). The effect of gap openers on 5-year citations turns from negative to positive, though it remains smaller than that of non-gap openers (with novel pair); the effects of both gap openers and non-gap openers (with novel pair) on Nobel-prize wins become significantly positive, with gap openers showing the larger coefficient; and non-gap openers (with novel pair) show a small positive top 1\% coefficient ($+0.099$, odds ratio 1.10) and a positive 5-year citation coefficient, where under Level 3 alone their coefficient is near zero and nonsignificant ($-0.001$, $p>0.05$; Table~\ref{table:s1}). The gap-opener coefficient remains more than five times larger ($+0.556$, odds ratio 1.74). These shifts share one cause, and the contrast that the paper rests on is unaffected by it. The Levels 3--5 criterion admits papers carrying fewer than two Level-3 concepts, so the estimation sample exceeds the 34,363,623-paper main corpus, and a first-time Level-5 pairing is far easier to achieve than a first-time Level-3 pairing, which narrows the baseline group to papers whose every concept pair already exists. Because the baseline is what each coefficient is measured against, the citation-count and top-percentile coefficients rise together, and the gap-opener and non-gap-opener top 1\% coefficients each increase by almost exactly $0.10$. Their difference---the increment attributable to opening a gap rather than merely introducing a novel pair---is $0.455$ under Level 3 and $0.457$ under Levels 3--5, a gap of $0.002$ against a standard error of $0.021$. The reference category moves; the topological increment does not. We therefore do not interpret the top 1\% shift as evidence that recombination alone increases the probability of top 1\% most-cited status, and we rest the topological claim on the nested and within-sample specifications, which hold the novel pair fixed and estimate that increment directly (Tables~\ref{tab:nested-model}--\ref{tab:within-sample}).

\begin{table}[p]
\centering
\caption{Robustness of gap-opener effects across Microsoft Academic Graph (MAG) taxonomy levels.}\label{tab:taxonomy-robustness}
\begin{adjustbox}{max width=\textwidth,center}
\begin{threeparttable}\footnotesize
\begin{tabular}{lcc}
\toprule
 & (1) & (2) \\
 & \makecell{Top 1\%\\ most cited\\ (Logistic)} & \makecell{CD\\ percentile\\ (OLS)} \\
\midrule
Gap openers (with novel pair) & \makecell{0.556***\\ (0.0172)} & \makecell{0.013***\\ (0.0006)} \\
Non-gap openers (with novel pair) & \makecell{0.099***\\ (0.0033)} & \makecell{0.002***\\ (0.0001)} \\
Non-gap openers (without novel pair) & Baseline & Baseline \\
\midrule
Controls & \checkmark & \checkmark \\
Year fixed effects & \checkmark & \checkmark \\
Field fixed effects & \checkmark & \checkmark \\
\midrule
R$^2$ & 0.1090 & 0.1655 \\
Observations & 38,902,733 & 25,729,819 \\
\bottomrule
\end{tabular}
\begin{tablenotes}
\footnotesize
\item \emph{Notes:} This table compares gap-opening papers, papers that introduce novel concept pairs without opening gaps, and papers that introduce no novel concept pairs on top 1\% most-cited status and CD percentile after re-identifying gaps using MAG concepts from Levels 3--5 rather than Level 3 alone. The top 1\% most-cited model is logistic, and the CD percentile model is OLS. Both models are re-estimated using MAG concepts at taxonomy levels 3--5. Because the levels 3--5 vocabulary assigns qualifying concepts to more papers than level 3 alone, the estimation sample is larger than in the main analysis. The unit of analysis is the paper; each paper contributes one observation. Controls: Team size, Number of references (log). Robust standard errors in parentheses. For logistic models, R$^2$ denotes pseudo-R$^2$.
\item \checkmark\ indicates the variable or fixed effect is included.
\item *$p<0.05$, **$p<0.01$, ***$p<0.001$.
\end{tablenotes}
\end{threeparttable}
\end{adjustbox}
\end{table}
\clearpage
\begin{table}[p]
\centering
\caption{Gap-opener effects across all outcomes using MAG Levels 3--5 concepts.}\label{tab:taxonomy-full}
\begin{adjustbox}{max width=\textwidth,center}
\begin{threeparttable}\footnotesize
\begin{tabular}{lccccccc}
\toprule
 & (1) & (2) & (3) & (4) & (5) & (6) & (7) \\
 & \makecell{C5\\ (Poisson)} & \makecell{C20\\ (Poisson)} & \makecell{SB index (log)\\ (OLS)} & \makecell{Top 1\%\\ most cited\\ (Logistic)} & \makecell{CD\\ percentile\\ (OLS)} & \makecell{A index\\ (OLS)} & \makecell{Nobel\\ prize\\ wins\\ (Logistic)} \\
\midrule
Gap openers (with novel pair) & \makecell{0.086***\\ (0.0092)} & \makecell{0.429***\\ (0.0185)} & \makecell{0.110***\\ (0.0014)} & \makecell{0.556***\\ (0.0172)} & \makecell{0.013***\\ (0.0006)} & \makecell{1.142***\\ (0.0112)} & \makecell{2.547***\\ (0.6302)} \\
Non-gap openers (with novel pair) & \makecell{0.177***\\ (0.0015)} & \makecell{0.211***\\ (0.0022)} & \makecell{0.001***\\ (0.0002)} & \makecell{0.099***\\ (0.0033)} & \makecell{0.002***\\ (0.0001)} & \makecell{0.639***\\ (0.0014)} & \makecell{1.686***\\ (0.2808)} \\
Non-gap openers (without novel pair) & Baseline & Baseline & Baseline & Baseline & Baseline & Baseline & Baseline \\
\midrule
Controls & \checkmark & \checkmark & \checkmark & \checkmark & \checkmark & \checkmark & \checkmark \\
Year fixed effects & \checkmark & \checkmark & \checkmark & \checkmark & \checkmark & \checkmark & \checkmark \\
Field fixed effects & \checkmark & \checkmark & \checkmark & \checkmark & \checkmark & \checkmark & \checkmark \\
\midrule
R$^2$ & 0.2079 & 0.1924 & 0.2669 & 0.1090 & 0.1655 & 0.1949 & 0.1040 \\
Observations & 10,438,643 & 10,438,643 & 30,580,814 & 38,902,733 & 25,729,819 & 26,997,213 & 11,429,388 \\
\bottomrule
\end{tabular}
\begin{tablenotes}
\footnotesize
\item \emph{Notes:} This table compares gap-opening papers, papers that introduce novel concept pairs without opening gaps, and papers that introduce no novel concept pairs across citation impact, disruption, atypicality, Nobel Prize wins, and delayed recognition after re-identifying gaps using all MAG concepts from Levels 3--5 rather than Level 3 alone. The models use every MAG concept from Levels 3--5. Because the levels 3--5 vocabulary assigns qualifying concepts to more papers than level 3 alone, the estimation sample is larger than in the main analysis. The unit of analysis is the paper; each paper contributes one observation. Controls: Team size, Number of references (log). Robust standard errors in parentheses. For logistic and Poisson models, R$^2$ denotes pseudo-R$^2$.
\item \checkmark\ indicates the variable or fixed effect is included.
\item *$p<0.05$, **$p<0.01$, ***$p<0.001$.
\end{tablenotes}
\end{threeparttable}
\end{adjustbox}
\end{table}

\clearpage
\subsection*{Replication across databases and concept systems}
We next test whether the citation-impact and disruption results replicate beyond MAG and its concept taxonomy. Tables~\ref{tab:replication-openalex}--\ref{tab:replication-patents} report the full regressions underlying the replication analysis summarized in main-text Fig.~\ref{fig:replication} for OpenAlex (topics), the PubMed Knowledge Graph (MeSH descriptors), Dimensions (text-derived concepts, no curated taxonomy), American Physical Society (APS) journals (text-derived concepts and Physics and Astronomy Classification Scheme [PACS] codes), and U.S. patents (Cooperative Patent Classification [CPC] codes at two levels of granularity). Data sources, inclusion criteria, and corpus-specific processing are documented in the section ``Replication data sources and network construction'' below. Two measurement sensitivities are worth noting explicitly. First, in APS journals the disruption effect is robust under both vocabularies, whereas the association between gap opening and top 1\% most-cited status is larger under text-derived concepts (odds ratio 3.20) than under PACS codes (odds ratio 1.45; both $p<0.01$; Table~\ref{tab:replication-aps}). Second, in the PubMed Knowledge Graph the gap-opener disruption effect ($+0.8$ percentile points) does not exceed that of non-gap openers (with novel pair) ($+1.1$; Table~\ref{tab:replication-pubmed})---the only database in which this ordering reverses.

\begin{table}[p]
\centering
\caption{Replication of gap-opener effects in OpenAlex.}\label{tab:replication-openalex}
\begin{adjustbox}{max width=\textwidth,center}
\begin{threeparttable}
\begin{tabular}{lcc}
\toprule
 & (1) & (2) \\
 & Top 1\% most cited & CD percentile \\
 & (Logistic) & (OLS) \\
\midrule
Gap openers (with novel pair) & 0.413*** & 0.030*** \\
 & (0.0176) & (0.0006) \\
Non-gap openers (with novel pair) & 0.115*** & 0.010*** \\
 & (0.0042) & (0.0001) \\
Non-gap openers (without novel pair) & Baseline & Baseline \\
\midrule
Controls & \checkmark & \checkmark \\
Year fixed effects & \checkmark & \checkmark \\
Discipline fixed effects & \checkmark & \checkmark \\
\midrule
R$^2$ & 0.095 & 0.165 \\
Observations & 101,638,624 & 101,638,624 \\
\bottomrule
\end{tabular}
\begin{tablenotes}
\footnotesize
\item \emph{Notes:} This table compares gap-opening papers and papers that introduce novel concept pairs without opening gaps with papers that introduce no novel concept pairs on top 1\% most-cited status and CD percentile in OpenAlex, using the OpenAlex topic classification. Top 1\% most-cited status is estimated with logistic regression, and CD percentile is estimated with OLS. CD percentile is reported on the manuscript's 0--1 scale. The unit of analysis is the paper--discipline pair; a paper contributes one observation for each discipline it belongs to, so $N$ counts paper--discipline observations rather than unique papers. Controls: Team size, Number of references (log); the CD model adds Citations (log). Standard errors in parentheses are clustered by paper to account for the repeated observations. For logistic models, R$^2$ denotes pseudo-R$^2$.
\item \checkmark\ indicates the variable or fixed effect is included.
\item *$p<0.05$, **$p<0.01$, ***$p<0.001$.
\end{tablenotes}
\end{threeparttable}
\end{adjustbox}
\end{table}
\clearpage
\begin{table}[p]
\centering
\caption{Replication of gap-opener effects in the PubMed Knowledge Graph.}\label{tab:replication-pubmed}
\begin{adjustbox}{max width=\textwidth,center}
\begin{threeparttable}
\begin{tabular}{lcc}
\toprule
 & (1) & (2) \\
 & Top 1\% most cited & CD percentile \\
 & (Logistic) & (OLS) \\
\midrule
Gap openers (with novel pair) & 0.132*** & 0.008*** \\
 & (0.0372) & (0.0011) \\
Non-gap openers (with novel pair) & 0.014 & 0.011*** \\
 & (0.0145) & (0.0004) \\
Non-gap openers (without novel pair) & Baseline & Baseline \\
\midrule
Controls & \checkmark & \checkmark \\
Year fixed effects & \checkmark & \checkmark \\
Field fixed effects & \checkmark & \checkmark \\
\midrule
R$^2$ & 0.112 & 0.102 \\
Observations & 11,156,533 & 9,776,452 \\
\bottomrule
\end{tabular}
\begin{tablenotes}
\footnotesize
\item \emph{Notes:} This table compares gap-opening papers and papers that introduce novel concept pairs without opening gaps with papers that introduce no novel concept pairs on top 1\% most-cited status and CD percentile in the PubMed Knowledge Graph, using MeSH major-topic descriptors. Top 1\% most-cited status is estimated with logistic regression, and CD percentile is estimated with OLS. CD percentile is reported on the manuscript's 0--1 scale. The unit of analysis is the paper--field pair; a paper contributes one observation for each field it belongs to, so $N$ counts paper--field observations rather than unique papers. Controls: Team size, Number of references (log); the CD model adds Citations (log). Standard errors in parentheses are clustered by paper to account for the repeated observations. For logistic models, R$^2$ denotes pseudo-R$^2$.
\item \checkmark\ indicates the variable or fixed effect is included.
\item *$p<0.05$, **$p<0.01$, ***$p<0.001$.
\end{tablenotes}
\end{threeparttable}
\end{adjustbox}
\end{table}
\clearpage
\begin{table}[p]
\centering
\caption{Replication of gap-opener effects in Dimensions.}\label{tab:replication-dimensions}
\begin{adjustbox}{max width=\textwidth,center}
\begin{threeparttable}
\begin{tabular}{lcc}
\toprule
 & (1) & (2) \\
 & Top 1\% most cited & CD percentile \\
 & (Logistic) & (OLS) \\
\midrule
Gap openers (with novel pair) & 0.514*** & 0.015*** \\
 & (0.0058) & (0.0002) \\
Non-gap openers (with novel pair) & 0.147*** & 0.006*** \\
 & (0.0040) & (0.0001) \\
Non-gap openers (without novel pair) & Baseline & Baseline \\
\midrule
Controls & \checkmark & \checkmark \\
Year fixed effects & \checkmark & \checkmark \\
Field fixed effects & \checkmark & \checkmark \\
\midrule
R$^2$ & 0.131 & 0.208 \\
Observations & 69,040,797 & 60,293,158 \\
\bottomrule
\end{tabular}
\begin{tablenotes}
\footnotesize
\item \emph{Notes:} This table compares gap-opening papers and papers that introduce novel concept pairs without opening gaps with papers that introduce no novel concept pairs on top 1\% most-cited status and CD percentile in Dimensions, using concepts derived directly from text. Top 1\% most-cited status is estimated with logistic regression, and CD percentile is estimated with OLS. The concepts are derived directly from text rather than from a curated taxonomy. CD percentile is reported on the manuscript's 0--1 scale. The unit of analysis is the paper--field pair; a paper contributes one observation for each field it belongs to, so $N$ counts paper--field observations rather than unique papers. Controls: Team size, Number of references (log); the CD model adds Citations (log). Standard errors in parentheses are clustered by paper to account for the repeated observations. For logistic models, R$^2$ denotes pseudo-R$^2$.
\item \checkmark\ indicates the variable or fixed effect is included.
\item *$p<0.05$, **$p<0.01$, ***$p<0.001$.
\end{tablenotes}
\end{threeparttable}
\end{adjustbox}
\end{table}
\clearpage
\begin{table}[p]
\centering
\caption{Replication of gap-opener effects in APS journals under text-derived concepts and PACS codes.}\label{tab:replication-aps}
\begin{adjustbox}{max width=\textwidth,center}
\begin{threeparttable}
\begin{tabular}{lcccc}
\toprule
 & \multicolumn{2}{c}{Text-derived concepts} & \multicolumn{2}{c}{PACS codes} \\
\cmidrule(lr){2-3}\cmidrule(lr){4-5}
 & (1) & (2) & (3) & (4) \\
 & Top 1\% most cited & CD percentile & Top 1\% most cited & CD percentile \\
 & (Logistic) & (OLS) & (Logistic) & (OLS) \\
\midrule
Gap openers (with novel pair) & 1.162*** & 0.077*** & 0.372** & 0.048*** \\
 & (0.1510) & (0.0064) & (0.1241) & (0.0042) \\
Non-gap openers (with novel pair) & 0.180*** & 0.027*** & 0.308*** & 0.033*** \\
 & (0.0392) & (0.0011) & (0.0331) & (0.0009) \\
Non-gap openers (without novel pair) & Baseline & Baseline & Baseline & Baseline \\
\midrule
Controls & \checkmark & \checkmark & \checkmark & \checkmark \\
Year fixed effects & \checkmark & \checkmark & \checkmark & \checkmark \\
Field fixed effects & \checkmark & \checkmark & \checkmark & \checkmark \\
\midrule
R$^2$ & 0.061 & 0.097 & 0.055 & 0.107 \\
Observations & 479,904 & 418,540 & 380,366 & 342,923 \\
\bottomrule
\end{tabular}
\begin{tablenotes}
\footnotesize
\item \emph{Notes:} This table compares gap-opening papers, papers that introduce novel concept pairs without opening gaps, and papers that introduce no novel concept pairs on top 1\% most-cited status and CD percentile in APS journals, identifying gaps separately with concepts derived from abstracts and PACS codes. Under each concept vocabulary, top 1\% most-cited status is estimated with logistic regression and CD percentile with OLS; columns (1)--(2) use concepts derived from abstracts, and columns (3)--(4) use PACS classification codes. CD percentile is reported on the manuscript's 0--1 scale. The unit of analysis is the paper; each paper contributes one observation. Each APS paper has a single field assignment. Controls: Team size, Number of references (log); the CD model adds Citations (log). Heteroskedasticity-robust standard errors in parentheses. For logistic models, R$^2$ denotes pseudo-R$^2$.
\item \checkmark\ indicates the variable or fixed effect is included.
\item *$p<0.05$, **$p<0.01$, ***$p<0.001$.
\end{tablenotes}
\end{threeparttable}
\end{adjustbox}
\end{table}
\clearpage
\begin{table}[p]
\centering
\caption{Replication of gap-opener effects in U.S. patents under CPC codes at two levels of granularity.}\label{tab:replication-patents}
\begin{adjustbox}{max width=\textwidth,center}
\begin{threeparttable}
\begin{tabular}{lcccc}
\toprule
 & \multicolumn{2}{c}{CPC subgroups (granular)} & \multicolumn{2}{c}{CPC subclasses (coarse)} \\
\cmidrule(lr){2-3}\cmidrule(lr){4-5}
 & (1) & (2) & (3) & (4) \\
 & Top 1\% most cited & CD percentile & Top 1\% most cited & CD percentile \\
 & (Logistic) & (OLS) & (Logistic) & (OLS) \\
\midrule
Gap openers (with novel pair) & 1.378*** & 0.028*** & 1.046*** & 0.030*** \\
 & (0.0223) & (0.0008) & (0.1080) & (0.0047) \\
Non-gap openers (with novel pair) & 0.732*** & 0.027*** & 0.527*** & 0.011*** \\
 & (0.0084) & (0.0002) & (0.0639) & (0.0024) \\
Non-gap openers (without novel pair) & Baseline & Baseline & Baseline & Baseline \\
\midrule
Controls & \checkmark & \checkmark & \checkmark & \checkmark \\
Year fixed effects & \checkmark & \checkmark & \checkmark & \checkmark \\
CPC-section fixed effects & \checkmark & \checkmark & \checkmark & \checkmark \\
\midrule
R$^2$ & 0.092 & 0.193 & 0.080 & 0.191 \\
Observations & 9,142,863 & 7,493,883 & 9,142,863 & 7,493,883 \\
\bottomrule
\end{tabular}
\begin{tablenotes}
\footnotesize
\item \emph{Notes:} This table compares gap-opening patents, patents that introduce novel CPC-code pairs without opening gaps, and patents that introduce no novel CPC-code pairs on top 1\% most-cited status and CD percentile, identifying gaps separately with granular CPC subgroups and coarser CPC subclasses. At each level of CPC granularity, top 1\% most-cited status is estimated with logistic regression and CD percentile with OLS; columns (1)--(2) use CPC subgroups, and columns (3)--(4) use CPC subclasses. CD percentile is reported on the manuscript's 0--1 scale. The unit of analysis is the patent--section pair; a patent contributes one observation for each CPC section it belongs to, so $N$ counts patent--section observations rather than unique patents. Controls: Team size, Number of references (log); the CD model adds Citations (log). Standard errors in parentheses are clustered by patent to account for the repeated observations. For logistic models, R$^2$ denotes pseudo-R$^2$.
\item \checkmark\ indicates the variable or fixed effect is included.
\item *$p<0.05$, **$p<0.01$, ***$p<0.001$.
\end{tablenotes}
\end{threeparttable}
\end{adjustbox}
\end{table}

\clearpage
\subsection*{Decomposition of topological and dyadic contributions}

This analysis tests whether the gap-opener association reflects topological position rather than the introduction of a novel concept pair alone.

The three-category specification in Table~\ref{table:s1} compares gap openers with non-gap openers (with novel pair), holding the introduction of a novel concept pair constant. Table~\ref{tab:nested-model} reports the algebraically equivalent nested parameterization, which estimates the effect of introducing any novel pair ($\beta_1$) alongside the incremental effect of that pair occupying a gap-opening position ($\beta_2$). The incremental topological effect is positive and significant for 20-year citations, the SB index, top 1\% status, the CD percentile, and atypicality, and it exceeds the novel-pair effect on every outcome except atypicality and 5-year citations. Table~\ref{tab:within-sample} restricts the sample to papers that introduce a novel concept pair and estimates the gap-opener effect within this homogeneous group, removing all variation in novel-pair status; the topological effect remains significant on the same outcomes.

\clearpage
\begin{table}[p]
\centering
\caption{Decomposition of novel-pair and gap-opening contributions.}\label{tab:nested-model}
\begin{adjustbox}{max width=\textwidth,center}
\begin{threeparttable}
\begin{tabular}{lcccccc}
\toprule
 & (1) & (2) & (3) & (4) & (5) & (6) \\
 & C5 & C20 & SB index (log) & Top 1\% most cited & CD percentile & A percentile \\
 & (Poisson) & (Poisson) & (OLS) & (Logistic) & (OLS) & (OLS) \\
\midrule
Novel concept pair ($\beta_1$) & $-$0.019*** & 0.058*** & 0.022*** & $-$0.001 & 0.003*** & 0.065*** \\
 & (0.0017) & (0.0026) & (0.0002) & (0.0041) & (0.0001) & (0.0001) \\
Gap-opening position ($\beta_2$) & $-$0.044*** & 0.215*** & 0.099*** & 0.455*** & 0.018*** & 0.019*** \\
 & (0.0100) & (0.0220) & (0.0017) & (0.0207) & (0.0009) & (0.0009) \\
\midrule
Controls & \checkmark & \checkmark & \checkmark & \checkmark & \checkmark & \checkmark \\
Year fixed effects & \checkmark & \checkmark & \checkmark & \checkmark & \checkmark & \checkmark \\
Field fixed effects & \checkmark & \checkmark & \checkmark & \checkmark & \checkmark & \checkmark \\
\midrule
R$^2$ & 0.197 & 0.183 & 0.264 & 0.106 & 0.146 & 0.151 \\
Observations & 8,643,571 & 8,643,571 & 25,537,277 & 31,983,201 & 21,613,648 & 19,461,981 \\
\bottomrule
\end{tabular}
\begin{tablenotes}
\footnotesize
\item \emph{Notes:} Relative to papers that introduce no novel concept pairs, this table separates the association of introducing a novel concept pair from the additional association of opening a gap across short- and long-term citations, delayed recognition, top 1\% most-cited status, disruption, and atypicality. The models use the full analysis sample. The unit of analysis is the paper; each paper contributes one observation. Baseline is non-gap openers (without novel pair), as in Table S1. Because gap openers also carry the novel-pair indicator, for the outcomes shared with Table S1, $\beta_1$ equals the Table S1 coefficient on non-gap openers (with novel pair), $\beta_2$ is the additional effect of occupying a gap-opening position relative to them, and $\beta_1+\beta_2$ equals the Table S1 gap-opener coefficient. The atypicality column is estimated on a separate reference sample and has no Table S1 counterpart. The Nobel outcome is omitted because quasi-complete separation makes the estimate uninformative in this parameterization. Controls: Team size, Number of references (log); the SB and CD models add Citations (log). Heteroskedasticity-robust standard errors in parentheses. For logistic and Poisson models, R$^2$ denotes pseudo-R$^2$.
\item \checkmark\ indicates the variable or fixed effect is included.
\item *$p<0.05$, **$p<0.01$, ***$p<0.001$.
\end{tablenotes}
\end{threeparttable}
\end{adjustbox}
\end{table}
\clearpage
\begin{table}[p]\centering
\caption{Gap-opener effect among papers with novel concept pairs.}\label{tab:within-sample}
\begin{adjustbox}{max width=\textwidth,center}
\begin{threeparttable}
\begin{tabular}{lcccccc}
\toprule
 & (1) & (2) & (3) & (4) & (5) & (6) \\
 & \makecell{C5\\(Poisson)} & \makecell{C20\\(Poisson)} & \makecell{SB index (log)\\(OLS)} & \makecell{Top 1\% most\\cited\\(Logistic)} & \makecell{CD percentile\\(OLS)} & \makecell{A percentile\\(OLS)} \\
\midrule
Gap opener & $-$0.067*** & 0.195*** & 0.090*** & 0.359*** & 0.012*** & 0.030*** \\
 & (0.0101) & (0.0224) & (0.0017) & (0.0209) & (0.0009) & (0.0009) \\
\midrule
Controls & \checkmark & \checkmark & \checkmark & \checkmark & \checkmark & \checkmark \\
Year fixed effects & \checkmark & \checkmark & \checkmark & \checkmark & \checkmark & \checkmark \\
Field fixed effects & \checkmark & \checkmark & \checkmark & \checkmark & \checkmark & \checkmark \\
\midrule
R$^2$ & 0.185 & 0.173 & 0.261 & 0.100 & 0.141 & 0.174 \\
Observations & 2,921,791 & 2,921,791 & 7,027,599 & 8,684,885 & 5,847,581 & 5,224,432 \\
\bottomrule
\end{tabular}
\begin{tablenotes}
\footnotesize
\item \emph{Notes:} Restricting the sample to papers that introduce novel concept pairs, this table compares papers that open gaps with papers that do not across short- and long-term citations, delayed recognition, top 1\% most-cited status, disruption, and atypicality. Outcomes span 5- and 20-year citations, the sleeping-beauty index (log), top 1\% most-cited status, CD percentile and atypicality, measured as the within-year A percentile. The unit of analysis is the paper; each paper contributes one observation. The sample is restricted to papers that introduce a novel concept pair, so the gap-opener coefficient is the difference between gap openers and non-gap openers (with novel pair). The Nobel outcome is omitted because quasi-complete separation makes the estimate uninformative. Controls: Team size, Number of references (log); the SB and CD models add Citations (log). Heteroskedasticity-robust standard errors in parentheses. For logistic and Poisson models, R$^2$ denotes pseudo-R$^2$.
\item \checkmark\ indicates the variable or fixed effect is included.
\item *$p<0.05$, **$p<0.01$, ***$p<0.001$.
\end{tablenotes}
\end{threeparttable}
\end{adjustbox}\end{table}

\clearpage

\subsection*{Gap duration and established impact predictors}

We next examine whether the reception of gap-opening papers varies with the duration of the gaps they open and whether the gap-opener association survives controls for established predictors of scientific impact.

Among gap openers whose gaps have subsequently closed, reception tracks the gap's own history. A gap closes at the filtration value where its homology class becomes a boundary, which corresponds to the first paper whose new concept pairing completes the simplices that fill the cycle's interior; gap duration is the elapsed time between that closing paper and the opener. Table~\ref{tab:gap-duration} regresses citation and disruption outcomes on the (log) lifetime of the gap, controlling for team size, references, atypicality, and year and field fixed effects; gap duration has a significant negative effect on all four outcomes examined, so papers whose gaps closed quickly show greater impact and disruption than those whose gaps persisted. Table~\ref{tab:net-effect} re-estimates the citation models controlling for established impact predictors---CD index, atypicality, team mean age, team freshness, and geographical distance---alongside the original controls. Gap openers retain large, significant effects on 20-year citations and top 1\% status, indicating that the gap-opener effect is not absorbed by known predictors of impact. Under this fuller control set, the five-year citation coefficient turns positive rather than negative, indicating that the early-citation deficit reported in Table~\ref{table:s1} is specific to the baseline specification, whereas the positive long-run citation association remains.

\clearpage
\begin{table}[p]\centering
\caption{Effect of gap duration on citation impact and disruption.}\label{tab:gap-duration}
\begin{adjustbox}{max width=\textwidth,center}
\begin{threeparttable}
\begin{tabular}{lcccc}
\toprule
 & (1) & (2) & (3) & (4) \\
 & \makecell{C5\\(Poisson)} & \makecell{C20\\(Poisson)} & \makecell{Top 1\% most\\cited\\(Logistic)} & \makecell{CD percentile\\(OLS)} \\
\midrule
Gap duration (log) & $-$0.077*** & $-$0.054*** & $-$0.076*** & $-$0.002** \\
 & (0.0071) & (0.0103) & (0.0170) & (0.0008) \\
\midrule
Controls & \checkmark & \checkmark & \checkmark & \checkmark \\
Year fixed effects & \checkmark & \checkmark & \checkmark & \checkmark \\
Field fixed effects & \checkmark & \checkmark & \checkmark & \checkmark \\
\midrule
R$^2$ & 0.182 & 0.211 & 0.103 & 0.100 \\
Observations & 51,210 & 51,210 & 89,058 & 83,514 \\
\bottomrule
\end{tabular}
\begin{tablenotes}
\footnotesize
\item \emph{Notes:} Among gap-opening papers whose gaps subsequently close, this table examines how the length of time a gap remains open is associated with short- and long-term citations, top 1\% most-cited status, and disruption. Outcomes are 5- and 20-year citations (Poisson), top 1\% most-cited status (logistic) and CD percentile (OLS). The unit of analysis is the paper; each paper contributes one observation. For papers that open multiple gaps, gap duration is averaged across those gaps. The sample is restricted to gap-opening papers whose gap has closed, and gap duration is the (log) lifetime of the knowledge gap. Controls: Team size, Number of references (log), A index. Heteroskedasticity-robust standard errors in parentheses. For logistic and Poisson models, R$^2$ denotes pseudo-R$^2$.
\item \checkmark\ indicates the variable or fixed effect is included.
\item *$p<0.05$, **$p<0.01$, ***$p<0.001$.
\end{tablenotes}
\end{threeparttable}
\end{adjustbox}\end{table}
\clearpage

 \begin{table}[p]
 \centering
 \caption{Gap-opener effect after controlling for established impact predictors.}\label{tab:net-effect}
 \begin{adjustbox}{max width=\textwidth,center}
 \begin{threeparttable}\footnotesize
 \begin{tabular}{lccc}
 \toprule
 & (1) & (2) & (3) \\
 & \makecell{C5\\ (Poisson)} & \makecell{C20\\ (Poisson)} & \makecell{Top 1\%\\ most cited\\ (Logistic)} \\
 \midrule
 Gap openers (with novel pair) & \makecell{0.039*\\ (0.0173)} & \makecell{0.333***\\ (0.0321)} & \makecell{0.572***\\ (0.0440)} \\
 Non-gap openers (with novel pair) & \makecell{0.002\\ (0.0025)} & \makecell{0.058***\\ (0.0036)} & \makecell{0.144***\\ (0.0109)} \\
 Non-gap openers (without novel pair) & Baseline & Baseline & Baseline \\
 \midrule
 CD index & \checkmark & \checkmark & \checkmark \\
 A index & \checkmark & \checkmark & \checkmark \\
 Team mean age (log) & \checkmark & \checkmark & \checkmark \\
 Team freshness & \checkmark & \checkmark & \checkmark \\
 Geographical distance (log) & \checkmark & \checkmark & \checkmark \\
 Controls (Team size, Number of references (log)) & \checkmark & \checkmark & \checkmark \\
 Year fixed effects & \checkmark & \checkmark & \checkmark \\
 Field fixed effects & \checkmark & \checkmark & \checkmark \\
 \midrule
 Pseudo-R$^2$ & 0.1959 & 0.2053 & 0.0963 \\ 
 Observations & 2,990,328 & 2,990,328 & 2,990,328 \\
 \bottomrule
 \end{tabular}
 \begin{tablenotes}
\footnotesize
\item \emph{Notes:} This table compares gap-opening papers and papers that introduce novel concept pairs without opening gaps with papers that introduce no novel concept pairs on short- and long-term citations and top 1\% most-cited status after accounting for disruption, atypicality, team mean age, team freshness, and geographical distance. The 5- and 20-year citation models are Poisson, and the top 1\% most-cited model is logistic. The unit of analysis is the paper; each paper contributes one observation. Robust standard errors in parentheses.
\item \checkmark\ indicates the variable or fixed effect is included.
\item *$p<0.05$, **$p<0.01$, ***$p<0.001$.
\end{tablenotes}
 \end{threeparttable}
 \end{adjustbox}
 \end{table}

\clearpage

\subsection*{Robustness to matched sampling}

Gap openers are proportionally rare, but the full dataset contains 289,041 of them (0.84\% of articles), so their rarity raises little concern about statistical power or about estimates driven by a small number of observations. Their uneven distribution raises a distinct concern about comparability---the average gap opener may differ systematically from the average paper in the other categories in ways that year and field fixed effects do not fully absorb. To address that concern directly, we re-estimate the outcomes in main-text Fig.~4, other than Nobel Prize wins, on a coarsened-exact-matched sample with matched-set fixed effects, matching each gap opener to control papers of the same field, year, team-size bin, and reference-count bin (and prior-citation bin where the corresponding specification controls for citations). Between 97.8\% and 99.9\% of gap openers find matched controls, and matching eliminates the substantial pre-matching covariate imbalances---all post-matching standardized mean differences fall below 0.05. Table~\ref{tab:matched-sampling} reports the results for a comparison against all control papers (Panel A) and the more stringent comparison against non-gap openers (with novel pair) only (Panel B); estimation is within matched sets, with a linear probability model for the binary top-cited outcomes. Gap openers remain significantly more disruptive and more highly cited in the long run in both panels, with coefficients that are smaller than the unmatched estimates, as expected when comparing observably similar papers, and the reduction is largest for the CD percentile. We exclude the Nobel outcome, which is too rare for within-set estimation---only 89 Nobel-linked papers appear in the estimation universe, spread across roughly 20,000 matched sets, leaving no usable outcome variation. Together, the large absolute number of gap openers and the persistence of the results within closely matched sets indicate that the findings are not driven by the rarity of the focal group or by systematic differences in observable characteristics between gap openers and comparison papers.

\clearpage
\begin{table}[p]
\centering
\caption{Gap-opener effects under coarsened exact matching.}\label{tab:matched-sampling}
\begin{adjustbox}{max width=\textwidth,center}
\begin{threeparttable}
\begin{tabular}{lccccccccc}
\toprule
 & (1) & (2) & (3) & (4) & (5) & (6) & (7) & (8) & (9) \\
 & CD percentile (OLS) & C5 (Poisson) & C20 (Poisson) & SB index (log) (OLS) & Top 1\% most cited (LPM) & Top 5\% most cited (LPM) & Top 10\% most cited (LPM) & Top 15\% most cited (LPM) & Top 20\% most cited (LPM) \\
\midrule
\multicolumn{10}{l}{\textit{Panel A: Gap opener vs. all others}}\\
Gap opener & 0.006*** & $-$0.029** & 0.252*** & 0.072*** & 0.005*** & 0.011*** & 0.014*** & 0.017*** & 0.018*** \\
 & (0.0009) & (0.0101) & (0.0219) & (0.0017) & (0.0003) & (0.0006) & (0.0007) & (0.0009) & (0.0009) \\
Matched-set fixed effects & \checkmark & \checkmark & \checkmark & \checkmark & \checkmark & \checkmark & \checkmark & \checkmark & \checkmark \\
Observations & 587,302 & 614,002 & 614,002 & 788,509 & 1,047,122 & 1,047,140 & 1,047,140 & 1,047,140 & 1,047,140 \\
\midrule
\multicolumn{10}{l}{\textit{Panel B: Gap opener vs. novel-pair non-openers}}\\
Gap opener & 0.003** & $-$0.039*** & 0.197*** & 0.060*** & 0.004*** & 0.009*** & 0.011*** & 0.013*** & 0.013*** \\
 & (0.0010) & (0.0102) & (0.0223) & (0.0017) & (0.0003) & (0.0006) & (0.0008) & (0.0009) & (0.0010) \\
Matched-set fixed effects & \checkmark & \checkmark & \checkmark & \checkmark & \checkmark & \checkmark & \checkmark & \checkmark & \checkmark \\
Observations & 555,477 & 598,459 & 598,459 & 748,768 & 1,031,428 & 1,031,273 & 1,031,273 & 1,031,273 & 1,031,273 \\
\bottomrule
\end{tabular}
\begin{tablenotes}
\footnotesize
\item \emph{Notes:} This table compares gap-opening papers with observably similar papers that do not open gaps---first all non-gap-opening papers and then only those introducing novel concept pairs---across disruption, short- and long-term citations, delayed recognition, and top 1\%--20\% most-cited status. Coarsened exact matching with matched-set fixed effects; each panel reports the gap-opener coefficient against the indicated comparison group. Each gap opener is matched with up to five control papers, and comparisons are made within matched sets. The unit of analysis is the paper; each paper contributes one observation. The sample is restricted to matched papers, so $N$ counts gap openers and their matched controls. LPM denotes a linear probability model, and the SB index is log-transformed. Heteroskedasticity-robust standard errors in parentheses.
\item \checkmark\ indicates the variable or fixed effect is included.
\item *$p<0.05$, **$p<0.01$, ***$p<0.001$.
\end{tablenotes}
\end{threeparttable}
\end{adjustbox}
\end{table}

\clearpage

\subsection*{Decomposition of the joint atypicality--disruption result}

Table~\ref{table:s3} reports the joint atypicality--disruption profiles of the three paper types but was not designed to isolate which margin drives the joint result. Table~\ref{tab:multinomial} estimates the same multinomial specification with citation controls added, on the subset of papers for which atypicality is measurable under the reference-based definition used here, which is why its sample is somewhat smaller than that of Table~\ref{table:s3}. Table~\ref{tab:wald} reports Wald tests comparing the joint (atypical-and-disruptive) coefficient with each single-margin coefficient. For gap openers, the joint coefficient (0.747) significantly exceeds both the atypicality-only coefficient (0.640) and the disruption-only coefficient (0.150), so the joint result is not reducible to either margin alone. Table~\ref{tab:margins} estimates the margins separately. Gap openers exert a stronger effect on atypicality than on disruption, but the disruption margin is distinctly a gap-opener phenomenon---gap openers raise the CD percentile by roughly seven times the margin of non-gap openers (with novel pair).

\clearpage
\begin{table}[p]\centering
\caption{Joint atypicality--disruption profiles after controlling for citations.}\label{tab:multinomial}
\begin{adjustbox}{max width=\textwidth,center}
\begin{threeparttable}
\begin{tabular}{lcccc}
\toprule
 & (1) & (2) & (3) & (4) \\
 & \makecell{P(Atypical)=1\\P(Disruptive)=1} & \makecell{P(Atypical)=1\\P(Disruptive)=0} & \makecell{P(Atypical)=0\\P(Disruptive)=1} & \makecell{P(Atypical)=0\\P(Disruptive)=0} \\
\midrule
Gap openers (with novel pair) & 0.747*** & 0.640*** & 0.150*** & Baseline \\
 & (0.0112) & (0.0091) & (0.0106) & \\
Non-gap openers (with novel pair) & 0.473*** & 0.483*** & 0.056*** & \\
 & (0.0018) & (0.0013) & (0.0019) & \\
Non-gap openers (without novel pair) & Baseline & Baseline & Baseline & \\
\midrule
Controls & \checkmark & \checkmark & \checkmark & \\
Year fixed effects & \checkmark & \checkmark & \checkmark & \\
Field fixed effects & \checkmark & \checkmark & \checkmark & \\
\midrule
Pseudo-R$^2$ & 0.102 & & & \\
Observations & 19,461,981 & & & \\
\bottomrule
\end{tabular}
\begin{tablenotes}
\footnotesize
\item \emph{Notes:} After accounting for citation count, this table compares gap-opening papers, papers that introduce novel concept pairs without opening gaps, and papers that introduce no novel concept pairs across four outcomes---being both atypical and disruptive, being atypical only, being disruptive only, or being neither atypical nor disruptive. Multinomial (unordered) logit predicting a paper's joint atypicality--disruption profile; the ``neither'' category (atypical and disruptive both 0) is the base outcome. The unit of analysis is the paper; each paper contributes one observation. This specification extends the joint atypicality--disruption model in Table S3 by adding the Citations (log) control and provides the estimates used for the Wald tests in Table~\ref{tab:wald}. Controls: Team size, Number of references (log), Citations (log). Conventional model-based standard errors in parentheses.
\item \checkmark\ indicates the variable or fixed effect is included.
\item *$p<0.05$, **$p<0.01$, ***$p<0.001$.
\end{tablenotes}
\end{threeparttable}
\end{adjustbox}\end{table}
\clearpage
\begin{table}[p]\centering
\caption{Wald comparisons of the joint atypicality--disruption coefficient with each single-margin coefficient.}\label{tab:wald}
\begin{adjustbox}{max width=\textwidth,center}
\begin{threeparttable}
\begin{tabular}{lcc}
\toprule
 & (1) & (2) \\
 & Joint vs.\ Atypical-only & Joint vs.\ Disruptive-only \\
\midrule
Gap openers (with novel pair) & \makecell{0.747 vs.\ 0.640\\ $\chi^2=92.01$***} & \makecell{0.747 vs.\ 0.150\\ $\chi^2=2{,}379.62$***} \\
Non-gap openers (with novel pair) & \makecell{0.473 vs.\ 0.483\\ $\chi^2=37.16$***} & \makecell{0.473 vs.\ 0.056\\ $\chi^2=36{,}340.62$***} \\
\bottomrule
\end{tabular}
\begin{tablenotes}
\footnotesize
\item \emph{Notes:} For gap-opening papers and papers that introduce novel concept pairs without opening gaps, this table tests whether the coefficient for being both atypical and disruptive differs from the coefficients for being atypical only or disruptive only. Each cell reports the two coefficients compared and the Wald $\chi^2$. This table reports Wald tests derived from the paper-level multinomial model in Table~\ref{tab:multinomial}; the underlying unit of analysis is the paper. The tests use that model's conventional model-based covariance estimates.
\item *$p<0.05$, **$p<0.01$, ***$p<0.001$.
\end{tablenotes}
\end{threeparttable}
\end{adjustbox}\end{table}
\clearpage
\begin{table}[p]
\centering
\caption{Separate contributions of atypicality and disruption to the gap-opener effect.}\label{tab:margins}
\begin{adjustbox}{max width=\textwidth,center}
\begin{threeparttable}
\begin{tabular}{lcccc}
\toprule
 & (1) & (2) & (3) & (4) \\
 & P(Atypical)=1 & P(Disruptive)=1 & A percentile & CD percentile \\
 & (Logistic) & (Logistic) & (OLS) & (OLS) \\
\midrule
Gap openers (with novel pair) & 0.619*** & 0.163*** & 0.084*** & 0.020*** \\
 & (0.0074) & (0.0080) & (0.0009) & (0.0010) \\
Non-gap openers (with novel pair) & 0.468*** & 0.045*** & 0.065*** & 0.003*** \\
 & (0.0011) & (0.0013) & (0.0001) & (0.0001) \\
Non-gap openers (without novel pair) & Baseline & Baseline & Baseline & Baseline \\
P(Atypical)=1 & -- & 0.302*** & -- & -- \\
 & & (0.0012) & & \\
A percentile & -- & -- & -- & 0.095*** \\
 & & & & (0.0002) \\
\midrule
Controls & \checkmark & \checkmark & \checkmark & \checkmark \\
Year fixed effects & \checkmark & \checkmark & \checkmark & \checkmark \\
Field fixed effects & \checkmark & \checkmark & \checkmark & \checkmark \\
\midrule
Observations & 19,461,981 & 19,461,981 & 19,461,981 & 19,461,981 \\
\bottomrule
\end{tabular}
\begin{tablenotes}
\footnotesize
\item \emph{Notes:} Relative to papers that introduce no novel concept pairs, this table separately estimates how gap-opening papers and papers that introduce novel concept pairs without opening gaps are associated with atypicality and disruption and examines whether the association with disruption persists after accounting for atypicality. Each column is a separate margin/continuous-variable regression: P(Atypical) and P(Disruptive) are logistic, A percentile and CD percentile are OLS. The unit of analysis is the paper; each paper contributes one observation. The P(Atypical) and A percentile rows test whether atypicality transmits the gap-opener effect to disruption. Controls: Team size, Number of references (log), Citations (log). Heteroskedasticity-robust standard errors in parentheses.
\item \checkmark\ indicates the variable or fixed effect is included.
\item *$p<0.05$, **$p<0.01$, ***$p<0.001$.
\end{tablenotes}
\end{threeparttable}
\end{adjustbox}
\end{table}

\clearpage

\subsection*{Gap size and generative capacity}

We next examine whether the size of a gap matters. We measure gap size by the length of the cycle that forms the gap's boundary. For 96.7\% of the gaps in Table~\ref{tab:gap-size}, the cycle length also equals the number of distinct concepts on the boundary; every size-four gap is a simple four-concept cycle.

The boundary used to measure the gap is fixed at birth, but the subsequent research region is not limited to the concepts on that boundary. Later papers can combine boundary concepts with additional concepts as the network develops. The filling structures identified by the algorithm reflect this expansion---in 95.9\% of the sampled gaps, the structure extends beyond the concepts on the original boundary.

Here, follow-on papers are papers published after the gap opens that use at least two concepts from its boundary. In the raw counts, four-concept gaps are followed over the remainder of MAG's observation period by an average of 139 such papers, and 10\% are followed by at least 269. Gaps bounded by cycles of six or more links are followed by an average of 6,100 such papers.

Table~\ref{tab:gap-size} reports regressions of three measures of generative capacity on log gap size---follow-on research, the length of time the gap remains open, and the number of concept triangles in the filling structure identified when the gap closes. To make gaps of different sizes comparable, the follow-on-paper model divides the paper count by the number of possible pairs among the concepts on the boundary and log-transforms the resulting value. Larger gaps attract more follow-on work, remain open longer, and have larger filling structures when they close. The opening paper's own citation impact, by contrast, does not increase with gap size. Specifically, among the 26,436 gaps in the Table S20 sample, regressions of $\log(1+\text{citations})$ for the opening paper on log gap size, with field fixed effects and heteroskedasticity-robust standard errors, yield coefficients indistinguishable from zero for both 20-year citations ($\beta=+0.009$, SE $=0.024$, $p=0.710$) and full-window citations ($\beta=-0.007$, SE $=0.025$, $p=0.793$). Gap size is modestly correlated with concept age (0.27) and only weakly with prior concept popularity (0.08), so the evidence distinguishes the subsequent activity organized around a gap from both the opening paper's own citation impact and the standing of the concepts on the boundary.

\clearpage
\begin{table}[p]\centering
\caption{Generative capacity of knowledge gaps by gap size.}\label{tab:gap-size}
\begin{adjustbox}{max width=\textwidth,center}
\begin{threeparttable}
\begin{tabular}{lc}
\toprule
Effect of (log) gap size & (1) \\
\midrule
Follow-on papers per concept-pair & 0.511*** \\
 & (0.0186) \\
Gap longevity (log years) & 0.202*** \\
 & (0.0223) \\
Concept triangles in filling structure (log) & 0.287*** \\
 & (0.0157) \\
\midrule
Observations & 26,436 \\
\bottomrule
\end{tabular}
\begin{tablenotes}
\footnotesize
\item \emph{Notes:} This table examines how gap size is associated with the number of subsequent papers using concepts from the gap boundary, the length of time the gap remains open, and the size of the filling structure when the gap closes. Each cell is the OLS coefficient from a regression of the row measure on log gap size in MAG, with field fixed effects. The unit of analysis is the gap. Heteroskedasticity-robust standard errors in parentheses. Gap size is the length of the minimal bounding cycle, computed under a uniform year-based filtration so that size is measured identically in every field. Estimates use a fixed random sample of up to 1,500 gaps per level-0 field (27,000 sampled; 26,436 with all measures observed) across the 18 fields in which minimal bounding cycles could be computed. Follow-on papers are papers published after the gap opens that use at least two concepts from its boundary. The paper count is divided by the number of possible pairs among the boundary concepts. The filling-structure outcome counts the concept triangles in the filling structure identified by the algorithm when the gap closes. Because the algorithm does not minimize this structure, the outcome should not be interpreted as the minimum number of triangles needed to close the gap. The correlation between log gap size and the log number of concept triangles in the filling structure is 0.14, so the closure result is not mechanical.
\item *$p<0.05$, **$p<0.01$, ***$p<0.001$.
\end{tablenotes}
\end{threeparttable}
\end{adjustbox}\end{table}

\clearpage

\subsection*{Robustness to concept-level characteristics}

This analysis tests whether the gap-opener association is explained by the age, prior popularity, or atypicality of the concepts a paper joins.

Table~\ref{tab:concept-quality} adds concept-level characteristics---concept age, prior concept popularity, and atypicality---to the citation models alongside the original controls. The gap-opener effect on 20-year citations and top 1\% status persists and, if anything, strengthens under these controls, indicating that the effect reflects the topology of the gap rather than the quality of the concepts it joins.

\clearpage
\begin{table}[p]\centering
\caption{Gap-opener effect after controlling for concept-level characteristics.}\label{tab:concept-quality}
\begin{adjustbox}{max width=\textwidth,center}
\begin{threeparttable}
\begin{tabular}{lcc}
\toprule
 & (1) & (2) \\
 & C20 & Top 1\% most cited \\
 & (Poisson) & (Logistic) \\
\midrule
Gap openers (with novel pair) & 0.308*** & 0.467*** \\
 & (0.0237) & (0.0240) \\
Non-gap openers (with novel pair) & 0.053*** & $-$0.004 \\
 & (0.0029) & (0.0046) \\
Non-gap openers (without novel pair) & Baseline & Baseline \\
Concept age & $-$1.353*** & $-$1.829*** \\
 & (0.0228) & (0.0189) \\
Concept popularity (log) & 0.073*** & 0.150*** \\
 & (0.0015) & (0.0019) \\
A percentile & 0.103*** & 0.036*** \\
 & (0.0057) & (0.0084) \\
\midrule
Controls & \checkmark & \checkmark \\
Year fixed effects & \checkmark & \checkmark \\
Field fixed effects & \checkmark & \checkmark \\
\midrule
Pseudo-R$^2$ & 0.182 & 0.098 \\
Observations & 5,252,338 & 20,932,923 \\
\bottomrule
\end{tabular}
\begin{tablenotes}
\footnotesize
\item \emph{Notes:} This table compares gap-opening papers and papers that introduce novel concept pairs without opening gaps with papers that introduce no novel concept pairs on 20-year citations and top 1\% most-cited status after accounting for concept age, prior concept popularity, and paper atypicality. Outcomes are 20-year citations (Poisson) and top 1\% most-cited status (logistic). The unit of analysis is the paper; each paper contributes one observation. Concept age and concept popularity (log) are measured for each paper's level-3 concepts; atypicality is the within-year A percentile; the sample is restricted to papers with a measurable atypicality. Controls: Team size, Number of references (log). Heteroskedasticity-robust standard errors in parentheses.
\item \checkmark\ indicates the variable or fixed effect is included.
\item *$p<0.05$, **$p<0.01$, ***$p<0.001$.
\end{tablenotes}
\end{threeparttable}
\end{adjustbox}\end{table}

\clearpage

\section*{Replication data sources and network construction}

\revision{This section documents the data, concept vocabularies, field definitions, network construction, outcome measures, and estimation samples used in each replication. The shared analytical procedure applies across all corpora, with corpus-specific differences described in the subsections that follow.}

\subsection*{Shared analytical procedure}
\noindent All replication pipelines identify gaps with the same persistent-homology procedure as the main analysis, computing dimension-one features of the clique filtration of the growing concept network (implemented in the GUDHI \citep{maria2014gudhi} and OAT \citep{oat2026} libraries). All pipelines estimate the same canonical regression specification for the paper's primary measures of citation impact and disruption, in which each outcome is regressed on the gap-opener indicators with controls for log reference count, team size, and, for the CD percentile, the log of one plus total citations, with year and field fixed effects (logistic regression for the top 1\% indicator and least squares for the CD percentile). All regression samples apply the sample filters of the main analysis---team size of at most 20, and publication years 1980--2020 except where a corpus's coverage is narrower, in which case the realized window is stated in that corpus's subsection below---together with the outcome-specific restrictions stated in each corpus subsection.

\revision{For corpora in which a paper or patent contributes observations to multiple fields or CPC sections, we use cluster-robust standard errors with the conventional CRV1 finite-sample correction, clustered by paper or patent. Models with one observation per paper use heteroskedasticity-robust standard errors with the conventional HC1 correction.}

\subsection*{OpenAlex}
We replicate the analysis in OpenAlex \citep{priem2022openalex} using the SciSciNet-v2 distribution \citep{lin2023scisci}. Paper metadata, discipline assignments, topic annotations, and the citation graph come from this distribution, which contains 249,803,129 records and 2,487,738,082 citation edges after excluding 150 records with inconsistent citation data. Concepts are OpenAlex topics, and each paper is assigned to one or more of 26 disciplines. Concept networks cover papers published 1900--2020, as in the main analysis. Within each discipline, nodes are topics, an edge appears when a paper first receives both topics, and the edge's filtration value is the rank-normalized publication date of that paper, with gap openers identified by the same persistent-homology procedure as the main analysis. We compute citation counts and the CD index from the distribution's citation graph. The top 1\% indicator and the CD percentile are rank percentiles of citations and CD within publication year and discipline. We estimate the canonical specification with year and discipline fixed effects and standard errors clustered by paper; N = 101,638,624 for both the top 1\% and CD-percentile models. The CD percentile is defined for every OpenAlex paper meeting the reference requirement, so the disruption and citation models share the same estimation sample; the main analysis additionally requires at least one forward citation, which is why the MAG disruption sample is smaller than its citation sample.

\subsection*{PubMed Knowledge Graph}
We replicate the analysis in the PubMed Knowledge Graph 2.0 (PKG) \citep{xu2020pubmed}, using the PKG24S4 release, a July 2025 refresh built from the 2025 PubMed Baseline. We include publications from 1985 through 2018, the period with consistent MeSH indexing, yielding 21,900,943 records, of which 9,461,549 have at least one qualifying descriptor. Concepts are MeSH descriptors flagged as major topics, restricted to the substantive trees A--M; 25,603 distinct descriptors appear. The 13 top-level MeSH tree letters define the fields, and a paper indexed in several trees contributes to each. Within each field, we retain descriptors appearing on between 0.0001 and 0.001 of the field's papers with at least two descriptors, and we construct a network in which an edge appears at the year of first co-occurrence, with publication year as the filtration parameter. A paper is a gap opener when it introduces a concept pair in the pair's first co-occurrence year and that pair is a birth pair. Citations come from PKG's integrated citation table, with 807,055,715 within-corpus edges after removing self-citations. The top 1\% indicator and the CD percentile are rank percentiles within publication year and field. We estimate the canonical specification with year and field fixed effects and standard errors clustered by paper; N = 11,156,533 for the top 1\% model and 9,776,452 for the CD-percentile model.

\subsection*{Dimensions}
We replicate the analysis in the Dimensions database \citep{herzog2020dimensions}, using an archived snapshot dated 1 January 2025.
The extract contains 163,924,684 publications and 2,321,638,334 citation edges; all publications with a recorded year enter the network stage, with no document-type filter. Concepts are Dimensions' text concepts extracted by natural language processing from titles and abstracts. We retain concepts with relevance scores of at least 0.7 and normalize the strings (Unicode lowercasing, whitespace collapse, removal of entries shorter than two characters or lacking alphabetic characters), leaving 9,925,373 unique concepts. Level-2 Fields of Research codes from the 2020 Australian and New Zealand Standard Research Classification define the fields (171 fields, spanning divisions 30 through 52); papers with several codes contribute to each.
Within each field, we retain concepts appearing on between 0.0001 and 0.001 of the field's papers with at least two concepts, with edges at the year of first co-occurrence and publication year as the filtration parameter. Persistent homology identifies 2,766,734 dimension-one features, and classification labels 2,492,294 gap openers (1.48\% of paper--field observations). The top 1\% indicator is a rank percentile within publication year among publications with at least one citation; the CD percentile is a rank percentile within publication year among papers with non-missing CD. We estimate the canonical specification with year and field fixed effects and standard errors clustered by paper.
N = 69,040,797 for the top 1\% model and 60,293,158 for the CD-percentile model.

\subsection*{American Physical Society journals}
We replicate the analysis in the corpus of American Physical Society (APS) journals (the Physical Review family and Reviews of Modern Physics, 1893--2018), comprising 636,294 records and 7,642,495 within-corpus citation edges. We run two parallel pipelines that differ only in the concept vocabulary. The first uses full PACS codes (e.g., 89.75.Hc); PACS codes are recorded for 441,340 papers, 7,458 distinct codes appear, and no frequency filter is applied. Because PACS was discontinued, code coverage ends in 2015 and the realized regression window is 1980--2015. The second uses concept phrases extracted from abstracts following \citet{kedrick2024cp}---noun-chunk extraction with spaCy, lemmatization, and standardization of punctuation and casing---yielding 113,683 distinct phrases (coverage ends in 2017, so the realized window is 1980--2017), retaining phrases appearing on between 0.001 and 0.01 of the field's papers with at least two concepts.
The APS database records a PACS domain---one of the ten top-level PACS divisions---for 582,679 papers; the domain defines the field in both pipelines, which construct per-field networks with edges at first co-occurrence and publication year as the filtration parameter. Persistent homology yields 8,268 birth pairs under PACS codes (gap-opener share 1.99\%) and 6,156 under text concepts (0.88\%). Outcomes are computed once on the shared paper set from the within-APS citation graph. The top 1\% indicator is a rank percentile within publication year, excluding papers with zero citations; the CD percentile is a rank percentile within publication year. Because each paper has a single field, we estimate the canonical specification with year and field fixed effects and heteroskedasticity-robust standard errors. For the top 1\% models, N = 380,366 (PACS) and 479,904 (text concepts); for the CD-percentile models, N = 342,923 and 418,540.

\subsection*{U.S. patents}
We replicate the analysis in U.S. patents using PatentsView bulk data files \citep{patentsview2025} from the U.S. Patent and Trademark Office, from the release issued 20 November 2025, which covers grants through 30 September 2025.
The corpus contains 9,361,444 granted patents (1976--2025) and 135,583,358 within-corpus citation edges after removing self-citations. Concepts are Cooperative Patent Classification (CPC) codes at two granularities, run in parallel on the identical patent population and citation graph; the granular pipeline uses leaf CPC subgroups (245,134 distinct codes) and the coarse pipeline uses CPC subclasses (677 distinct codes). The nine CPC sections (A--H and Y) define the fields. Within each section, we retain codes appearing on between 0.0001 and 0.001 of the section's patents with at least two codes and construct networks with edges at first co-occurrence and grant year as the filtration parameter (1976--2020). The granular run yields 155,427 dimension-one features and 147,905 gap openers (1.15\% of patent--section observations); the coarse run yields 3,747 features and 4,024 gap openers (0.03\%; several patents can introduce the same pair in its first co-occurrence year, so openers can outnumber features), so the paired runs test sensitivity to vocabulary granularity on a fixed population. Team size is the number of distinct disambiguated inventors. The top 1\% indicator and the CD percentile are rank percentiles within grant year and CPC section. We estimate the canonical specification with year and section fixed effects and standard errors clustered by patent; in both runs, N = 9,142,863 for the top 1\% model and 7,493,883 for the CD-percentile model.

\ifdefined\SIinmain\else
\bibliographystyle{plainnat}
\bibliography{bibliography}

@article{cole1994,
 ISSN = {08848971, 15737861},
 URL = {http://www.jstor.org/stable/685039},
 author = {Stephen Cole},
 journal = {Sociological Forum},
 number = {2},
 pages = {133--154},
 publisher = {[Wiley, Springer]},
 title = {Why Sociology Doesn't Make Progress like the Natural Sciences},
 volume = {9},
 year = {1994}
}

@article{marcus1993,
  author = {Marcus, Mitchell P. and Santorini, Beatrice and Marcinkiewicz, Mary Ann},
  title = {Building a large annotated corpus of {English}: The {Penn Treebank}},
  journal = {Computational Linguistics},
  volume = {19},
  number = {2},
  pages = {313--330},
  year = {1993}
}

@article{engle1982,
  author = {Engle, Robert F.},
  title = {Autoregressive conditional heteroscedasticity with estimates of the variance of {United Kingdom} inflation},
  journal = {Econometrica},
  volume = {50},
  number = {4},
  pages = {987--1007},
  year = {1982}
}

@article{adleman1994,
  author = {Adleman, Leonard M.},
  title = {Molecular computation of solutions to combinatorial problems},
  journal = {Science},
  volume = {266},
  number = {5187},
  pages = {1021--1024},
  year = {1994}
}

@article{chen2000,
  author = {Chen, Junghuei and Wood, David H.},
  title = {Computation with biomolecules},
  journal = {Proceedings of the National Academy of Sciences},
  volume = {97},
  number = {4},
  pages = {1328--1330},
  year = {2000}
}

@article{iijima1991,
  author = {Iijima, Sumio},
  title = {Helical microtubules of graphitic carbon},
  journal = {Nature},
  volume = {354},
  number = {6348},
  pages = {56--58},
  year = {1991}
}

@article{binnig1986,
  author = {Binnig, G. and Quate, C. F. and Gerber, Ch.},
  title = {Atomic force microscope},
  journal = {Physical Review Letters},
  volume = {56},
  number = {9},
  pages = {930--933},
  year = {1986}
}

@book{cole1995,
  title={Making science: Between nature and society},
  author={Cole, Stephen},
  year={1992},
  publisher={Harvard University Press}
}

@article{hansen1999search,
  title={The search-transfer problem: The role of weak ties in sharing knowledge across organization subunits},
  author={Hansen, Morten T},
  journal={Administrative science quarterly},
  volume={44},
  number={1},
  pages={82--111},
  year={1999},
  publisher={SAGE Publications}
}

@book{burt1992structural,
  title={Structural Holes: The Social Structure of Competition},
  author={Burt, Ronald S},
  year={1992},
  publisher={Harvard University Press}
}

@article{hargadon1997technology,
  title={Technology brokering and innovation in a product development firm},
  author={Hargadon, Andrew and Sutton, Robert I},
  journal={Administrative science quarterly},
  pages={716--749},
  year={1997},
  publisher={JSTOR}
}

@article{burt2004structural,
  title={Structural holes and good ideas},
  author={Burt, Ronald S},
  journal={American journal of sociology},
  volume={110},
  number={2},
  pages={349--399},
  year={2004},
  publisher={The University of Chicago Press}
}

@article{obstfeld2005social,
  title={Social networks, the tertius iungens orientation, and involvement in innovation},
  author={Obstfeld, David},
  journal={Administrative science quarterly},
  volume={50},
  number={1},
  pages={100--130},
  year={2005},
  publisher={SAGE Publications Sage CA: Los Angeles, CA}
}

@article{fleming2007collaborative,
  title={Collaborative brokerage, generative creativity, and creative success},
  author={Fleming, Lee and Mingo, Santiago and Chen, David},
  journal={Administrative science quarterly},
  volume={52},
  number={3},
  pages={443--475},
  year={2007},
  publisher={SAGE Publications Sage CA: Los Angeles, CA}
}

@article{granovetter1973strength,
  title={The strength of weak ties},
  author={Granovetter, Mark S},
  journal={American journal of sociology},
  volume={78},
  number={6},
  pages={1360--1380},
  year={1973},
  publisher={University of Chicago Press}
}

@article{ahuja2000collaboration,
  title={Collaboration networks, structural holes, and innovation: A longitudinal study},
  author={Ahuja, Gautam},
  journal={Administrative science quarterly},
  volume={45},
  number={3},
  pages={425--455},
  year={2000},
  publisher={SAGE Publications}
}

@article{Torunsky2025,
	author = {Torunsky, Nathan T. and Kedrick, Kara and Vilares, Iris},
	date = {2025/02/19},
	doi = {10.1038/s41598-025-89781-9},
	id = {Torunsky2025},
	isbn = {2045-2322},
	journal = {Scientific Reports},
	number = {1},
	pages = {6096},
	title = {Information seeking and the expected utility of information about COVID-19 can be associated with uncertainty and related attitudes},
	url = {https://doi.org/10.1038/s41598-025-89781-9},
	volume = {15},
	year = {2025}}

@article{kedrick2024cp,
	author = {Kedrick, Kara and Levitskaya, Ekaterina and Funk, Russell J.},
	date = {2024/10/01},
	doi = {10.1038/s41562-024-01957-x},
	id = {Kedrick2024},
	isbn = {2397-3374},
	journal = {Nature Human Behaviour},
	number = {10},
	pages = {1915--1923},
	title = {Conceptual structure and the growth of scientific knowledge},
	url = {https://doi.org/10.1038/s41562-024-01957-x},
	volume = {8},
	year = {2024}}

@article{hofstra2020paradox,
	author = {Bas Hofstra and Vivek V. Kulkarni and Sebastian Munoz-Najar Galvez and Bryan He and Dan Jurafsky and Daniel A. McFarland},
	doi = {10.1073/pnas.1915378117},
	eprint = {https://www.pnas.org/doi/pdf/10.1073/pnas.1915378117},
	journal = {Proceedings of the National Academy of Sciences},
	number = {17},
	pages = {9284-9291},
	title = {The Diversity--Innovation Paradox in Science},
	url = {https://www.pnas.org/doi/abs/10.1073/pnas.1915378117},
	volume = {117},
	year = {2020}}

@article{weitzman1998recombinant,
  title={Recombinant growth},
  author={Weitzman, M. L.},
  journal={The Quarterly Journal of Economics},
  volume={113},
  pages={331--360},
  year={1998}
}

@article{dey2017,
	author = {Dey, Ratnadeep and Roy, Anurag and Chakraborty, Tanmoy and Ghosh, Saptarshi},
	date = {2017/12/01},
	doi = {10.1007/s11192-017-2543-3},
	id = {Dey2017},
	isbn = {1588-2861},
	journal = {Scientometrics},
	number = {3},
	pages = {1645--1663},
	title = {Sleeping beauties in Computer Science: characterization and early identification},
	url = {https://doi.org/10.1007/s11192-017-2543-3},
	volume = {113},
	year = {2017}}

@article{glaeser2023,
	author = {Glaeser, Chloe Kim and Glaeser, Stephen and Labro, Eva},
	doi = {10.1287/mnsc.2022.4469},
	eprint = {https://doi.org/10.1287/mnsc.2022.4469},
	journal = {Management Science},
	number = {5},
	pages = {3080-3099},
	title = {Proximity and the Management of Innovation},
	url = {https://doi.org/10.1287/mnsc.2022.4469},
	volume = {69},
	year = {2023}}

@article{brucks2022,
	author = {Brucks, Melanie S. and Levav, Jonathan},
	date = {2022/05/01},
	doi = {10.1038/s41586-022-04643-y},
	id = {Brucks2022},
	isbn = {1476-4687},
	journal = {Nature},
	number = {7908},
	pages = {108--112},
	title = {Virtual communication curbs creative idea generation},
	url = {https://doi.org/10.1038/s41586-022-04643-y},
	volume = {605},
	year = {2022}}

@article{lin2023,
	author = {Lin, Yiling and Frey, Carl Benedikt and Wu, Lingfei},
	date = {2023/11/01},
	doi = {10.1038/s41586-023-06767-1},
	id = {Lin2023},
	isbn = {1476-4687},
	journal = {Nature},
	number = {7989},
	pages = {987--991},
	title = {Remote collaboration fuses fewer breakthrough ideas},
	url = {https://doi.org/10.1038/s41586-023-06767-1},
	volume = {623},
	year = {2023}}

@article{ollinger2008,
	author = {{\"O}llinger, Michael and Jones, Gary and Knoblich, G{\"u}nther},
	doi = {10.1027/1618-3169.55.4.269},
	journal = {Experimental psychology},
	month = {01},
	pages = {269-82},
	title = {Investigating the Effect of Mental Set on Insight Problem Solving},
	volume = {55},
	year = {2008}}

@article{li2019dataset,
  title={A dataset of publication records for Nobel laureates},
  author={Li, Jichao and Yin, Yian and Fortunato, Santo and Wang, Dashun},
  journal={Scientific data},
  volume={6},
  number={1},
  pages={33},
  year={2019},
  publisher={Nature Publishing Group UK London}
}

@article{van2004sleeping,
  title={Sleeping beauties in science},
  author={Van Raan, Anthony FJ},
  journal={Scientometrics},
  volume={59},
  number={3},
  pages={467--472},
  year={2004},
  publisher={Springer}
}

@article{blech2020,
	author = {Blech, Christine and Gaschler, Robert and Bilali{\'c}, Merim},
	doi = {10.1080/13546783.2019.1685001},
	eprint = {https://doi.org/10.1080/13546783.2019.1685001},
	journal = {Thinking \& Reasoning},
	number = {4},
	pages = {552--580},
	publisher = {Routledge},
	title = {Why do people fail to see simple solutions? Using think-aloud protocols to uncover the mechanism behind the Einstellung (mental set) effect},
	url = {https://doi.org/10.1080/13546783.2019.1685001},
	volume = {26},
	year = {2020}}

@article{wiley1998,
	author = {Wiley, Jennifer},
	date = {1998/07/01},
	doi = {10.3758/BF03211392},
	id = {Wiley1998},
	isbn = {1532-5946},
	journal = {Memory \& Cognition},
	number = {4},
	pages = {716--730},
	title = {Expertise as mental set: The effects of domain knowledge in creative problem solving},
	url = {https://doi.org/10.3758/BF03211392},
	volume = {26},
	year = {1998}}

@article{deJong2010,
	author = {de Jong, Ton},
	date = {2010/03/01},
	doi = {10.1007/s11251-009-9110-0},
	id = {de Jong2010},
	isbn = {1573-1952},
	journal = {Instructional Science},
	number = {2},
	pages = {105--134},
	title = {Cognitive load theory, educational research, and instructional design: some food for thought},
	url = {https://doi.org/10.1007/s11251-009-9110-0},
	volume = {38},
	year = {2010}}

@article{wang2017bias,
  title={Bias against novelty in science: A cautionary tale for users of bibliometric indicators},
  author={Wang, Jian and Veugelers, Reinhilde and Stephan, Paula},
  journal={Research Policy},
  volume={46},
  number={8},
  pages={1416--1436},
  year={2017},
  publisher={Elsevier}
}

@book{berlyne1960,
  title={Conflict, Arousal and Curiosity},
  author={Berlyne, D. E.},
  year={1960},
  publisher={McGraw-Hill},
  address={New York},
  doi={10.1037/11164-000}
}

@article{Sharot2020,
author = {Sharot, T. and Sunstein, C. R.},
year = {2020},
pages = {14–19},
title = {How people decide what they want to know},
volume = {4},
journal = {Nature Human Behaviour},
doi = {doi: 10.1038/s41562-019-0793-1}
}

@article{kedrick2023,
	author = {Kedrick, Kara and Schrater, Paul and Koutstaal, Wilma},
	doi = {https://doi.org/10.1111/cogs.13253},
	eprint = {https://onlinelibrary.wiley.com/doi/pdf/10.1111/cogs.13253},
	journal = {Cognitive Science},
	number = {4},
	pages = {e13253},
	title = {The Multifaceted Role of Self-Generated Question Asking in Curiosity-Driven Learning},
	url = {https://onlinelibrary.wiley.com/doi/abs/10.1111/cogs.13253},
	volume = {47},
	year = {2023}}

@article{Golman2021,
author = {Golman, Russell and Loewenstein, George and Molnar, Andras and Saccardo, Silvia},
title = {The Demand for, and Avoidance of, Information},
journal = {Management Science},
volume = {68},
number = {9},
pages = {6454-6476},
year = {2022},
doi = {10.1287/mnsc.2021.4244},

URL = { 
    
        https://doi.org/10.1287/mnsc.2021.4244
    
    

},
eprint = { 
    
        https://doi.org/10.1287/mnsc.2021.4244
    
    

}

}

@article{Golman2018,
title = {Information gaps: A theory of preferences regarding the presence and absence of information.},
journal = {Decision},
volume = {5},
number = {3},
pages = {143–164},
year = {2018},
doi = {https://doi-org.ezp3.lib.umn.edu/10.1037/dec0000068},
author = {Golman, R. and Loewenstein, G.},
}

@article{fleming2001,
	author = {Fleming, Lee},
	doi = {10.1287/mnsc.47.1.117.10671},
	eprint = {https://doi.org/10.1287/mnsc.47.1.117.10671},
	journal = {Management Science},
	number = {1},
	pages = {117-132},
	title = {Recombinant Uncertainty in Technological Search},
	url = {https://doi.org/10.1287/mnsc.47.1.117.10671},
	volume = {47},
	year = {2001}}

@article{Zhou2018,
	author = {Zhou, Xiaoping and Liang, Xun and Zhao, Jichao and Zhang, Shusen},
	date = {2018/08/06},
	doi = {10.1038/s41598-018-30249-4},
	id = {Zhou2018},
	isbn = {2045-2322},
	journal = {Scientific Reports},
	number = {1},
	pages = {11749},
	title = {Cycle Based Network Centrality},
	url = {https://doi.org/10.1038/s41598-018-30249-4},
	volume = {8},
	year = {2018}}

@article{Fan2021,
	author = {Fan, Tianlong and L{\"u}, Linyuan and Shi, Dinghua and Zhou, Tao},
	date = {2021/12/20},
	doi = {10.1038/s42005-021-00781-3},
	id = {Fan2021},
	isbn = {2399-3650},
	journal = {Communications Physics},
	number = {1},
	pages = {272},
	title = {Characterizing cycle structure in complex networks},
	url = {https://doi.org/10.1038/s42005-021-00781-3},
	volume = {4},
	year = {2021}}

@article{li2025innovation,
  title={Innovation by Displacement},
  author={Li, Linzhuo and Lin, Yiling and Wu, Lingfei},
  journal={arXiv preprint arXiv:2512.03723},
  year={2025}
}

@article{cui2022aging,
  title={Aging scientists and slowed advance},
  author={Cui, Haochuan and Wu, Lingfei and Evans, James A},
  journal={arXiv preprint arXiv:2202.04044},
  year={2022}
}

@article{bhaskar2020,
  author = {Bhaskar, Dhananjay and Zhang, William Y. and Wong, Ian Y.},
  title = {Topological data analysis of collective and individual epithelial cells using persistent homology of loops},
  journal = {Soft Matter},
  volume = {17},
  number = {17},
  pages = {4653--4664},
  year = {2021},
  doi = {10.1039/d1sm00072a}
}

@article{myers2019,
  title = {Persistent homology of complex networks for dynamic state detection},
  author = {Myers, Audun and Munch, Elizabeth and Khasawneh, Firas A.},
  journal = {Phys. Rev. E},
  volume = {100},
  issue = {2},
  pages = {022314},
  numpages = {14},
  year = {2019},
  month = {Aug},
  publisher = {American Physical Society},
  doi = {10.1103/PhysRevE.100.022314},
  url = {https://link.aps.org/doi/10.1103/PhysRevE.100.022314}
}

@article{huang1991optical,
  title={Optical coherence tomography},
  author={Huang, David and Swanson, Eric A and Lin, Charles P and Schuman, Joel S and Stinson, William G and Chang, Warren and Hee, Michael R and Flotte, Thomas and Gregory, Kenton and Puliafito, Carmen A and others},
  journal={science},
  volume={254},
  number={5035},
  pages={1178--1181},
  year={1991},
  publisher={American Association for the Advancement of Science}
}

@article{LaVail1972,
	author = {Jennifer H. LaVail and Matthew M. LaVail},
	doi = {10.1126/science.176.4042.1416},
	eprint = {https://www.science.org/doi/pdf/10.1126/science.176.4042.1416},
	journal = {Science},
	number = {4042},
	pages = {1416-1417},
	title = {Retrograde Axonal Transport in the Central Nervous System},
	url = {https://www.science.org/doi/abs/10.1126/science.176.4042.1416},
	volume = {176},
	year = {1972}}

@article{Loewenstein1994,
title = {The psychology of curiosity: A review and reinterpretation},
journal = {Psychological Bulletin},
volume = {116},
pages = {75–98},
year = {1994},
doi = {doi:10.1037/0033-2909.116.1.75},
author = {Loewenstein, G.},
}

@article{bollerslev1986generalized,
  title={Generalized autoregressive conditional heteroskedasticity},
  author={Bollerslev, Tim},
  journal={Journal of econometrics},
  volume={31},
  number={3},
  pages={307--327},
  year={1986},
  publisher={Elsevier}
}

@article{tversky1992advances,
  title={Advances in prospect theory: Cumulative representation of uncertainty},
  author={Tversky, Amos and Kahneman, Daniel},
  journal={Journal of Risk and uncertainty},
  volume={5},
  number={4},
  pages={297--323},
  year={1992},
  publisher={Springer}
}

@article{Kang2009,
title = {The wick in the candle of learning: Epistemic curiosity activates reward circuitry and enhances memory.},
journal = {Psychological Science},
volume = {20},
pages = {963–973},
year = {2009},
doi = {https://doi.org/10.1111/j.1467-9280.2009.02402.x},
author = {Kang, M. J and Hsu, M. and Krajbich, I. M. and Loewenstein, G. and McClure, S. M. and Wang, J. T. and Camerer, C. F.},
}

@article{bodelier2004,
author = {Bodelier, Paul and Laanbroek, Hendrikus (Riks)},
year = {2004},
month = {04},
pages = {265-77},
title = {Nitrogen as regulatory factor of methane oxidation in soils and sediments},
volume = {47},
journal = {FEMS microbiology ecology},
doi = {10.1016/S0168-6496(03)00304-0}
}

@article{hall2002,
author = {Hall, Mark and Dugan, Elizabeth and Zheng, Beiyao and Mishra, Aneil},
year = {2001},
month = {06},
pages = {613 - 639},
title = {Trust in Physicians and Medical Institutions: What Is It, Can It Be Measured, and Does It Matter?},
volume = {79},
journal = {Milbank Quarterly},
doi = {10.1111/1468-0009.00223}
}

@article{costin1971,
	author = {Frank Costin and William T. Greenough and Robert J. Menges},
	doi = {10.3102/00346543041005511},
	eprint = {https://doi.org/10.3102/00346543041005511},
	journal = {Review of Educational Research},
	number = {5},
	pages = {511-535},
	title = {Student Ratings of College Teaching: Reliability, Validity, and Usefulness},
	url = {https://doi.org/10.3102/00346543041005511},
	volume = {41},
	year = {1971}}

@book{kuhn1962,
  title={The structure of scientific revolutions},
  author={Kuhn, Thomas S.},
  year={1962},
  publisher={University of Chicago Press},
  place={Chicago}
}

@article{wu2026innovation,
  title={Is Innovation Becoming Less Disruptive? An Inventory of the Literature},
  author={Wu, Xiangting and Wu, Linhui and Park, Michael and Leahey, Erin and Funk, Russell J},
  journal={arXiv preprint arXiv:2602.05140},
  year={2026}
}

@article{becher1994significance,
  title={The significance of disciplinary differences},
  author={Becher, Tony},
  journal={Studies in Higher education},
  volume={19},
  number={2},
  pages={151--161},
  year={1994},
  publisher={Taylor \& Francis}
}

@book{popper1963,
  title={Conjectures and Refutations: {The} Growth of Scientific Knowledge},
  author={Popper, Karl R.},
  year={1963},
  note={Reprinted 2002},
  publisher={Routledge \& Kegan Paul},
  place={New York}
}

@article{cole2006,
    title = {Disciplinary knowledge revisited: The social construction of sociology},
    author={Cole, Stephen},
    journal = {American Sociologist},
    volume = {37},
    pages = {41–56},
    year = {2006},
}

@incollection{kolb1981,
  author    = {Kolb, David A.},
  title     = {Learning Styles and Disciplinary Differences},
  booktitle = {The Modern American College},
  editor    = {Chickering, Arthur W.},
  publisher = {Jossey-Bass},
  address   = {San Francisco, CA},
  year      = {1981}
}

@article{biglan1973,
  author  = {Biglan, Anthony},
  title   = {The Characteristics of Subject Matter in Different Academic Areas},
  journal = {Journal of Applied Psychology},
  volume  = {57},
  number  = {3},
  pages   = {195--203},
  year    = {1973}
}

@book{laudan1978,
  title={Progress and its problems: {Toward} a theory of scientific growth},
  author={Laudan, Larry},
  year={1977},
  publisher={University of California Press},
  place={Berkeley}
}

@book{lakatos_musgrave_1970, place={Cambridge}, title={Criticism and the growth of knowledge: Proceedings of the international colloquium in the philosophy of science, London, 1965}, volume={4}, publisher={Cambridge University Press}, year={1970}, author={Lakatos, Imre and Musgrave, Alan}}

@article{otter2017roadmap,
  title={A roadmap for the computation of persistent homology},
  author={Otter, Nina and Porter, Mason A and Tillmann, Ulrike and Grindrod, Peter and Harrington, Heather A},
  journal={EPJ Data Science},
  volume={6},
  number={1},
  pages={17},
  year={2017},
  publisher={Springer}
}

@techreport{cowen2019rate,
  author = {Cowen, Tyler and Southwood, Ben},
  series = {Working Paper},
  title = {Is the Rate of Scientific Progress Slowing Down?},
  year = {2019},
  institution={George Mason University}
}

@article{bloom2020ideas,
  title={Are ideas getting harder to find?},
  author={Bloom, Nicholas and Jones, Charles I and Van Reenen, John and Webb, Michael},
  journal={American Economic Review},
  volume={110},
  number={4},
  pages={1104--44},
  year={2020}
}

@article{jones2009burden,
  title={The burden of knowledge and the “death of the renaissance man”: Is innovation getting harder?},
  author={Jones, Benjamin F},
  journal={The Review of Economic Studies},
  volume={76},
  number={1},
  pages={283--317},
  year={2009},
  publisher={Wiley-Blackwell}
}

@article{arbesman2011quantifying,
  title={Quantifying the ease of scientific discovery},
  author={Arbesman, Samuel},
  journal={Scientometrics},
  volume={86},
  number={2},
  pages={245--250},
  year={2011},
  publisher={Springer}
}

@book{cowen2011great,
  title={The great stagnation: How America ate all the low-hanging fruit of modern history, got sick, and will (eventually) feel better: A Penguin eSpecial from Dutton},
  author={Cowen, Tyler},
  year={2011},
  publisher={Penguin}
}

@book{gordon2017rise,
  title={The rise and fall of American growth: The US standard of living since the civil war},
  author={Gordon, Robert J},
  volume={70},
  year={2017},
  publisher={Princeton University Press}
}

@article{wu2019large,
  title={Large teams develop and small teams disrupt science and technology},
  author={Wu, Lingfei and Wang, Dashun and Evans, James A},
  journal={Nature},
  volume={566},
  number={7744},
  pages={378--382},
  year={2019},
  publisher={Nature Publishing Group}
}

@article{uzzi2013atypical,
  title={Atypical combinations and scientific impact},
  author={Uzzi, Brian and Mukherjee, Satyam and Stringer, Michael and Jones, Ben},
  journal={Science},
  volume={342},
  number={6157},
  pages={468--472},
  year={2013},
  publisher={American Association for the Advancement of Science}
}

@article{rzhetsky2015choosing,
  title={Choosing experiments to accelerate collective discovery},
  author={Rzhetsky, Andrey and Foster, Jacob G and Foster, Ian T and Evans, James A},
  journal={Proceedings of the National Academy of Sciences},
  volume={112},
  number={47},
  pages={14569--14574},
  year={2015},
  publisher={National Acad Sciences}
}

@article{foster2015tradition,
  title={Tradition and innovation in scientists’ research strategies},
  author={Foster, Jacob G and Rzhetsky, Andrey and Evans, James A},
  journal={American Sociological Review},
  volume={80},
  number={5},
  pages={875--908},
  year={2015},
  publisher={Sage Publications Sage CA: Los Angeles, CA}
}

@article{wuchty2007increasing,
  title={The increasing dominance of teams in production of knowledge},
  author={Wuchty, Stefan and Jones, Benjamin F and Uzzi, Brian},
  journal={Science},
  volume={316},
  number={5827},
  pages={1036--1039},
  year={2007},
  publisher={American Association for the Advancement of Science}
}

@article{gardner2022toroidal,
  title={Toroidal topology of population activity in grid cells},
  author={Gardner, Richard J and Hermansen, Erik and Pachitariu, Marius and Burak, Yoram and Baas, Nils A and Dunn, Benjamin A and Moser, May-Britt and Moser, Edvard I},
  journal={Nature},
  volume={602},
  number={7895},
  pages={123--128},
  year={2022},
  publisher={Nature Publishing Group UK London}
}

@article{reimann2017cliques,
  title={Cliques of neurons bound into cavities provide a missing link between structure and function},
  author={Reimann, Michael W and Nolte, Max and Scolamiero, Martina and Turner, Katharine and Perin, Rodrigo and Chindemi, Giuseppe and D{\l}otko, Pawe{\l} and Levi, Ran and Hess, Kathryn and Markram, Henry},
  journal={Frontiers in computational neuroscience},
  volume={11},
  pages={48},
  year={2017},
  publisher={Frontiers}
}

@inproceedings{carriere2020perslay,
  title={Perslay: A neural network layer for persistence diagrams and new graph topological signatures},
  author={Carri{\`e}re, Mathieu and Chazal, Fr{\'e}d{\'e}ric and Ike, Yuichi and Lacombe, Th{\'e}o and Royer, Martin and Umeda, Yuhei},
  booktitle={International Conference on Artificial Intelligence and Statistics},
  pages={2786--2796},
  year={2020},
  organization={PMLR}
}

@article{love2023topological,
  title={Topological Convolutional Layers for Deep Learning},
  author={Love, Ephy R and Filippenko, Benjamin and Maroulas, Vasileios and Carlsson, Gunnar},
  journal={Journal of Machine Learning Research},
  volume={24},
  number={59},
  pages={1--35},
  year={2023}
}

@article{Zeng2021FreshTA,
  title={Fresh teams are associated with original and multidisciplinary research},
  author={An Zeng and Ying Fan and Zengru Di and Yougui Wang and Shlomo Havlin},
  journal={Nature Human Behaviour},
  year={2021},
  volume={5},
  pages={1314 - 1322},
  url={https://api.semanticscholar.org/CorpusID:233036821}
}

@article{Wagner2018InternationalRC,
  title={International research collaboration: Novelty, conventionality, and atypicality in knowledge recombination},
  author={Wagner, Caroline S. and Whetsell, Travis A. and Mukherjee, Satyam},
  journal={Research Policy},
  volume={48},
  number={5},
  pages={1260--1270},
  year={2019},
  doi={10.1016/j.respol.2019.01.002}
}

@article{Yang2024ExploringTC,
  title={Exploring team creativity: The nexus between freshness and experience},
  author={Wenlong Yang and Yang Wang},
  journal={J. Informetrics},
  year={2024},
  volume={18},
  pages={101588},
  url={https://api.semanticscholar.org/CorpusID:272438232}
}

@article{Ke2015DefiningAI,
  title={Defining and identifying Sleeping Beauties in science},
  author={Qing Ke and Emilio Ferrara and Filippo Radicchi and Alessandro Flammini},
  journal={Proceedings of the National Academy of Sciences},
  year={2015},
  volume={112},
  pages={7426 - 7431},
  url={https://api.semanticscholar.org/CorpusID:617138}
}

@article{Funk2017ADN,
  title={A Dynamic Network Measure of Technological Change},
  author={Russell J. Funk and Jason Owen-Smith},
  journal={Manag. Sci.},
  year={2017},
  volume={63},
  pages={791-817},
  url={https://api.semanticscholar.org/CorpusID:5981473}
}

@article{Park2023PapersAP,
  title={Papers and patents are becoming less disruptive over time},
  author={Michael Park and Erin Leahey and Russell J. Funk},
  journal={Nature},
  year={2023},
  volume={613},
  pages={138-144},
  url={https://api.semanticscholar.org/CorpusID:255466666}
}

@article{gebhart2020emergence,
  title={The Emergence of Higher-Order Structure in Scientific and Technological Knowledge Networks},
  author={Gebhart, Thomas and Funk, Russell J.},
  journal={arXiv preprint arXiv:2009.13620},
  year={2020}
}

@article{watts1999small,
  author = {Watts, Duncan J.},
  title = {Networks, Dynamics, and the Small-World Phenomenon},
  journal = {American Journal of Sociology},
  volume = {105},
  number = {2},
  pages = {493--527},
  year = {1999}
}

@article{uzzi2005collaboration,
  author = {Uzzi, Brian and Spiro, Jarrett},
  title = {Collaboration and Creativity: The Small World Problem},
  journal = {American Journal of Sociology},
  volume = {111},
  number = {2},
  pages = {447--504},
  year = {2005}
}

@article{cronin2001hyperauthorship,
  author  = {Cronin, Blaise},
  title   = {Hyperauthorship: A postmodern perversion or evidence of a structural shift in scholarly communication practices?},
  journal = {Journal of the American Society for Information Science and Technology},
  volume  = {52},
  number  = {7},
  pages   = {558--569},
  year    = {2001}
}

@article{milojevic2014principles,
  author  = {Milojevi{\'c}, Sta{\v s}a},
  title   = {Principles of scientific research team formation and evolution},
  journal = {Proceedings of the National Academy of Sciences},
  volume  = {111},
  number  = {11},
  pages   = {3984--3989},
  year    = {2014}
}

@article{azoulay2011incentives,
  author  = {Azoulay, Pierre and Graff Zivin, Joshua S. and Manso, Gustavo},
  title   = {Incentives and creativity: Evidence from the academic life sciences},
  journal = {The RAND Journal of Economics},
  volume  = {42},
  number  = {3},
  pages   = {527--554},
  year    = {2011}
}

@article{bikard2015exploring,
  author  = {Bikard, Micha{\"e}l and Murray, Fiona and Gans, Joshua S.},
  title   = {Exploring trade-offs in the organization of scientific work: Collaboration and scientific reward},
  journal = {Management Science},
  volume  = {61},
  number  = {7},
  pages   = {1473--1495},
  year    = {2015}
}

@article{lu2020contributorship,
  author  = {Lu, Chao and Zhang, Yingyi and Ahn, Yong-Yeol and Ding, Ying and Zhang, Chenwei and Ma, Dandan},
  title   = {Co-contributorship network and division of labor in individual scientific collaborations},
  journal = {Journal of the Association for Information Science and Technology},
  volume  = {71},
  number  = {10},
  pages   = {1162--1178},
  year    = {2020}
}

@article{li2025recombination,
  title   = {Can recombination displace dominant scientific ideas?},
  author  = {Li, Linzhuo and Lin, Yiling and Wu, Lingfei},
  journal = {arXiv preprint arXiv:2506.15959},
  year    = {2025}
}

@article{priem2022openalex,
  title={{OpenAlex}: A fully-open index of scholarly works, authors, venues, institutions, and concepts},
  author={Priem, Jason and Piwowar, Heather and Orr, Richard},
  journal={arXiv preprint arXiv:2205.01833},
  year={2022}
}

@article{lin2023scisci,
  title={{SciSciNet}: A large-scale open data lake for the science of science research},
  author={Lin, Zihang and Yin, Yian and Liu, Lu and Wang, Dashun},
  journal={Scientific Data},
  volume={10},
  pages={315},
  year={2023},
  doi={10.1038/s41597-023-02198-9}
}

@article{xu2020pubmed,
  title={Building a {PubMed} knowledge graph},
  author={Xu, Jian and Kim, Sunkyu and Song, Min and Jeong, Minbyul and Kim, Donghyeon and Kang, Jaewoo and Rousseau, Justin F. and Li, Xin and Xu, Weijia and Torvik, Vetle I. and Bu, Yi and Chen, Chongyan and Ebeid, Islam Akef and Li, Daifeng and Ding, Ying},
  journal={Scientific Data},
  volume={7},
  pages={205},
  year={2020},
  doi={10.1038/s41597-020-0543-2}
}

@article{herzog2020dimensions,
  title={Dimensions: Bringing down barriers between scientometricians and data},
  author={Herzog, Christian and Hook, Daniel and Konkiel, Stacy},
  journal={Quantitative Science Studies},
  volume={1},
  number={1},
  pages={387--395},
  year={2020},
  doi={10.1162/qss_a_00020}
}

@misc{patentsview2025,
  title={{PatentsView} bulk data files},
  author={{United States Patent and Trademark Office}},
  year={2025},
  howpublished={\url{https://patentsview.org}}
}

@incollection{maria2014gudhi,
  title={The {Gudhi} library: Simplicial complexes and persistent homology},
  author={Maria, Cl{\'e}ment and Boissonnat, Jean-Daniel and Glisse, Marc and Yvinec, Mariette},
  booktitle={Mathematical Software -- ICMS 2014},
  series={Lecture Notes in Computer Science},
  volume={8592},
  pages={167--174},
  publisher={Springer},
  year={2014},
  doi={10.1007/978-3-662-44199-2_28}
}

@misc{oat2026,
  title={Open Applied Topology ({OAT}): {oat\_python} and {oat\_rust}},
  author={{Open Applied Topology Project}},
  year={2026},
  howpublished={\url{https://github.com/OpenAppliedTopology}},
  note={Accessed August 2026}
}

@inproceedings{ogihara1997,
  author = {Ogihara, Mitsunori and Ray, Animesh},
  title = {Simulating {B}oolean circuits on a {DNA} computer},
  booktitle = {Proceedings of the First Annual International Conference on Computational Molecular Biology ({RECOMB} '97)},
  pages = {226--231},
  year = {1997},
  publisher = {ACM Press},
  address = {Santa Fe, New Mexico},
  doi = {10.1145/267521.267759}
}

@article{devolder2013,
  author = {De Volder, Michael F. L. and Tawfick, Sameh H. and Baughman, Ray H. and Hart, A. John},
  title = {Carbon Nanotubes: Present and Future Commercial Applications},
  journal = {Science},
  volume = {339},
  number = {6119},
  pages = {535--539},
  year = {2013},
  doi = {10.1126/science.1222453}
}

@article{giessibl2003,
  author = {Giessibl, Franz J.},
  title = {Advances in atomic force microscopy},
  journal = {Reviews of Modern Physics},
  volume = {75},
  number = {3},
  pages = {949--983},
  year = {2003},
  doi = {10.1103/RevModPhys.75.949}
}
\fi
\ifdefined\SIinmain\expandafter \fi

\end{document}